\documentclass[pra]{revtex4}
\usepackage[ansinew]{inputenc}
\usepackage{graphicx}
\usepackage{subfig}
\usepackage{amsmath}
\usepackage{amsthm}
\usepackage{bm}
\usepackage{layout}
\usepackage{float}
\usepackage{amsfonts}
\usepackage{amssymb}
\usepackage{txfonts}%
\usepackage{caption}
\usepackage[colorlinks]{hyperref}
\usepackage{color}
\theoremstyle{plain}

\theoremstyle{definition}
\newtheorem{thm}{Theorem}[section]
\newtheorem{defn}{Definition}[section]
\newtheorem{lem}{Lemma}[section]
\newtheorem{prop}{Proposition}[section]
\newtheorem{conj}{Conjecture}[section]

\newtheorem{pr}{Proof}[section]
\theoremstyle{remark}

\newcommand\id{\leavevmode\hbox{\small1\kern-3.3pt\normalsize1}}

\newcommand{\Tr}{\mbox{Tr}}

\newcommand{\RN}[1]{\textup{\uppercase\expandafter{\romannumeral#1}}}

\begin{document}

\newcommand{\braket}[2]{\langle #1|#2\rangle}
\newcommand{\cp}{\text{\footnotesize{$\prec$}}}
\newcommand{\cf}{\text{\footnotesize{$\succ$}}}
\newcommand{\ncp}{\text{\footnotesize{$\npreceq$}}}
\newcommand{\ncf}{\text{\footnotesize{$\nsucceq$}}}
\newcommand{\ind}{\text{\footnotesize{$\npreceq \nsucceq$}}}
\newcommand{\rarr}{\rightarrow}

\title{Causal and causally separable processes}

\author{Ognyan Oreshkov$^1$ and Christina Giarmatzi$^{1,2}$}

\affiliation{
$^1$QuIC, Ecole Polytechnique de Bruxelles, C.P. 165, Universit\'e Libre de Bruxelles, 1050 Brussels, Belgium\\
$^{2}$Centre for Engineered Quantum Systems, Centre for Quantum Computer and Communication Technology, School of Mathematics and Physics, University of Queensland, Brisbane, Queensland 4072, Australia}

\begin{abstract} 

The idea that events are equipped with a partial causal order is central to our understanding of physics in the tested regimes: given two pointlike events $A$ and $B$, either $A$ is in the causal past of $B$, $B$ is in the causal past of $A$, or $A$ and $B$ are space-like separated. Operationally, the meaning of these order relations corresponds to constraints on the possible correlations between experiments performed in the vicinities of the respective events: if $A$ is in the causal past of $B$, an experimenter at $A$ could signal to an experimenter at $B$ but not the other way around, while if $A$ and $B$ are space-like separated, no signaling is possible in either direction. In the context of a concrete physical theory, the correlations compatible with a given causal configuration may obey further constraints. For instance, space-like correlations in quantum mechanics arise from local measurements on joint quantum states, while time-like correlations are established via quantum channels. Similarly to other variables, however, the causal order of a set of events could be random, and little is understood about the constraints that causality implies in this case. A main difficulty concerns the fact that the order of events can now generally depend on the operations performed at the locations of these events, since, for instance, an operation at A could influence the order in which B and C occur in A's future. So far, no formal theory of causality compatible with such dynamical causal order has been developed. Apart from being of fundamental interest in the context of inferring causal relations, such a theory is imperative for understanding recent suggestions that the causal order of events in quantum mechanics can be indefinite. Here, we develop such a theory in the general multipartite case. Starting from a background-independent definition of causality, we derive an iteratively formulated canonical decomposition of multipartite causal correlations. For a fixed number of settings and outcomes for each party, these correlations form a polytope whose facets define causal inequalities. The case of quantum correlations in this paradigm is captured by the process matrix formalism. We investigate the link between causality and the closely related notion of causal separability of quantum processes, which we here define rigorously in analogy with the link between Bell locality and separability of quantum states. We show that causality and causal separability are not equivalent in general by giving an example of a physically admissible tripartite quantum process that is causal but not causally separable. We also show that there are causally separable quantum processes that become non-causal if extended by supplying the parties with entangled ancillas. This motivates the concepts of extensibly causal and extensibly causally separable (ECS) processes, for which the respective property remains invariant under extension. We characterize the class of ECS quantum processes in the tripartite case via simple conditions on the form of the process matrix. We show that the processes realizable by classically controlled quantum circuits are ECS and conjecture that the reverse also holds.
\end{abstract}

%\pacs{1315, 9440T}

\maketitle
\section{Introduction}

The possibility for dynamical and indefinite causal structures in quantum theory and more general probabilistic theories has recently attracted a great deal of interest, both from a foundational point of view and in the context of quantum information processing \cite{hardyqg, QGcomputers, Chiribella12, OCB, Colnaghi, Chiribella12b, Baumeler1, Nakago, Baumeler2, Araujo, brukner, Morimae, Ibnouhsein, brukner2, OC, OC2, Procopio, Lee, Araujo3, Baumeler3, Araujo4, Portmann, Cyril}. Motivated by the long standing search for a theory of quantum gravity, where the causal structure is expected to be dynamical as in General Relativity but fundamentally probabilistic in nature, as well as by the exploration of novel quantum architectures beyond the standard circuit model, operational ways of thinking about causal order in a probabilistic setting have provided new perspectives on quantum mechanics, its possible applications, and routes for potential extensions. 

A general framework for the study of correlations between local experiments without the assumption of a predefined causal order between them was proposed in Ref. \cite{OCB}. In this so called \textit{process framework}, each experiment is associated with an input and an output system between which an experimenter can perform different operations, but no specific assumption about the existence of a causal structure in which the experiments are embedded is made. When the experiments take place at fixed locations in a background space-time in circumstances defined without post-selection, the causal structure of space-time imposes signaling constraints on the correlations between the experiments. For example, there can be signaling from one experiment to another only if the former takes place in the past light cone of the latter, but no signaling between space-like separated locations or from the future to the past is possible. In Ref. \cite{OCB}, it was shown that if the local operations are described by quantum mechanics, it is possible to conceive correlations that are incompatible with any underlying causal structure. Such correlations allow two parties, Alice and Bob, to establish correlations that violate a \textit{causal inequality}, which is impossible if their operations take place in a causal order, even if that order is random. A similar possibility was subsequently shown to exist in a multipartite setting even when the local operations are purely classical \cite{Baumeler2}, which in the bipartite case is not possible \cite{OCB}. It is not known at present whether such joint processes could have a physical realization without post-selection, that is, whether one could prepare a setup that leads to correlations violating causal inequalities between separate experimenters who locally experience the validity of standard quantum mechanics.

Another peculiar effect that seems at odds with causality, which has a physical realization without post-selection, arises when local quantum operations are applied in an order that depends on the value of a variable prepared in a quantum superposition \cite{Chiribella12, Colnaghi, Chiribella12b, Araujo, Procopio}, a technique known as `quantum switch' \cite{Chiribella12}. This approach allows achieving certain tasks that are impossible if the quantum operations are applied in a definite causal order. In contrast to the violation of a causal inequality, however, this conclusion depends on the assumed description of the local operations and is {theory-dependent}. %Does this imply that the validity of quantum mechanics, not only locally but even globally, is incompatible with causality? Obviously the answer depends on the precise meaning of the question and in particular of the word `causality'.

So far, the analysis of these effects has relied on semi-rigorous considerations about what it means for a process to be compatible with `definite causal order'. A fully rigorous argument requires such considerations to be rooted in a clear notion of causality, which, however, in this background-independent setting has been lacking. Such a notion is expected to have a universal expression which can be applied in the context of any number of parties, but how to formulate it turns out to be a nontrivial problem. Simple considerations in the multipartite case show that the causal order of a set of local experiments should most generally be considered to be a random variable that can depend on the settings of these experiments. The latter possibility cannot be excluded since compatibly with our intuition of causality we can conceive of scenarios in which the setting in a given local experiment can influence the order in which other experiments take place in the future. In other words, causality should be expressed as a rule that constrains the joint conditional probabilities for the events in the local experiments and the causal order between them, allowing for the possibility that causal configurations unfold as a result of events in the past. A formal theory of such dynamical causal order is essential not only for understanding the subject of indefinite causal order in quantum mechanics or more general theories, but also for the problem of inferring causal structure beyond the classic paradigm of underlying deterministic variables and static causal relations \cite{Pearl}.

In this paper, we develop rigorous theory-independent and theory-dependent notions of causality in the process framework and characterize the structure and relations between the corresponding classes of processes they define. Section II is devoted to the theory-independent perspective, which contains our core result. We formalize the process framework in theory-independent terms and propose a definition of causality which allows for the possibility of dynamical causal order. We develop a number of concepts, such as multipartite signaling, reduced and conditional processes, and derive necessary and sufficient conditions for a process to be causal, which are expressed in the form of an iteratively defined canonical decomposition of the probabilities in the process. This decomposition can be understood as describing a causal `unraveling' of the events in the experiment in a sequence, showing that the proposed notion of causality yields the structure expected from intuition. Apart from being logically non-trivial, this result has important conceptual implications -- it presents us with an understanding of causal order as a random function on random events rather than the ordering of underlying locations in which events happen. This perspective is in the spirit of the idea of background independence in general relativity, according to which there are no underlying locations, but only events and the relations between them. In Section III, we focus on the quantum process framework, where we develop different theory-dependent notions of causality, which in principle have analogues in more general process theories too. Specifically, we investigate several possible generalizations of the bipartite notion of causal separability, which was previously defined heuristically in the bipartite case by postulating a particular form of the quantum process matrix \cite{OCB}. We show that this form can be understood as arising from the canonical decomposition of causal processes under the condition that each process in this decomposition is a valid quantum process. We define the multipartite concept based on this principle. We show that the sets of causal and causally separable processes are not equivalent in the multipartite case, by giving an explicit example of a class of processes that are causal but not causally separable. This example is based on the `quantum switch' technique discussed earlier. We also show that, surprisingly, there exist causally separable (and hence causal) quantum processes that become non-causal if extended by supplying the parties with an entangled input ancilla. This example of `activation of non-causality' is constructed based on a suitable modification of the non-causal process matrix of Ref.~\cite{OCB}. This observation motivates the concepts of \textit{extensibly causal} and \textit{extensibly causally separable} (ECS) processes, for which the respective property remains invariant under extension with arbitrary input ancillas. We derive a characterization of the class of ECS quantum processes in the tripartite case in terms of simple conditions on the form of the process matrix, which generalize the known form of bipartite causally separable process matrices. In the bipartite case, causal separability and extensible causal separability are equivalent, hence the class of ECS processes can be regarded as another possible multipartite generalization of the previously known bipartite concept. Finally, we consider the class of processes realizable by classically controlled quantum circuits, which we show is inside the class of ECS processes. These, too, are equivalent to the causally separable processes in the bipartite case and provide a possible multipartite generalization based on a different principle. We conjecture that the processes that can be obtained by classically controlled quantum circuits are equivalent to the ECS processes, and hence are described by process matrices obeying the simple conditions we have derived. We provide arguments in favor of this conjecture based on analysis in the tripartite case. In Section IV, we summarize our results and discuss future research directions. 

\section{The process framework} 

\subsection{General processes}

The process framework introduced in Ref.~\cite{OCB} describes probabilities for the outcomes of local experiments associated with different parties, Alice, Bob, Charlie, etc., performed in abstract circumstances defined without assuming the existence of a global causal order between the experiments, but only a local order of the events in each of them. Each local experiment can be thought of as performed inside an isolated laboratory, where, at a given instant, an input system is received in the laboratory, it is subject to some operation that yields one of a set of possible outcomes, and, at a given later instant, an output system is sent out of the laboratory. The input and output systems are assumed to provide the only means of information exchange between events in the laboratory and any events in the rest of the whole experiment. The framework in Ref.~\cite{OCB} was developed for the case where the local experiments are described by standard quantum mechanics, under a set of specific assumptions. These assumptions are that the joint probabilities of the outcomes of the local experiments are non-contextual functions of the transformations (described by completely positive (CP) maps) associated with the local outcomes, and that the local experiments can be extended to act on ancillas prepared in any joint quantum state. %However, for the purposes of understanding the operational essence of the idea of local experiments without predefined causal order, and possibly applying it in the context of more general probabilistic theories, it is desirable to formulate the framework in purely operational terms without reference to a concrete theory.

There is a straightforward way in which an analogous theory can be formulated starting from any generalized operational probabilistic theory that has a formulation in the circuit framework \cite{HardyCircuit,CDP2, CDP, Hardy2} following the construction in Ref.~\cite{OCB}. Indeed, the concepts of \textit{transformation} and \textit{state} are defined for any such theory, and so is the idea of a composite system that is employed in the notion of adding an ancilla. [Note that the representation of the quantum process framework \cite{OCB} in terms of \textit{process matrices} (see Section \ref{SectionII}) is built around the Choi-Jamio{\l}kowski isomorphism \cite{jam,choi}, which may not be available for arbitrary theories, but this concerns the representation of the framework.] However, the above assumptions underlying the extension from a circuit theory to a process theory, albeit arguably natural, are by no means mandatory. For example, one can conceive of extensions of quantum theory in which the joint probability distributions are contextual, but nevertheless for each single party the marginal probabilities are non-contextual and consistent with standard quantum mechanics. One can also conceive of theories in which the allowed non-signaling ancillary resources are not quantum states, although they give valid non-contextual probabilities for the outcomes of any combination of local quantum measurements \cite{popt}. It is therefore of interest to formulate a general process framework in operational terms without additional assumptions about how that framework may be related to theories expressed in the circuit framework. This is important also for the question of understanding the concept of \textit{causal inequality} introduced in Ref.~\cite{OCB}, which tests the compatibility of a process with an underlying causal structure in theory-independent terms. 

To this end, we will describe each local experiment, say that of Alice, by two variables -- a \textit{setting} $s^A$, and an \textit{outcome} $o^A_s$ for that setting. What these variables are supposed to correspond to in practice will be discussed below. The possible settings for a given local experiment are assumed to belong to some set $S^A$, and the outcomes for each value $s^A$ of the setting to a set $O^A_s$. Since we can formally extend the possible outcomes for each setting with fictitious outcomes that never occur, without loss of generality we can assume that the sets $O^A_s$ are identical for all $s^A\in S^A$, i.e., $O^A_s\equiv O^A$. A particular \textit{event} in Alice's laboratory is thus described by a pair of variables $(s^A,o^A)\in S^A\times O^A$. An operation is a collection of possible events $\{(s^A,o^A)\}_{o^A\in O^A}$ for a fixed value of $s^A\in S^A$. The very occurrence of the local experiments, as well as the circumstances in which they take place, would be conditional on some variable that we will denote by $w^{A,B,C,\cdots}$, which belongs to some set $\Omega^{A,B,C, \cdots}$ of possible such variables. What the variable $w^{A,B,C,\cdots}$ is supposed to correspond to in practice will also be discussed below. \\

%\comment{This is how it looks if I change the remaining parts where still A, B notation is used for the parties. However, it's not necessary to change the notation from this part of the paper on. We can only change it starting from the multipartite signaling, as it is at the moment. Also, I haven't touched the Appendix (although most notations might have changed automatically from the Find&Replace function).}

\begin{defn}
\textbf{(Process):} Mathematically, we define a process $\mathcal{W}^{A,B, \cdots}$ for a set of local experiments (or parties) $\mathcal{S}=\{A,B, \cdots\}$ as the collection of conditional probabilities
\end{defn}

\begin{gather}
\label{eq:process}
\mathcal{W}^{A,B,\cdots}\equiv\{{P}(o^A,o^B , ...|s^A,s^B , ... , w^{A,B,\cdots})\}, \\
o^X \in O^X, s^X\in S^X, \hspace{0.1cm} X\in \mathcal{S}, \nonumber
\end{gather}
for a given value of $w^{A,B,\cdots}\in \Omega^{A,B,\cdots}$.\\

\begin{defn}
\textbf{(Trivial process):} For the purposes of expressing more succinctly certain conditions later, it is convenient to allow the set of local experiments $\mathcal{S}=\{A,B, \cdots\}$ in the definition to be the empty set $\{  \}$ as a special case. In that case, the corresponding process will be referred to as the \textit{trivial process}. We define it to consist of a single probability -- that for the trivial outcome given the trivial setting -- which is equal to $1$.
\end{defn}

In abstract terms, a theory in the process framework is specified by listing the different types of input and output systems, all possible settings and outcomes between input and output systems of specific types, all possible variables  $w^{A,B,\cdots}$ for which we have a valid occurrence of a set of local experiments $\mathcal{S}=\{A,B, \cdots\}$, and the corresponding processes (\ref{eq:process}). Similarly to operational probabilistic theories in the circuit framework \cite{HardyCircuit, CDP2, CDP, Hardy2}, it is understood that equivalence classes of the variables $s^X$, $o^X$, and $w^{A,B,\dots}$, with regard to the probabilities (\ref{eq:process}) are taken, and these variables are identified with their equivalence classes.

But what are these variables supposed to describe in practice? In Refs.~\cite{OC,OC2}, it was argued that there are two main ideas that underlie the concept of operation in the standard circuit framework for operational probabilistic theories \cite{HardyCircuit, CDP2, CDP, Hardy2}. The first one, termed the \textit{closed-box assumption}, is the idea that the input and output systems of an operation are the only means of information exchange responsible for the correlations between the outcomes of that operation and the outcomes of other operations in the global experiment. The second idea,  termed the \textit{no-post-selection criterion}, which makes sense assuming a predefined notion of temporal ordering as in the standard circuit formulation, is that the variable that defines an operation, or the setting $s^X$, can be known with certainty before the time of interaction with the input system unconditionally on any events in the future. 

Since no predefined global time is assumed in our picture, the latter condition will be imagined to hold only with respect to the local temporal sequence of events observed by each experimenter. Furthermore, we will assume that the variable $w^{A,B,\cdots}$ that defines the global setup in which the individual experiments take place is also obtained without post-selection. We can make sense of this idea by imagining that the variable is associated with an event that fits within each of the local temporal frames of the experimenters and is such that it occurs before any of them receives the input system. %For example, we can imagine that each experimenter receives a notification that the setup is prepared and the experiment will be performed prior to receiving the input system according to her/his local time. This does not constitute any exchange of information that could violate the closed-box assumption, because that assumption is required to hold in the context of the global experiment, i.e., conditionally on variables that define that the experiment is performed. 
We will call processes that describe experiments of this kind \textit{pre-selected processes}. (For a generalization that admits post-selection, see Ref.~\cite{OC}).

For the rest of this paper, we will consider only pre-selected processes. We will drop the explicit specification `pre-selected' for brevity, and will refer to them simply as processes, unless we want to explicitly emphasize the assumption of pre-selection. We will also drop the explicit specification of the variable $w^{A,B,\cdots}$ on which the joint experiment is conditioned, and we will simply write $\mathcal{W}^{A,B, \cdots}\equiv\{p(o^A,o^B , ...|s^A,s^B , ... )\}$, keeping in mind that every process describes circumstances defined by such a variable and hence all probabilities we consider are implicitly conditional on such a variable.

\subsection{Causal processes}

In the circuit framework for operational probabilistic theories, causality is defined as the property that the probability distribution over the outcomes of a given operation in a circuit do not depend on what operations take place in the absolute future or absolute elsewhere \cite{Eddington} of that operation as defined by the \textit{strict partial order} (SPO) of the circuit composition \cite{CDP2, CDP}. More specifically, every circuit describes a set of operations taking place at the vertices of a directed acyclic graph, whose directed edges (the circuit `wires') correspond to systems that go from one operation to another. Such a graph defines a SPO on the operations in a circuit (a precise definition of SPO is given below) -- one operation is in the \textit{absolute past} of another (equivalently, the latter is in the \textit{absolute future} of the former) if there exists a directed path from the former to the latter through the graph. If there is no directed path connecting two operations, we say that one is the \textit{absolute elsewhere} of the other.  If we imagine that there is a local experiment taking place at every vertex of such a graph, the property of causality says that the probabilities for the outcomes of local experiments that are in the causal past or causal elsewhere of a given local experiment cannot depend on the setting of that experiment. A circuit theory that obeys this condition, such as standard quantum theory, is called causal, and for such a theory the SPO defined by the circuit composition can be interpreted as causal order \cite{CDP2,CDP}. This interpretation corresponds to the intuitive idea that, if the setting of a local experiment is regarded as up to the `free choice' of an experimenter, then any correlations between that setting and other variables must indicate a causal influence of the setting on those variables. From this perspective, causality can be understood as the condition that a variable can influence only variables in its immediate location or in its absolute future. 

In the process framework, we do not assume the existence of a given circuit in which the local experiments are embedded. Thus, there is no natural SPO with respect to which to define causality. Nevertheless, we may ask whether the probabilities described by a given process are compatible with the existence of a SPO with respect to which causality is satisfied. How to formulate this precisely, however, is not immediately clear because the process framework can describe situations in which the SPO may be random. For instance, it can describe the correlations between local experiments that can be embedded in different circuits according to some probability distribution. Clearly, if the SPO between the local experiments is random, it must be the case that conditionally on that SPO taking any particular value, the probabilities of the outcomes of the parties given their settings must obey the above notion of causality. This condition, however, is not sufficient to capture the idea of causality. For example, consider the local experiments of two parties, Alice and Bob, which are embedded at random in one of two possible causal circuits where they occur in different orders. The probabilities for all events and the specific circuit could be such that, conditionally on any particular circuit being realized, the joint probabilities of the outcomes of the parties given their settings obey the above notion of causality, but nevertheless the setting of Alice could be correlated with the circuit in which her experiment is embedded, and thereby with the SPO on the two local experiments. Intuitively, such a situation should be in conflict with causality, because if Alice's setting could not influence events that occur in the past, it should not influence whether or not Bob performs an operation in the past. The circuit notion of causality cannot be used to define such an independence from the past, because there the past is defined assuming a fixed circuit. This indicates that we need a more general notion of causality that imposes constraints on how the SPO on the local experiments can depend on the parties' settings. A simple possibility is to require that the SPO on the local experiments must be independent of the parties' setting. This condition, however, is too restrictive, because, compatibly with the idea of causality, we can conceive of scenarios where the setting of a given party influences the order in which other parties perform their experiments in that party's absolute future. Thus, a more sophisticated definition of causality is needed for the process framework. We next develop such a definition.

First, let us review the properties of SPO and introduce some terminology. A SPO on a nonempty set of local elements $\mathcal{S}= \{A,B,C, \cdots\}$ is a binary relation $\cp$ which satisfies the following conditions: \textit{(1)} \textit{irreflexivity} -- not $A\cp A$; \textit{(2)} \textit{transitivity} -- if $A\cp B$ and $B\cp C$, then $A\cp C$; \textit{(3)} \textit{anti-symmetry} -- if $A\cp B$, then not $B\cp A$. When two local experiments $A$ and $B$ satisfy $A \cp B$ (equivalently, $B \cf A$), we will say that $A$ is in the \textit{absolute past} of $B$, or that $B$ is in the \textit{absolute future} of $A$ \cite{Eddington}. It will be convenient to introduce the notation $A\ncp B$ (equivalently, $B\ncf A$), which means $A$$\neq$$B$ and not $A\cp B$, that is, $A$ and $B$ are different and $A$ is not in the absolute past of $B$ (equivalently, $B$ is not in the absolute future of $A$). We will also introduce the notation $A \ind B$, which means $A\ncp B$ and $A\ncf B$, that is, $A$ and $B$ are different and $A$ is neither in the absolute past nor in the absolute future of $B$ (and hence, $B$ is neither in the absolute past nor in the absolute future of $A$). In the case when $A \ind B$, we will say that $A$ and $B$ are \textit{absolutely independent}, or that $A$ is in the \textit{absolute elsewhere} \cite{Eddington} of $B$ (and similarly, $B$ is in the absolute elsewhere of $A$). A prototypical example of these relations is the causal order between the points in a Minkowski space-time -- the absolute past/future of a given point corresponds to the points in the past/future light-cone of this point, excluding the point itself, while the absolute elsewhere consists of the points that are space-like separated from the point.  

Note that if a set of elements $\mathcal{S}=\{A, B, \cdots\}$ is equipped with a SPO, the elements $X$ and $Y$ in any pair $(X,Y)\in \mathcal{S}\times \mathcal{S}$ are related by $X\cp Y$, $X \cf Y$, $X\ind Y$, or $X{\text=}Y$. The SPO on the set $\mathcal{S}=\{A, B, \cdots\}$ is equivalently described by the list of respective relations for each such pair, which we will denote by  $\kappa(A, B, \cdots)$. (This list obviously must respect the properties of SPO listed above.) Since for pairs $(X,X)$ of identical elements this relation is trivially $X=X$, when we explicitly describe $\kappa(A, B, \cdots)$, we will only list the pairwise relations for all pairs of distinct elements of the set (if any). Note that this description is generally redundant due to the transitivity of SPO. %Typically, the term `(partial) order' is used for the relations $\cp$ or $\cf$, while $\ind$ is said to correspond to a lack of order. To avoid confusion, we will use the term  \textit{causal configuration} to signify the list of binary relations $\cp$, $\cf$, or $\ind$, for every pair in a set of local experiments $\{A, B, \cdots\}$. We will denote the causal configuration of a set of local experiments by $\kappa(A, B, \cdots)$, noting that specifying $\kappa(A, B, \cdots)$ is  equivalent to specifying the partial order on the experiments. 
If we are given the pairwise relations for a set $\mathcal{S}=\{A, B, \cdots\}$, we have, in particular, pairwise relations for any nonempty subset $\mathcal{S}'=\{X, Y, \cdots\}\subset \mathcal{S}$, i.e., a SPO $\kappa(A, B, \cdots)$ on $\mathcal{S}$ implies a SPO $\kappa(X, Y, \cdots)$ on $\mathcal{S}'\subset \mathcal{S}$, $\mathcal{S}'\neq \{ \}$.

As discussed above, the SPO $\kappa(A, B, \cdots)$ on a set of local experiments $\mathcal{S}=\{A, B, \cdots\}$ in terms of which causality would be defined can most generally be random and correlated with the events in these experiments. The notion of causality would impose constraints on the possible correlations. We want these constraints to formalize the following intuition about causality: \\

\textit{The choice of setting in a local experiment cannot affect the occurrence of events in the absolute past or absolute elsewhere of that experiment, nor the SPO on such events and the experiment in question.} \\

%\textit{The choice of setting in a local experiment cannot affect the outcomes of other local experiments that occur in the absolute past or absolute elsewhere of that experiment, nor the SPO on such experiments and the experiment in question.} \\

%In other words, we want to formalize the idea that at the time of a local experiment there exists a `past' (consisting of all events in the absolute past and absolute elsewhere of that experiment, as well as of the SPO on these events and the experiment in question), which cannot be thought of as having already happened and which is thus impossible to alter by any operation that the experimenter may perform at present. 
Since a process is defined by the conditional probabilities for the outcomes of the local experiments given their settings and does not assume the existence of probabilities for the settings, we will formulate the above constraint at the level of probabilities conditional on the settings. We define this as follows.\\

%Let $\Pi_{X}  = \{Y, Z, \cdots  \in \{A, B, \cdots \}:  Y\ncf X, Z\ncf X, \cdots\}$ denote the set of all local experiments that are not in the future of $X$, $X\in \{A, B, \cdots \}$ (this set in general depends on the causal configuration $\kappa(A, B, \cdots) $ of the local experiments). \\

%Note that above we defined the partial order to be associated with the local events in a given joint event, and not with the local experiments \textit{per se}, which are defined as collections of possible local events. Of course, the causal order between the local events in any particular joint event that occurs can be said to imply that the local experiments occur in that order. We can therefore speak about the causal order between experiments too, which will be useful in formulating the constraint of causality, but it is important to emphasize that the so defined causal order between the local experiments is an \textit{a priori} random dynamical variable that most generally can be correlated with the concrete combination of events that occur in the local experiments. This is the case even if the causal order between the local events in each joint event is a deterministic function of the joint event, since the joint event is itself random. A nontrivial example is the case of three local experiments, where the order between two of them may depend deterministically on the random event that takes place in the third local experiment in their past.

\begin{defn}
\textbf{(Causal process):} A process $\mathcal{W}^{A,B,\cdots} \equiv \{p(o^A,o^B, \cdots|s^A, s^B, \cdots) \}$ for a nonempty set of local experiments $\mathcal{S}=\{A, B, \cdots\}$ is called \textit{causal} if and only if there exists  a probability distribution
$p(\kappa(A,B,\cdots), o^A,o^B, \cdots|s^A, s^B, \cdots)$,  $\sum_{\kappa(A,B,\cdots)} p(\kappa(A,B,\cdots), o^A,o^B, \cdots|s^A, s^B, \cdots)= p(o^A,o^B, \cdots|s^A, s^B, \cdots)$, where the random variable $\kappa (A, B, \cdots)$ takes values in the possible SPOs on $\mathcal{S}=\{A, B, \cdots\}$, such that for every local experiment, e.g. $A$, every subset $\mathcal{X}= \{X, Y, \cdots \}$ of the rest of the local experiments, and every SPO $\kappa(A, X, Y, \cdots) \equiv \kappa(A, \mathcal{X})$ on the local experiment in question and that subset, we have
\end{defn}
\begin{gather}
p(\kappa(A, \mathcal{X}), A\ncp \mathcal{X}, o^{\mathcal{X}}| s^A, s^B, \cdots ) 
= p(\kappa(A, \mathcal{X}), A\ncp \mathcal{X}, o^{\mathcal{X}} |s^B, \cdots ).\label{causalorderDEF}
\end{gather} 
Here, $o^{\mathcal{X}}$ denotes collectively the outcomes of all local expriments in $\mathcal{X}$, and $A\ncp \mathcal{X}$ denotes the condition that all these local experiments are in the causal past or causal elsewhere of $A$ (i.e., $A\npreceq X,  A\npreceq Y, \cdots$, for all $X, Y, \cdots \in \mathcal{X}$). [The probability $p(\kappa(A, \mathcal{X}), A\ncp \mathcal{X}, o^{\mathcal{X}}| s^A, s^B, \cdots )$ is understood obtained from $p(\kappa(A,B,\cdots), o^A,o^B, \cdots|s^A, s^B, \cdots)$ by summing over all cases in which $\kappa(A,B,\cdots)$ is compatible with $\kappa(A, \mathcal{X})$ and $A\ncp \mathcal{X}$ (obviously, if $\kappa(A, \mathcal{X})$ itself is not compatible with $A\ncp \mathcal{X}$, the respective probability is zero) and over all possible outcomes of the local experiments in the complement of $\mathcal{X}$].\\

\textit{Remark.} A monopartite process is trivially causal. \\

For a process  $\mathcal{W}^{A,B,\cdots}$ that is causal, the binary relation $\cp$ of the SPO $\kappa (A, B, \cdots)$ can be interpreted as \textit{causal order}. In that case, we will use the terms `\textit{causal past}', `\textit{causal future}', `\textit{causal elsewhere}' and `\textit{causally independent}' in the place of `absolute past', `absolute future', `absolute elsewhere' and `absolutely independent', respectively. We will also refer to the list of pairwise relations $\kappa (A, B, \cdots)$ as the \textit{causal configuration} of the local experiments (in the case of a monopartite process, the causal configuration is trivial). 

Our goal next is to understand the structure of causal processes that arises from this definition and show that it corresponds exactly to what one expects from intuition. 

\subsection{Fixed-order causal processes, (no) signaling, reduced and conditional processes}

Before we consider the case of general causal processes, it will be instructive to investigate the special case of causal processes for which the causal configuration of the local experiments is fixed. As we will show, the constraints on such processes can be expressed via the concept of signaling, which we develop below. We also introduce several related concepts that will be of use later.\\

\begin{defn}
\textbf{(Fixed-order causal process):} A process $\mathcal{W}^{A,B,\cdots} \equiv \{p(o^A,o^B, \cdots|s^A, s^B, \cdots) \}$ is called \textit{fixed-order causal} if it is \textit{compatible} with a deterministic causal configuration, i.e., if it satisfies condition (\ref{causalorderDEF}) for a SPO $\kappa(A,B,\cdots)$ that takes a particular value $\kappa(A,B,\cdots)=\kappa_*(A,B,\cdots)$ with unit probability for all possible settings of the parties:
\end{defn}
\begin{gather}
p(\kappa(A,B,\cdots), o^A,o^B, \cdots|s^A, s^B, \cdots)= 0,\notag\\
\textrm{iff} \hspace{0.3cm}\kappa(A,B,\cdots)\neq \kappa_*(A,B,\cdots),\nonumber\\
 \forall s^A\in S^A, \forall s^B\in S^B, \cdots, \forall o^A\in O^A, \forall o^B\in O^B, \cdots.
\end{gather}\\

Since our definition of causal process implies that the setting of a local experiment cannot be correlated with the outcomes of local experiments that are in the absolute past or absolute elsewhere of that experiment, one may expect that for any fixed causal configuration of the local experiments, causality would impose constraints on the possibility for signaling between them, similarly to the case in the circuit framework. In the case of two experiments, signaling can be defined as follows:\\

\begin{defn}
\textbf{(Bipartite signaling):} We say that there is \textit{no signaling} from Alice ($A$) to Bob ($B$) in a bipartite process $\mathcal{W}^{A,B}$ if and only if the probabilities of the process satisfy
\end{defn}
\begin{gather}
p(o^B|s^B, s^A) \equiv  \sum_{o^A\in O^A} p(o^A, o^B|s^B, s^A) = p(o^B|s^B), \label{eq:signaling} \\
\forall  s^A\in S^A, s^B\in S^B, o ^B\in O^B, \nonumber
\end{gather}
i.e., the marginal probabilities for the outcomes of Bob are independent of the setting of Alice for any possible setting of Bob. Equivalently, we say that there \textit{is} signaling from Alice to Bob in the process $\mathcal{W}^{A,B}$ if and only if this condition is not satisfied.  \\

%\textit{Note:} This notion of signaling assumes pre-selected processes. \\

For a fixed-order causal process $\mathcal{W}^{A,B}$, where one of the relations $A\cp B$, $B\cp A$, or $A\ind B$ holds with unit probability for all settings of the parties, we can see that signaling is possible from one experiment to the other only if the former is in the causal past of the latter, which agrees with the notion of causality in the circuit framework \cite{CDP2, CDP}. Indeed, assume for example that $B\cp A$, i.e., $p(\kappa(A,B) =B\cp A|s^A,s^B)=1$, $\forall s^A\in S^A$, $\forall s^B\in S^B$ (and hence $p(\kappa(A,B) = A\cp B|s^A,s^B  )=0$ and $p(\kappa(A,B) = A\ind B|s^A,s^B)=0$, $\forall s^A\in S^A$, $\forall s^B\in S^B$). Then, we have
\begin{equation}
\begin{split}
p(o^B | s^A, s^B) &=  p({ A\cp B}, o^B | s^A, s^B )\ +\  p({ B\cp A} , o^B| s^A, s^B)\ +\ p({ A\ind B} , o^B| s^A, s^B ) \\
&=  p({ A\ncp B}, o^B | s^A, s^B ) = p({ A\ncp B}, o^B | s^B )= p(o^B| s^B), \\
&\forall  s^A\in S^A, s^B\in S^B, o ^B\in O^B, 
\end{split}
\end{equation}
i.e., there is no signaling from Alice to Bob. In a similar way, we see that if $A\cp B$, there is no signaling from Bob to Alice, while if $A\ind B$, there is no signaling from Alice to Bob and no signaling from Bob to Alice. 

In the case of more than two local experiments, the relevant generalization of the above notion of signaling may not be immediately obvious. Notice that if a given bipartite process $\mathcal{W}^{A,B}$ involves no signaling between $A$ and $B$, such a process is in principle compatible with the causal configuration $A\ind B$ (in fact, it is compatible with any causal configuration of the two parties). However, in the case of processes for more than two local experiments, even if there is lack of signaling between any pair of experiments for all possible settings of the rest of the experiments, the process may not be compatible with a causal configuration in which all experiments are causally independent.

\captionsetup[subfigure]{labelformat=empty}
\begin{figure}
\centering
\subfloat[a)]{\includegraphics[scale=.3]{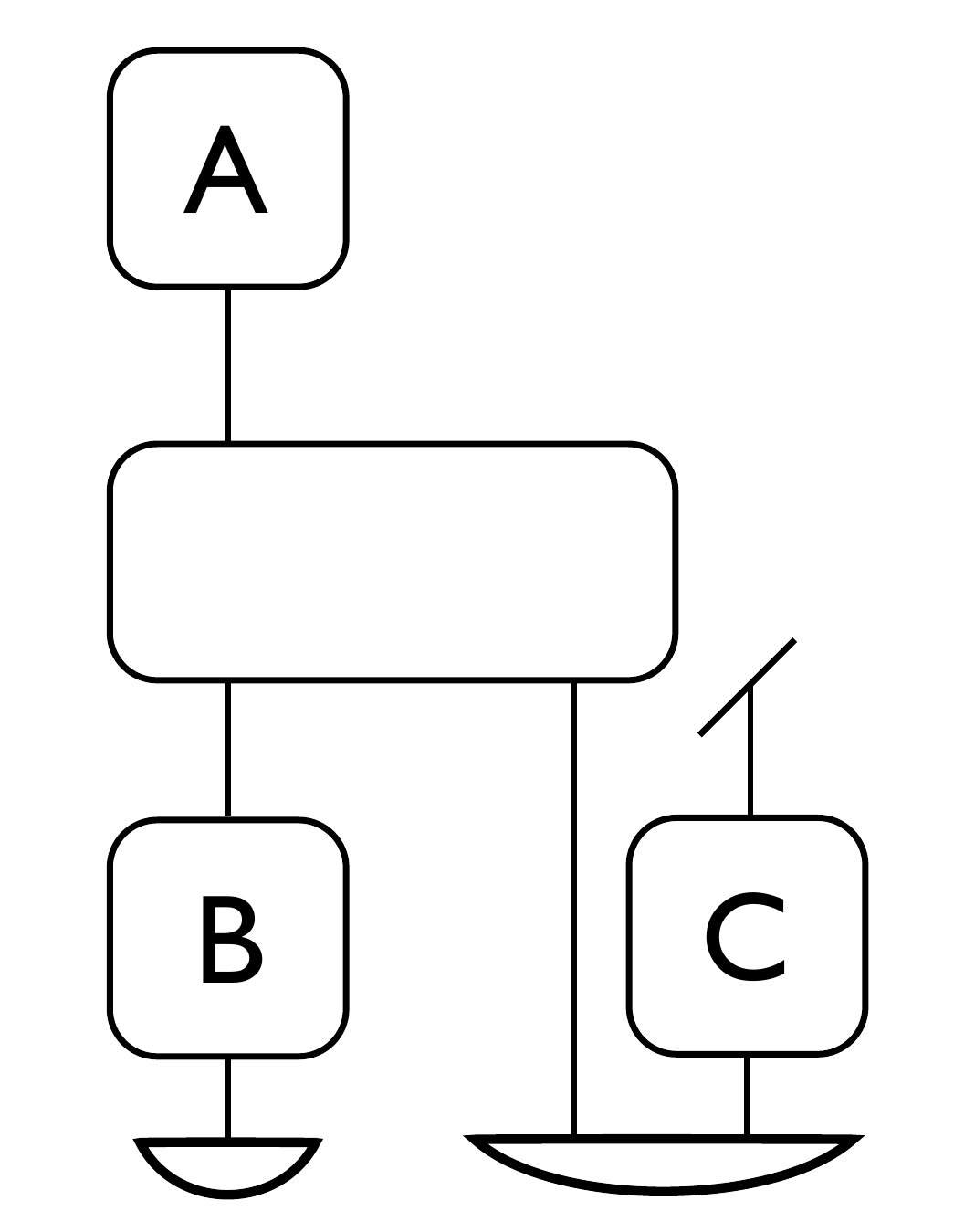}{\label{subfig:ex1}}}
\hspace{1cm}%
\subfloat[b)]{\includegraphics[scale=.3]{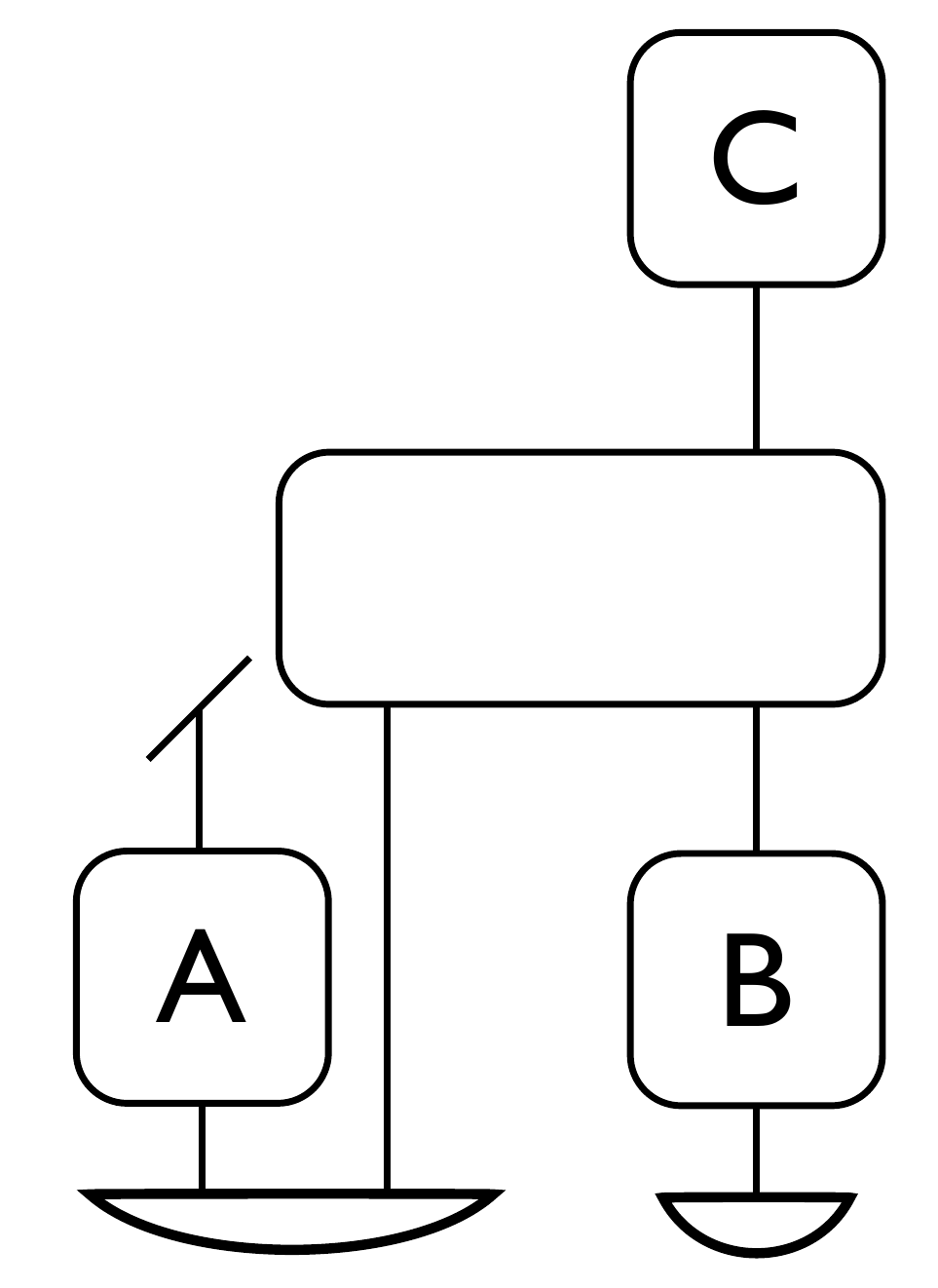}{\label{subfig:ex2}}}
\hspace{1cm}%
\subfloat[c)]{\includegraphics[scale=.3]{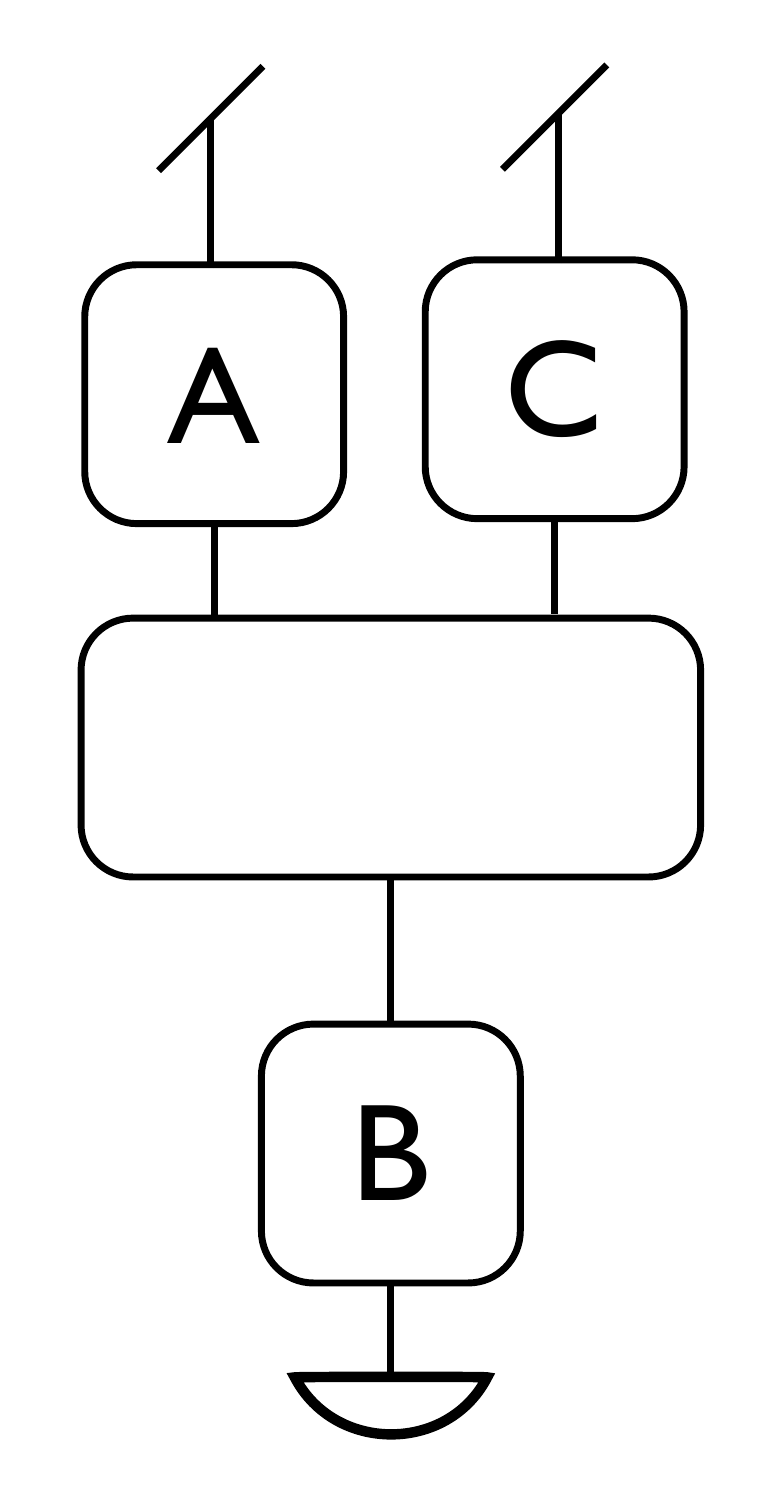}{\label{subfig:ex3}}}
\caption{Certain types of multipatite signaling correlations do not involve bipartite signaling and do not imply the existence of a causal connection between any particular pairs of channels. The example discussed in the text could arise from any of the mechanisms sketched here.}
\end{figure}

\begin{figure}[htb!]
\centering
\includegraphics[scale=.3]{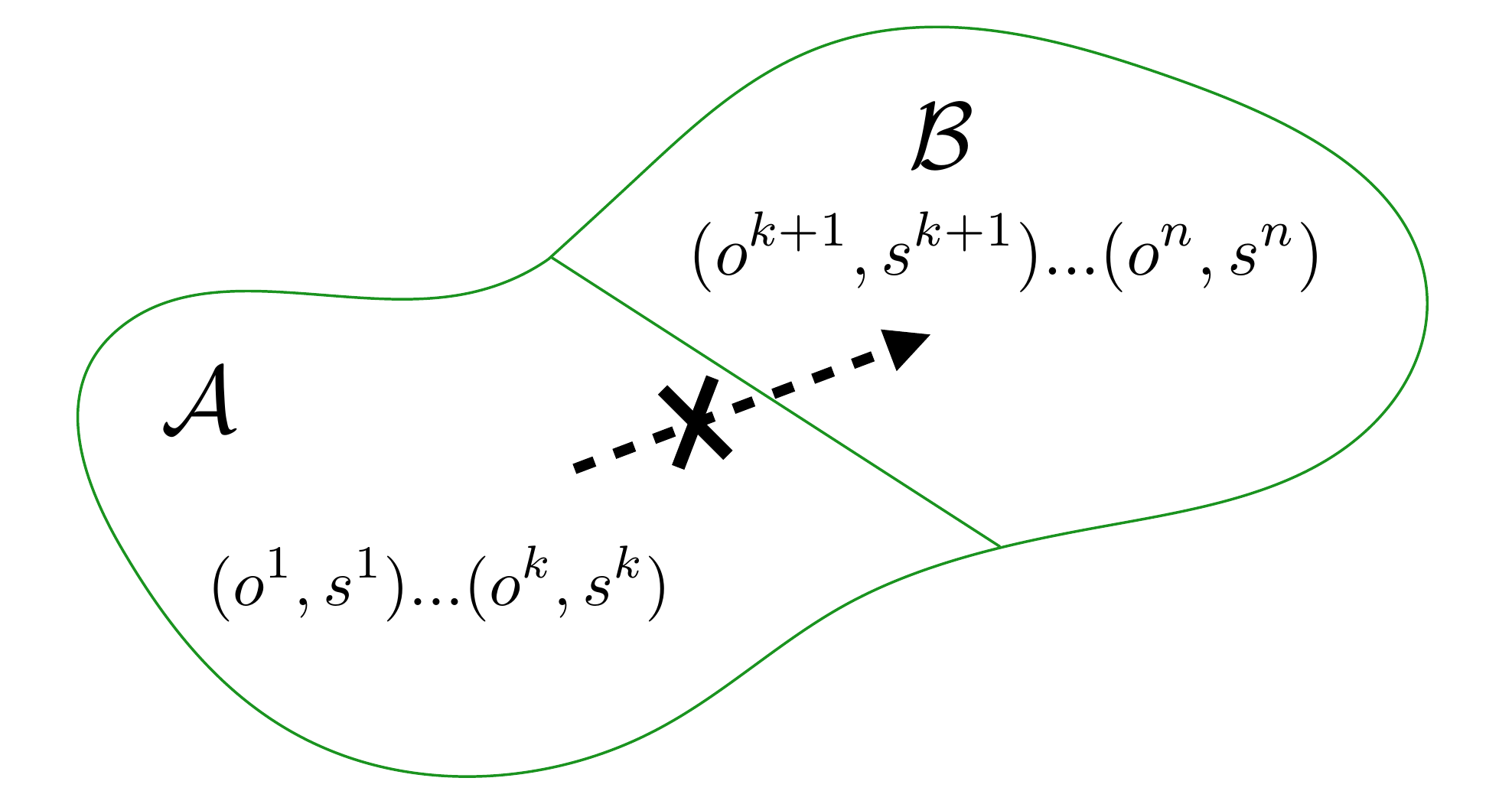}
\caption{Pictorial representation of the definition of multipartite signaling.}
\label{fig:multisignal}
\end{figure}

To see this, consider three local experiments performed by Alice, Bob, and Charlie, where each party's input and output systems are classical bits, and each party is allowed to perform any classical stochastic operation from the input bit to the output bit. Let the experiments of Bob and Charlie be causally independent, and let Alice's experiment be in the absolute future of Bob's experiment, but in the absolute elsewhere of Charlie's experiment (i.e., the causal configuration of the three parties is [$B\cp A$, $A\ind C$, $B\ind C$]). Imagine that Charlie receives his input system in one of the two possible states $0$ or $1$ with probability $1/2$, and depending on that state, Alice and Bob are in one of the following two scenarios. In the first scenario (say, when Charlie receives $0$), Bob receives a random bit as an input system, his output bit is sent unaltered into the input system of Alice, and Alice's output bit is discarded. In the second scenario (when Charlie receives $1$), Bob again receives a random input bit, but this time his output bit is flipped before sending it into Alice's input, and Alice's output bit is again discarded. In both cases, the output system of Charlie is discarded. Clearly, the described situation can be realized in agreement with a fixed causal configuration of the parties -- all we need to do is supply Bob with a random bit and correlate the channel from Bob to Alice with the input system of Charlie, discarding the outcomes of Alice and Charlie. The mechanism realizing this is sketched in Fig. \ref{subfig:ex1}. Note that the tripartite process corresponding to this scenario would involve no signaling from Bob to Alice in spite of the existence of a channel from Bob to Alice. This is the case irrespectively of what operation Charlie performs. Obviously, there can be no signaling from Alice to Bob either, since Alice operates in the future of Bob, nor can there be signaling between Alice and Charlie, or between Bob and Charlie, since Charlie is causally independent of both Alice and Bob. Thus, we have no signaling between any pair of parties, no matter what the setting of the third party is. Yet, the possible correlations between the parties cannot be realized if all parties are causally independent because if Alice and Charlie measure their input bits and collect the results of their measurements, they can infer the bit sent out by Bob, which is impossible if all parties are causally independent. We might say that in this case we have signaling from Bob to Alice and Charlie together. But intuitively, given the described scenario, this signaling should be from Bob to Alice only, since there is no channel connecting Bob's output system to Charlie's input. However, the latter conclusion is based on knowledge about the mechanism by means of which the correlations are established, or about the causal configuration of the parties, and does not follow solely from the correlations between them. Indeed, the tripartite joint probabilities for the outlined scenario are symmetric with respect to interchanging the roles of Alice and Charlie, and thus they could arise from a different mechanism in a situation where Alice is causally independent of both Bob and Charlie, and Charlie is in the causal future of Bob (Fig. \ref{subfig:ex2}). They could also arise from a channel from Bob to both Alice and Charlie (Fig. \ref{subfig:ex3}) which transforms Bob's output bit into either correlated or anti-correlated random input bits for Alice and Charlie. We therefore see that, at the level of the joint probabilities for the parties' experiments, there is no way of distinguishing between these different mechanism of information transmission, and hence no way of giving a definition of signaling among a proper subset of the parties that unambiguously captures the existence of such a mechanism. We can, however, give an unambiguous definition of lack of signaling between two complementary subsets of the parties (Fig. \ref{fig:multisignal}), as well as an associated notion of multipartite signaling, generalizing the bipartite case.\\\\

%\comment{I changed the notation of the parties from $1,\cdots, n$ to $1, \cdots, n$, to avoid the problem of super-superscript. I like it a lot better. The original notations can be taken from older versions.}

\begin{defn}
\textbf{(Multipartite signaling):} Consider an $n$-partite process $\mathcal{W}^{1,\cdots, n}$ for a set of local experiments $\mathcal{S}=\{1,\cdots, n\}$, $n=0,1,\cdots$ (in the case of $n=0$, this is understood as the empty set, and correspondingly the process is the trivial process). Let $\mathcal{A} = \{1, \cdots, k\}  $ and $\mathcal{B}=\{k+1, \cdots, n\}$, $0\leq k \leq n$, be two complementary subsets of the experiments, $\mathcal{A}\cup \mathcal{B} = \mathcal{S}$, $\mathcal{A}\cap \mathcal{B} = \{ \}$ (for simplicity, we take them to be the first $k$ and the next $n-k$ experiments, which can always be ensured by relabeling). We say that there is \textit{no signaling} from the subset $\mathcal{A}$ to the complementary subset $\mathcal{B}$ in the process $\mathcal{W}^{1,\cdots, n}$ if and only if
\end{defn}
\begin{gather}
p(o^{k+1}, \cdots, o^n|s^{1}, \cdots, s^n) \equiv  p(o^{k+1}, \cdots, o^n|s^{k+1}, \cdots, s^{n}), \label{Defsignaling}\\
\forall s^{j}\in S^{j},  o^{j}\in O^{j}, j=1,\cdots, n.\nonumber
\end{gather}
Equivalently, we say that there \textit{is} signaling from ($1$ {or}  $\cdots$ {or} $k$) to ($k+1$ {or} $\cdots$ {or} $n$) if and only if this condition is not satisfied.\\
%\comment{if you like the picture we can keep it and refer to it somehow}

%\begin{defn}
%\textbf{(Multipartite signaling):} Consider an $n$-partite process $\mathcal{W}^{1,\cdots n}$ for a set of local experiments $\mathcal{S}=\{1,\cdots, n\}$, $n=0,1,\cdots$ (in the case of $n=0$, this is understood as the empty set, and correspondingly the process is the trivial process). Let $\mathcal{A} = \{1, \cdots, k\}  $ and $\mathcal{B}=\{k+1, \cdots, n\}$, $0\leq k \leq n$, be two complementary subsets of the experiments, $\mathcal{A}\cup \mathcal{B} = \mathcal{S}$, $\mathcal{A}\cap \mathcal{B} = \{ \}$ (for simplicity, we take them to be the first $k$ and the next $n-k$ experiments, which can always be ensured by relabeling). We say that there is \textit{no signaling} from the subset $\mathcal{A}$ to the complementary subset $\mathcal{B}$ in the process $\mathcal{W}^{1,\cdots n}$ if and only if
%\end{defn}
%\begin{gather}
%p(o^{k+1}, \cdots, o^{n}|s^{1}, \cdots, s^{n}) \equiv  p(o^{k+1}, \cdots, o^{n}|s^{k+1}, \cdots, s^{n}), \label{Defsignaling}\\
%\forall s^{X^j}\in S^{X^j},  o^{X^j}\in O^{X^j}, j=1,\cdots, n.\nonumber
%\end{gather}
%Equivalently, we say that there \textit{is} signaling from ($1$ {or}  $\cdots$ {or} $k$) to ($k+1$ {or} $\cdots$ {or} $n$) if and only if this condition is not satisfied.\\

\textit{Remark.} There is no signaling from or to the empty subset. \\

Note that this definition only says whether there is signaling from one or more local experiments from a given subset to one or more local experiments from the complementary subset, but in the general case it does not identify pairs of experiments between which there is signaling. In the case of two experiments, the definition reduces to the notion of bipartite signaling defined earlier.\\ 

\begin{defn}
\textbf{(Non-signaling process):} A process $\mathcal{W}^{1,\cdots, n}$ for a set of local experiments $\mathcal{S}=\{1,\cdots, n\}$, $n=0,1,\cdots$, is called \textit{non-signaling} if and only if there is no signaling from $\mathcal{A}$ to $\mathcal{B}$ for any pair of complementary subsets $\mathcal{A}$ and $\mathcal{B}$ of $\mathcal{S}$.
\end{defn}

\textit{Remark.} Monopartite processes and the trivial process are non-signaling. \\

From the definition of causal process, one easily obtains the following relation between the existence of multipartite signaling among the local experiments described by a given process and the causal configuration of these experiments. \\

\begin{prop}\label{prop:2.1}
In an $n$-partite fixed-order process $\mathcal{W}^{1,\cdots, n}$, $n\geq 1$, compatible with a deterministic causal configuration $\kappa_*(1,\cdots,n)$, there can be signaling from ($1$ {or}  $\cdots$ {or} $k$) to ($k+1$ {or} $\cdots$ {or} $n$), only if at least one of $\{1, \cdots, k\}$ is in the absolute past of at least one of  $\{k+1, \cdots, n\}$ according to $\kappa_*(1,\cdots,n)$. 
\end{prop}

It turns out that we can formulate necessary and sufficient conditions for a process to be fixed-order causal, which are expressed entirely in terms of the condition stated in Proposition \ref{prop:2.1} applied to different subsets of the experiments. To formulate the conditions precisely, we will need to introduce the concept of \textit{reduced process}.\\

\begin{defn}
\textbf{(Reduced process):} Consider an $n$-partite process $\mathcal{W}^{1,\cdots, n}$, $n\geq 0$, for a set of local experiments $\mathcal{S}=\{1,\cdots n\}$. Let $ \mathcal{A} =\{1,  \cdots, k\} $ and $\mathcal{B} = \{k+1, \cdots, n\} $, $0\leq k < n$, be two complementary subsets of the experiments (specified up to relabeling), such that there is no signaling from $\mathcal{B}$ to $\mathcal{A}$. This means that
\end{defn}
\begin{gather}
p(o^{1}, \cdots, o^{k}| s^{1}, \cdots, s^{n})
= p(o^{1},  \cdots, o^{k}| s^{1},  \cdots, s^{k}),\\
\forall s^{j}\in S^{j},  o^{j}\in O^{j}, j=1,\cdots n,\nonumber
\end{gather}
i.e., we have well defined conditional probabilities $p(o^{1}, \cdots, o^{k}| s^{1}, \cdots, s^{k})$ for the experiments in $\mathcal{A}$. The collection of these probabilities will be called \textit{reduced process} for $\mathcal{A}$ and will be denoted by $\mathcal{W}^{\mathcal{A}}\equiv \mathcal{W}^{1,\cdots, k}$.\\

Note that if a multipartite process is a valid pre-selected process, any of its reduced processes is also a valid pre-selected process because it is defined conditionally on the same pre-selected event. Note also that a general multipartite process need not admit any reduced processes apart from the trivial process and itself, since it may involve signaling from every proper subset of the local experiments to its complementary subset.

Before we state the conditions for a process to be fixed-order causal, we introduce another concept that will be needed later. \\

\begin{defn}
\textbf{(Conditional process):} Consider an $n$-partite process $\mathcal{W}^{1,\cdots, n}$, $n\geq 0$, for a set of local experiments $\mathcal{S}=\{1,\cdots, n\}$. Let $ \mathcal{A} =\{1,  \cdots, k\} $ and $\mathcal{B} = \{k+1, \cdots, {n}\} $, $0\leq k < n$, be two complementary subsets of the experiments (specified up to relabeling), such that there is no signaling from $\mathcal{B}$ to $\mathcal{A}$ (and hence we can define a reduced process $\mathcal{W}^{\mathcal{A}}\equiv \mathcal{W}^{1,\cdots, k}$). For each fixed event $(s^1, o^1, \cdots s^k, o^k)$ in $\mathcal{A}$ for which $p(o^1, \cdots, o^k| s^1, \cdots, s^k)\neq 0$, consider the collection of conditional probabilities $\{p(o^{k+1}, \cdots, o^{n}| s^{k+1}, \cdots, s^{n},  s^1, o^1, \cdots, s^{k}, o^k ) \}$. These can be thought of as an $(n-k)$-partite process for $\mathcal{B}$ dependent on the event $(s^{1}, o^{1}, \cdots, s^k, o^k)$ in $\mathcal{A}$. The collection of these processes for all values of $(s^{1}, o^{1}, \cdots, s^k, o^k)$ for which $p(o^{1}, \cdots, o^k| s^{1}, \cdots, s^k)\neq 0$ will be called \textit{conditional process} and will be denoted by $\mathcal{W}^{\mathcal{B}|\mathcal{A}}\equiv \mathcal{W}^{k+1, \cdots ,n|1, \cdots, k}$. The relation between the whole process and the reduced and conditional processes can be written in the compact form
\end{defn}
\begin{gather}
\mathcal{W}^{\mathcal{A},\mathcal{B}} \equiv \mathcal{W}^{1,\cdots ,n} = \mathcal{W}^{k+1, \cdots ,n|1, \cdots, k} \circ\mathcal{W}^{1,\cdots ,k}\nonumber\\
 \equiv \mathcal{W}^{\mathcal{B}|\mathcal{A}}\circ \mathcal{W}^{\mathcal{A}},
\end{gather}
where the product $\circ$ between $\mathcal{W}^{\mathcal{B}|\mathcal{A}}$ and $ \mathcal{W}^{\mathcal{A}}$ denotes multiplication of the respective probabilities of these processes, when defined, for the same value of the event in $\mathcal{A}$: 
\begin{gather}
p(o^{1},  \cdots, o^{n}| s^{1},  \cdots, s^{n}) = p(o^{k+1}, \cdots, o^{n}| s^{k+1}, \cdots, s^{n},  s^{1}, o^{1}, \cdots, s^k, o^k ) \  p(o^{1},  \cdots, o^k| s^{1},  \cdots, s^k), 
\end{gather}
for $p(o^{1}, \cdots, o^k| s^{1}, \cdots, s^k)\neq 0$, and 
\begin{gather}
p(o^{1},  \cdots, o^{n}| s^{1},  \cdots, s^{n}) =0, 
\end{gather}
for $p(o^{1}, \cdots, o^k| s^{1}, \cdots, s^k)= 0$.\\

\begin{prop}\label{prop:2.2}
A process $\mathcal{W}^{1,\cdots, n}$ for a set of local experiments $\mathcal{S}=\{1,\cdots, n\}$, $n\geq 1$, is compatible with a deterministic causal configuration $\kappa_*(1,\cdots,n)$ of these experiments (and is thereby fixed-order causal) if and only if, for the assumed causal configuration, Proposition \ref{prop:2.1}   holds for the full process and all of its reduced processes for all bipartitions of the local experiments into two complementary subsets. The Proof \ref{pr:prop2.2} is given in the Appendix.
\end{prop}

We next turn to general causal processes, beginning with the bipartite case.

\subsection{Bipartite causal processes}

Consider a process $\mathcal{W}^{A,B}$ describing the local experiments of two parties, Alice and Bob. If the process is causal, there exist probabilities $p({ A\cp B}|s^A, s^B)$, $p(B\cp A|s^A, s^B)$, $p(A \ind B|s^A, s^B)$, with $p({A\cp B}|s^A, s^B)+p(B\cp A|s^A, s^B)+p(A \ind B|s^A, s^B) =1$. We can therefore write the joint probabilities of the process in the form
%\comment{I added cdot between the probabilities, and used "equation" and "split" environments for alignment, for the equations to look better. I did this for the whole subsection D. I like it. If you agree, I can do it for the whole paper. Also, added spaces after + throughout the whole text}
\begin{equation}
\begin{split}
 p(o^A,o^B|s^A,s^B  ) &= p({ A\cp B}|s^A, s^B) \ p(o^A,o^B|s^A,s^B,   {A\cp B}) \\
 &+\ p(B\cp A|s^A, s^B) \ p(o^A,o^B|s^A,s^B    , B\cp A)\\
 &+\ p(A \ind B|s^A, s^B) \ p(o^A,o^B|s^A,s^B, { A\ind B}),
\end{split}
\end{equation}
where each of the probability distributions $p(o^A,o^B|s^A,s^B,   {A\cp B})$, $p(o^A,o^B|s^A,s^B   ,B\cp A) $, and $p(o^A,o^B|s^A,s^B   ,A \ind B)$, is defined assuming that $p(A\cp B|s^A, s^B)\neq 0$, $p(B\cp A|s^A, s^B)\neq 0$, and $p(A \ind B|s^A, s^B)\neq 0$, respectively, otherwise that term is absent from the expansion. The definition of causality (\ref{causalorderDEF}) implies that $p({ A\cp B}|s^A, s^B) = p({ A\cp B}|s^A)$, $p(B\cp A|s^A, s^B) = p(B\cp A| s^B)$, $p({A\ind B}|s^A, s^B) = p({ A\ind B})$. Since the sum of these probabilities must be unity, we obtain $p({ A\cp B}|s^A) = p({ A\cp B})$, $p(B\cp A|s^B)= p(B\cp A)$, i.e., the causal configuration of the local experiments is independent of the parties' settings. Thus, the probabilities of a bipartite causal process $\mathcal{W}_{c}^{A,B}$ have the form
\begin{equation}
\begin{split}
 p(o^A,o^B|s^A,s^B) &= p(A\cp B) \ p(o^A,o^B|s^A,s^B,  { A\cp B}) \\
 &+\ p(B\cp A) \ p(o^A,o^B|s^A,s^B,  B\cp A)\\
 &+\ p({ A\ind B}) \ p(o^A,o^B|s^A,s^B,  { A\ind B}),
\end{split}
\label{eq:biconstraint}
\end{equation}
where the probability distributions $p(o^A,o^B|s^A,s^B,  { A\cp B})\equiv p( A\cp B, o^A,o^B|s^A, s^B) /p({ A\cp B})  $, $ p(o^A,o^B|s^A,s^B, B\cp A)\equiv p(B\cp A, o^A,o^B|s^A, s^B) /p(B\cp A)  $, and $p(o^A,o^B|s^A,s^B,  { A\ind B})\equiv p( A\ind B, o^A,o^B|s^A, s^B) /p({ A\ind B}) $, whenever defined, describe processes, 
which we will denote by $\mathcal{W}^{A\cp B}$, $\mathcal{W}^{B\cp A}$, and $\mathcal{W}^{A\ind B}$, respectively. (Note that we can imagine that the causal configuration $\kappa({A,B})$ taking values $A\cp B$, $B\cp A$, or $A \ind B$, is associated with an event in the past of both $A$ and $B$, i.e., the processes $\mathcal{W}^{A\cp B}$, $\mathcal{W}^{B\cp A}$, and $\mathcal{W}^{A\ind B}$, can be thought of as proper pre-selected processes.) The assumption of causality imposes conditions on these processes too. Specifically, it can be seen that each of them must obey a no-signaling constraint compatible with the concrete causal configuration it is conditioned on: the first one must involve no signaling from Bob to Alice, $p(o^A|s^A,s^B, { A\cp B}) =   p(o^A|s^A,  { A\cp B})$; the second one must involve no signaling from Alice to Bob, $p(o^B|s^A,s^B, B\cp A) =   p(o^B|s^B,  B\cp A)$; and the third one must involve no signaling in either direction, $p(o^A|s^A,s^B, { A\ind B}) =   p(o^A|s^A, { A\ind B})$, $p(o^B|s^A,s^B, { A\ind B}) =   p(o^B|s^B, { A\ind B})$, i.e., these are fixed-order causal processes. In a compact form, we can write 
\begin{gather}
\mathcal{W}_{c}^{A,B} =
p({ A\cp B})\   \mathcal{W}^{A\cp B}+\ p(B\cp A)\  \mathcal{W}^{B\cp A} +\   p({ A\ind B})\  \mathcal{W}^{A\ind B},
\label{eq:biconstraintW}
\end{gather}
i.e., a bipartite causal process has the form of a probabilistic mixture of processes that are compatible with the different mutually exclusive causal configurations of the parties (and correspondingly involve only one-way signaling in the respective direction, or no signaling). This form is not only necessary but also sufficient for a process to be causal because it explicitly gives a joint probability distribution $p(\kappa(A,B), o^A,o^B|s^A, s^B) = p( \kappa (A,B)) p (o^A,o^B|s^A, s^B, \kappa(A,B))$  that obeys the condition for causality (\ref{causalorderDEF}) when each conditional distribution $p (o^A,o^B|s^A, s^B, \kappa(A,B))$ obeys the no-signaling constraints compatible with $\kappa(A,B)$. 
Indeed, we have 
%\comment{here I removed the equations from the text and put it in an equation mode to be more readable.}
\begin{equation}
\begin{split}
p(A\ncp B, o^B |s^A, s^B) & = p(B\cp A, o^B |s^A, s^B) \ +\ p(A\ind B, o^B |s^A, s^B)\\ 
& = p(B\cp A)\  p (o^B |s^A, s^B, B\cp A)\ +\ p(A\ind B)\ p(o^B |s^A, s^B, A\ind B)\\ 
& = p(B\cp A)\  p (o^B |s^B, B\cp A)\ +\ p(A\ind B)\ p(o^B |s^B, A\ind B) \ = \ p(A\ncp B, o^B |s^B),
\end{split}
\end{equation}
and similarly $p(B\ncp A, o^A |s^A, s^B) =p(B\ncp A, o^A |s^A)$.

Since the non-signaling probabilities $p(o^A,o^B|s^A,s^B,  {A \ind B})$ are compatible with the one-way signaling constraints for the cases $A\cp B$ or $B\cp A$, we can also write the probabilities (\ref{eq:biconstraint}) in the {non-unique} form
\begin{equation}
 p(o^A,o^B|s^A,s^B) = p(w^{A\ncp B})\   p(o^A,o^B|s^A,s^B, w^{A\ncp B}) \ 
+\ p(w^{B\ncp A}) \  p(o^A,o^B|s^A,s^B  , w^{B\ncp A}),\label{eq:biconstraint2}
\end{equation}
where $w^{A\ncp B}$ and $w^{B\ncp A}$ are two mutually exclusive variables for which the experiments of Alice and Bob respect the relations $A\ncp B$ and $B\ncp A$, respectively, with the probabilities of these variables satisfying $p(w^{A\ncp B})+p(w^{B\ncp A})=1$.  In a compact form, this can be written
\begin{gather}
\mathcal{W}_{c}^{A,B} = q \ \mathcal{W}^{A\ncp B}+\ (1-q)\ \mathcal{W}^{B\ncp A}, \hspace{0.2cm} 0\leq q\leq 1,\label{eq:biconstraintW2}
\end{gather}
where $\mathcal{W}^{Y\ncp X} $ is a process that involves no signaling from $Y$ to $X$, i.e.,
\begin{gather}
\mathcal{W}^{Y\ncp X} = \mathcal{W}^{Y|X}\circ \mathcal{W}^X.
\end{gather}

The constraint (\ref{eq:biconstraintW2}) (equivalently, (\ref{eq:biconstraint2})) provides a means of testing whether a given bipartite process theory is compatible with causal order. For every fixed number of settings and fixed number of outcomes for each party, the joint probabilities satisfying Eq.~(\ref{eq:biconstraint2}) form a convex polytope, which is the convex hull of the polytope of probabilities that involve no signaling from Alice to Bob, and the polytope of probabilities that involve no signaling from Bob to Alice \cite{Branciard}. The non-trivial facets of this `causal polytope' define bipartite causal inequalities, similar to the one in Ref.~\cite{OCB}, whose violation by a given process theory indicates that the theory is not compatible with causal order. Note that a causal inequality does not need to be a facet of the causal polytope -- it may correspond to an external plane. For instance, the causal inequality of Ref.~\cite{OCB}, which concerns the case where one party has a binary input and a binary output while the other one has a quaternary input and a binary output, is not a facet of the respective causal polytope ~\cite{Araujo4}. One way of seeing this is to note that the derivation of the inequality in Ref.~\cite{OCB} only used certain consequences of the requirement that the causal configuration of the parties must be independent of the parties' settings, but not the full requirement. The bipartite causal polytope for binary inputs and binary outputs has been characterized by Branciard \cite{Branciard} (see Ref.~\cite{Araujo4}).

\subsection{The tripartite and $n$-partite causal processes}

\begin{figure}[!htb]
\centering
\includegraphics[scale=.4]{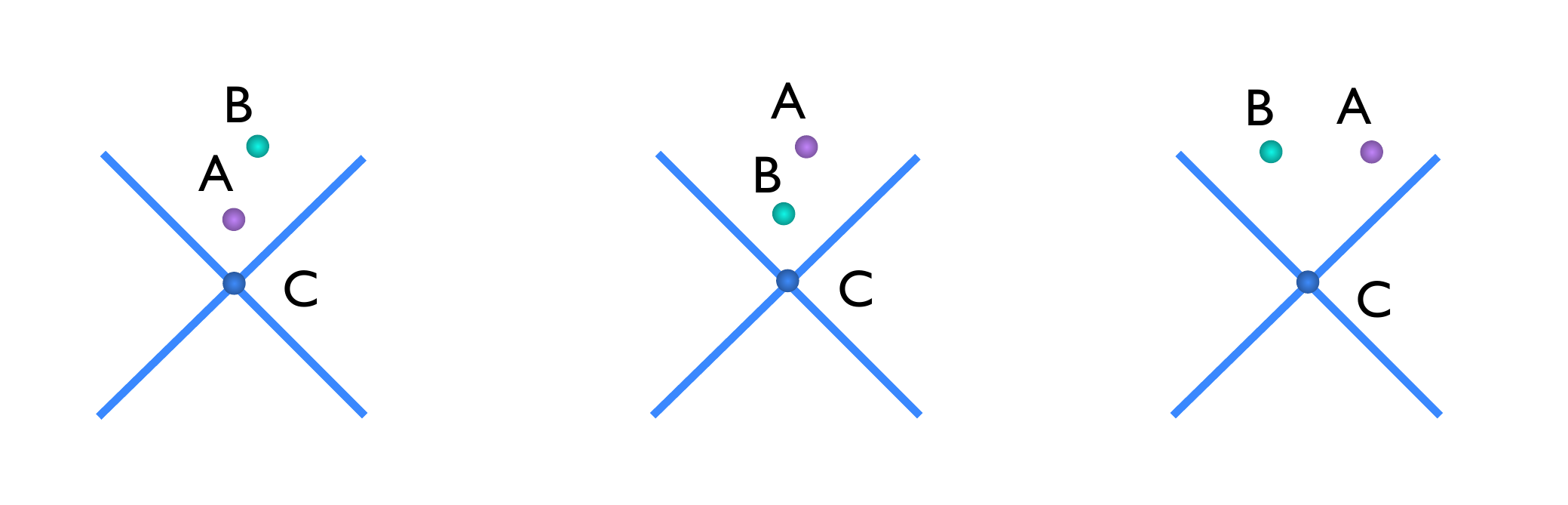}
%\captionsetup{width=0.8\textwidth}
\caption{In a causal setup where Charlie performs his experiment in the causal past of both Alice and Bob, the causal configuration of Alice and Bob may depend on the setting of Charlie.}
\label{fig:3Causal config}
\end{figure}

In the case of more than two parties, causal processes need not have the simple form of probabilistic mixtures of fixed-order causal processes with probability weights that are independent of the parties' settings. This is because, consistently with causality, we have the possibility that the causal configuration of a subset of the local experiments may depend on the settings of other local experiments in their past. For example, imagine that we have a tripartite experiment where the input and output systems of each party correspond to the internal (e.g., spin) degrees of freedom of a particle that enters the respective laboratory at a given instant and leaves it at a given later instant. The time at which each party receives her/his particle is determined by some predefined mechanism, which also governs any exchange of information taking place outside of the parties' laboratories. (Note that in order for the internal degrees of freedom of the particle to constitute the only means of information exchange between each local experiment and the rest of the experiment, the experiment should be so designed that no communication via the times of input or output of the parties is possible. For example, each party may be restricted not to possess any common time reference frame with the rest of the experiment and to perform her/his operation during a fixed time interval with a stopwatch.) In such a case, if Charlie receives a particle first, the operation that he applies on the system could affect the order in which Alice and Bob receive their particles afterwards, since we can conceive of a mechanism that selects different future scenarios for that order conditionally on the outcome of a measurement performed on the internal degrees of freedom of the particle coming out of Charlie's laboratory. This can result in the different scenarios depicted in Fig.~\ref{fig:3Causal config}. By construction, the outlined setup is compatible with the condition that the setting of each local experiment can be chosen independently of events in the causal past and causal elsewhere of that experiment, as well as of the causal configuration of such events and the experiment in question, so it would be associated with a valid causal process.   

Clearly, the  dependence of the causal configuration of the parties on the parties' settings cannot be arbitrary, because it must agree with causality. To formulate the constraints on this dependence, we will need to introduce some more terminology.

For any fixed causal configuration $\kappa(1, \cdots, n )$ of the local experiments $\mathcal{S}=\{1,\cdots,n\}$, there are local experiments that are in no-one else's causal future. The full set of such local experiments, $\{i,j,\cdots\} \subset \{1,\cdots, n\}$, will be referred to as the local experiments that are \textit{first}, or as the \textit{first consecutive set} and will be denoted by $[i, j, \cdots ]^{\RN{1}}$. Next, if the first consecutive set does not include all of the local experiments, there are local experiments whose causal past contains local experiments from $[i, j, \cdots ]^{\RN{1}}$ and only from $[i, j, \cdots ]^{\RN{1}}$. The full set of these will be referred to as the local experiments that are \textit{second}, or as the \textit{second consecutive set}, and will be denoted by $[k, l, \cdots ]^{\RN{2}}$. Then, if the first and second consecutive sets do not include all local experiments, there are local experiments whose causal past contains local experiments from both sets $[i, j, \cdots ]^{\RN{1}}$ and $[k, l, \cdots ]^{\RN{2}}$ and only from those sets. The full set of these will be referred to as the local experiments that are \textit{third}, or as the \textit{third consecutive set}, and will be denoted by $[p, q, \cdots ]^{\RN{3}}$, and so on.

The following proposition will play a central role in our derivation of the form of multipartite causal processes.

\begin{prop}\label{prop:2.3} 
Consider a causal process for $\mathcal{S}=\{1,\cdots, n\}$, $n\geq 1$, with an associated joint probability distribution $p(\kappa(1,\cdots, n), o^{1}, \cdots, o^n|s^1, \cdots, s^{n})$, where $\kappa(1,\cdots, n)$ are the causal configurations of the local experiments. The probability for the first $\textrm{K}$ consecutive sets to consist of specific local experiments, ${[1_{\RN{1}}, \cdots, n_{\RN{1}}]^{\RN{1}}}$, $\cdots$, ${[1_{{\textrm{K}}}, \cdots, n_{\textrm{K}}]^{{\textrm{K}}}}$, these experiments to have a specific causal configuration $\kappa( 1_{\RN{1}}, \cdots, {n_{\textrm{K}}})$, the experiments in the first ${\textrm{K}}-\RN{1}$ consecutive sets to have a specific set of outcomes $o^{1_{\RN{1}}}, \cdots, o^{n_{{\textrm{K}}-\RN{1}}}$, and a given (possibly empty) subset  $\{1_{\textrm{K}},\cdots,  g_{\textrm{K}}  \} \subset \{1_{\textrm{K}},\cdots,  n_{\textrm{K}} \}$ of  the local experiments in the ${\textrm{K}}^{th}$ set (given up to relabeling) to have specific outcomes $o^{1_{\textrm{K}}}, \cdots, o^{g_{\textrm{K}}}$, can depend non-trivially only on the settings of the local experiments indicated in the first ${\textrm{K}}-\RN{1}$ consecutive sets and the subset $\{1_{\textrm{K}},\cdots,  g_{\textrm{K}} \}$, 
\end{prop}
\begin{gather}
p(\kappa( 1_{\RN{1}}, \cdots, n_{\textrm{K}}), [1_{\RN{1}}, \cdots, n_{\RN{1}}]^{\RN{1}},
\cdots, [1_{\textrm{K}}, \cdots, n_{\textrm{K}}]^{{\textrm{K}}}, o^{1_{\RN{1}}}, \cdots, o^{g_{\textrm{K}}}|s^{1}, \cdots, s^{n}) \nonumber \\
= p(\kappa( 1_{\RN{1}}, \cdots, n_{\textrm{K}}), [1_{\RN{1}}, \cdots, n_{\RN{1}}]^{\RN{1}},
\cdots, [1_{\textrm{K}}, \cdots, n_{\textrm{K}}]^{{\textrm{K}}}, o^{1_{\RN{1}}}, \cdots, o^{g_{\textrm{K}}}|s^{1_{\RN{1}}}, \cdots, s^{g_{\textrm{K}}}),\label{inductive}
\end{gather}
where we define the $\textrm{0}^{th}$ set as the empty set. The Proof \ref{pr:prop:2.3} is given in the Appendix.

An important consequence of Proposition \ref{prop:2.3} is that the probability for a given set of local experiments to be first is independent of the settings of all parties (this is the case of ${\textrm{K}}=1$ and the subset $\{1_{\textrm{K}}, \cdots, g_{\textrm{K}}\}$ being empty). For example, consider the different causal configurations of three parties -- Alice ($A$), Bob ($B$), and Charlie ($C$) -- which are compatible with $[C]^{\RN{1}}$ (Fig.~\ref{fig:3Causal config}). Each of the individual configurations has a probability that may depend on the setting of Charlie, but the overall probability for Charlie to be first, i.e., for any one of these configurations to be realized (which is the sum of the probabilities for the individual configurations), is independent of the settings of all parties, including Charlie. This independence of the first consecutive set on the settings of all parties will play a key role in our characterization of the structure of multipartite causal processes. We will first develop the characterization for the case of three parties in order to illustrate the underlying principle, and then we will extend it to the general multipartite case.

\begin{table}
\begin{center}
\scalebox{1}{
  \begin{tabular}{| c | c |}
    \hline
    Groups of tripartite causal configurations whose probabilities are independent of the parties' settings,\\ defined by the set of parties that are first\\ \hline
     $[A]^{\RN{1}}$:  [$A\cp B$, $A\cp C$, $B\cp C$] or  [$A\cp B$, $A\cp C$, $C\cp B$] or [$A\cp B$, $A\cp C$, $B\ind C$] \\ \hline
     $[B]^{\RN{1}}$:  [$B\cp A$, $B\cp C$, $A\cp C$] or  [$B\cp A$, $B\cp C$, $C\cp A$] or [$B\cp A$, $B\cp C$, $A\ind C$] \\ \hline
     $[C]^{\RN{1}}$:  [$C\cp A$, $C\cp B$, $A\cp B$] or  [$C\cp A$, $C\cp B$, $B\cp A$] or [$C\cp A$, $C\cp B$, $A\ind B$] \\ \hline
     $[A,B]^{\RN{1}}$:  [$A\ind B$, $A\cp C$, $B\ind C$] or [$A\ind B$, $A\ind C$, $B\cp C$] or [$A\ind B$, $A\cp C$, $B\cp C$]\\ \hline
     $[A,C]^{\RN{1}}$:  [$A\ind C$, $A\cp B$, $B\ind C$] or [$A\ind C$, $A\ind B$, $C\cp B$] or [$A\ind C$, $A\cp B$, $C\cp B$]\\ \hline
     $[B,C]^{\RN{1}}$:  [$B\ind C$, $B\cp A$, $C\ind A$] or [$B\ind C$, $B\ind A$, $C\cp A$] or [$B\ind C$, $B\cp A$, $C\cp A$]\\ \hline
     $[A,B,C]^{\RN{1}}$:  [$A\ind B$, $B\ind C$, $A\ind C$] \\
    \hline
  \end{tabular}}
\end{center}
%\captionsetup{width=0.8\textwidth}
\caption{The mutually exclusive groups of tripartite causal configurations}
\label{tbl:the groups}
\end{table}

The groups of tripartite causal configurations compatible with the different possibilities for the first consecutive set of parties are listed in Table~\ref{tbl:the groups}. In terms of these possibilities, the probabilities of a tripartite causal process can be written 
\begin{equation}
\begin{split}
 p(o^A,o^B,o^C|s^A,s^B,s^C)  &=  p([A]^{\RN{1}})\ p(o^A,o^B,o^C|s^A,s^B,s^C, [A]^{\RN{1}}) \\
& +\ p([B]^{\RN{1}})\ p(o^A,o^B,o^C|s^A,s^B,s^C, [B]^{\RN{1}})\\
& +\ p([C]^{\RN{1}})\ p(o^A,o^B,o^C|s^A,s^B,s^C, [C]^{\RN{1}})  \\ 
& +\ p([A,B]^{\RN{1}})\ p(o^A,o^B,o^C|s^A,s^B,s^C, [A,B]^{\RN{1}}) \\
& +\ p([A,C]^{\RN{1}})\ p(o^A,o^B,o^C|s^A,s^B,s^C, [A,C]^{\RN{1}})  \\ 
& +\ p([B,C]^{\RN{1}})\ p(o^A,o^B,o^C|s^A,s^B,s^C, [B,C]^{\RN{1}}) \\ 
& +\ p([A,B,C]^{\RN{1}})\ p(o^A,o^B,o^C|s^A,s^B,s^C, [A,B,C]^{\RN{1}}), 
\end{split} 
\label{tripart1'}
\end{equation}
where
\begin{gather} 
p([A]^{\RN{1}})\ +\ p([B]^{\RN{1}})\ +\ p([C]^{\RN{1}})  \nonumber\\
 \ +\ p([A,B]^{\RN{1}}) \ +\ p([A,C]^{\RN{1}})\ +\ p([B,C]^{\RN{1}})\ +\ p([A,B,C]^{\RN{1}})=1,
\label{eq:indprob}
\end{gather}
and the probability distributions $p(o^A,...|s^A,..., [\cdots]^{\RN{1}}) $ for a given $[\cdots]^{\RN{1}}$, defined whenever $p([\cdots]^{\RN{1}})\neq 0$, describe processes which we will denote by $\mathcal{W}^{[\cdots]^{\RN{1}}}$. (Note that we can imagine that the variable $[\cdots]^{\RN{1}}$ is associated with an event in the past of all local experiments, i.e., these can be thought of as a proper pre-selected process.)

In a compact form, Eq.~(\ref{tripart1'}) can be written
\begin{equation}
\begin{split}
 \mathcal{W}_{c}^{A,B,C} &=  p([A]^{\RN{1}})\ \mathcal{W}^{[A]^{\RN{1}}}+\ p([B]^{\RN{1}})\ \mathcal{W}^{[B]^{\RN{1}}} +p([C]^{\RN{1}})\ \mathcal{W}^{[C]^{\RN{1}}}\\
  &+\ p([A,B]^{\RN{1}})\ \mathcal{W}^{[A,B]^{\RN{1}}}\ +\ p([A,C]^{\RN{1}}) \ \mathcal{W}^{[A,C]^{\RN{1}}} \ +\ p([B,C]^{\RN{1}})\ \mathcal{W}^{[B,C]^{\RN{1}}}\\
 &+\ p([A,B,C]^{\RN{1}})\ \mathcal{W}^{[A,B,C]^{\RN{1}}},
\end{split} 
\label{tripart2}
\end{equation}
i.e., the overall process is a mixture of processes defined conditionally on the different scenarios $[\cdots]^{\RN{1}}$. The processes $\mathcal{W}^{[\cdots]^{\RN{1}}}$ cannot be arbitrary but must be compatible with causality, the conditions for which we derive next. 

Consider the case in which one party is first, say $[C]^{\RN{1}}$ (Fig.~\ref{fig:3Causal config}). There are three distinct causal configurations compatible with this case, in which $A\cp B$, $B\cp A$, or $A\ind B$ (Table~\ref{tbl:the groups}). We can expand $p(o^A,o^B,o^C|s^A,s^B,s^C, [C]^{\RN{1}}) $ conditionally on these configurations as follows:
\begin{equation}
\begin{split} 
 p(o^A, o^B, o^C|s^A, s^B, s^C, [C]^{\RN{1}}) &=p(o^C| s^A, s^B, s^C,  [C]^{\RN{1}}) \times\\
& [p(A\cp B| s^A, s^B, s^C, o^C,  [C]^{\RN{1}}) \  p(o^A,o^B |s^A,s^B, s^C, o^C, A\cp B, [C]^{\RN{1}})\ +\  \\
 & p(B\cp A| s^A, s^B, s^C, o^C, [C]^{\RN{1}}) \ p(o^A,o^B |s^A,s^B, s^C, o^C,B\cp A, [C]^{\RN{1}})\ +\ \\
 & p(A\ind B| s^A, s^B, s^C, o^C,  [C]^{\RN{1}})  \ p(o^A,o^B |s^A,s^B, s^C, o^C, A\ind B, [C]^{\RN{1}})],
\end{split}
\label{tripartC1}
\end{equation}
where $p(o^A,o^B |s^A,s^B, s^C, o^C, \kappa(A,B), [C]^{\RN{1}})$ is defined when $p(\kappa(A,B)| s^A, s^B, s^C, o^C, [C]^{\RN{1}})\neq 0 $, and  
\begin{gather}
  p(A\cp B| s^A, s^B, s^C, o^C,   [C]^{\RN{1}})  +\ p(B\cp A|  s^A, s^B, s^C, o^C,  [C]^{\RN{1}}) \nonumber\\
 +p(A\ind B| s^A, s^B, s^C, o^C,  [C]^{\RN{1}}) =1.\label{tripartC11}
\end{gather}
%\comment{Here I put the following equations into an equation environment instead of being in the text - it was very hard to follow}

From Proposition \ref{prop:2.3}, we have that 
\[
p(o^C| s^A, s^B, s^C,  [C]^{\RN{1}}) \equiv p([C]^{\RN{1}}, o^C| s^A, s^B, s^C)/ p([C]^{\RN{1}})=  p([C]^{\RN{1}}, o^C| s^C)/ p([C]^{\RN{1}})= p(o^C| s^C,  [C]^{\RN{1}}).
\]
 Similarly, we have 
 \begin{equation}
 \begin{split}
p(A\cp B| s^A, s^B, s^C, o^C,  [C]^{\RN{1}})&= p(A\cp B| s^A, s^C, o^C,   [C]^{\RN{1}}), \\
p(B\cp A|  s^A, s^B, s^C, o^C,  [C]^{\RN{1}})  &= p(B\cp A|  s^B, s^C, o^C,  [C]^{\RN{1}}), \\ 
p(A\ind B| s^A, s^B, s^C, o^C,  [C]^{\RN{1}})&= p(A\ind B| s^C, o^C,  [C]^{\RN{1}}), 
\end{split}
\end{equation}
which together with Eq.~(\ref{tripartC11}) implies 
%(that the probability of any causal configuration of A and B, given that C is first, will depend on the event on C and the fact that C is first)
\begin{equation}
\begin{split}
p(A\cp B| s^A, s^B, s^C, o^C,   [C]^{\RN{1}})&= p(A\cp B| s^C, o^C,   [C]^{\RN{1}}),\\
p(B\cp A|  s^A, s^B, s^C, o^C,  [C]^{\RN{1}})  &= p(B\cp A|  s^C, o^C,  [C]^{\RN{1}}), \\
p(A\ind B| s^A, s^B, s^C, o^C,  [C]^{\RN{1}})&= p(A\ind B| s^C, o^C, [C]^{\RN{1}}).
\end{split}
\end{equation}
Substituting this in Eq.~\eqref{tripartC1}, we obtain
\begin{gather} 
 p(o^A, o^B, o^C|s^A, s^B, s^C, [C]^{\RN{1}}) =p(o^C| s^C,  [C]^{\RN{1}}) \times \notag\\
  [p(A\cp B| s^C,o^C, [C]^{\RN{1}})  \ p(o^A,o^B |s^A,s^B, s^C, o^C, A\cp B, [C]^{\RN{1}}) 
\ +\ p(B\cp A| s^C, o^C,  [C]^{\RN{1}})  \ p(o^A,o^B |s^A,s^B, s^C, o^C,B\cp A, [C]^{\RN{1}}) \notag\\ 
 \ +\ p(A\ind B| s^C,o^C, [C]^{\RN{1}})  \ p(o^A,o^B |s^A,s^B, s^C, o^C, A\ind B, [C]^{\RN{1}})],\label{tripartC}
\end{gather}
with 
\begin{gather} 
p(A\cp B| s^C, o^C,   [C]^{\RN{1}})  +\ p(B\cp A|  s^C, o^C,  [C]^{\RN{1}}) 
 +p(A\ind B| s^C, o^C,  [C]^{\RN{1}}) =1,\label{tripartC111}
\end{gather}
where the probability distributions $p(o^A,o^B |s^A,s^B, s^C, o^C, A\cp B, [C]^{\RN{1}}) $, $p(o^A,o^B |s^A,s^B, s^C, o^C,B\cp A, [C]^{\RN{1}}) $, and $ p(o^A,o^B |s^A,s^B, s^C, o^C, A\ind B, [C]^{\RN{1}})$ describe bipartite processes for Alice and Bob for every fixed value of $(s^C,o^C)$. The assumption of causality implies conditions for these processes too. They must respect the no-signaling constraints imposed by the causal configuration $\kappa(A,B)$ they are conditioned on -- the first one must involve no signaling from Bob to Alice, the second one must involve no signaling from Alice to Bob, and the third one must involve no signaling between Alice and Bob in either direction. This follows from the fact that 
\begin{gather}
p(o^A,o^B |s^A,s^B, s^C, o^C,\kappa(A,B), [C]^{\RN{1}}) = 
\frac{p([C]^{\RN{1}}, \kappa(A,B), o^A, o^B, o^C|s^A,s^B, s^C )}{ p( [C]^{\RN{1}}) \ p(o^C| s^C,  [C]^{\RN{1}}) \ p(\kappa(A,B)| s^C, o^C,  [C]^{\RN{1}}) },\label{numerator}
\end{gather}
%\comment{shouldn't the 'or' in the following line be 'and'}
and the observation that since only the numerator on the right-hand side depends on $s^A$, $o^A$, $s^B$, and $o^B$, the respective no-signaling constraints on the quantity on the left-hand side follow from the requirement that the numerator is compatible with Eq.~ (\ref{causalorderDEF}).

Notice that the probabilities $p(o^C| s^C,  [C]^{\RN{1}}) $ in Eq.~(\ref{tripartC}) define a reduced monopartite process for Charlie, $\mathcal{W}^{C}$, while the probabilities enclosed by the square brackets define a conditional bipartite process $\mathcal{W}_c^{A,B|C}$, which is causal (indicated by the subscript $c$) for every fixed $(s^C,o^C)$. In a compact form, this can be written
\begin{equation}
\mathcal{W}^{[C]^{\RN{1}}} = \mathcal{W}_{c}^{A,B|C}\circ  \mathcal{W}^C. \label{tripartCshort}
\end{equation}

The form (\ref{tripartCshort}) is necessary for a causal process for which all causal configurations that have non-zero probabilities respect ${[C]^{\RN{1}}}$ (in that case, a causal process of the general form (\ref{tripart2}) reduces to the term $\mathcal{W}^{[C]^{\RN{1}}}$). It is also sufficient, because this form provides an explicit joint probability distribution $p^{[C]^{\RN{1}}}(\kappa(A,B,C), o^A, o^B, o^C|s^A,s^B,s^C)$   -- equal to $p([C]^{\RN{1}}, \kappa(A,B), o^A, o^B, o^C|s^A,s^B, s^C ) =  p(o^C| s^C,  [C]^{\RN{1}}) \ p(\kappa(A,B)| s^C, o^C, [C]^{\RN{1}})\  p(o^A,o^B |s^A,s^B, s^C, o^C, \kappa(A,B), [C]^{\RN{1}})$ when $\kappa(A,B,C)$ is compatible with $[C]^{\RN{1}}$, and to zero otherwise -- for which condition (\ref{causalorderDEF}) is satisfied with respect to every party. Indeed, condition (\ref{causalorderDEF}) is satisfied with respect to $C$ since the probability for any party being in the causal past or causal elsewhere of $C$ is zero. It is also satisfied with respect to $A$ (similarly for $B$) since the no-signaling constraints respected by $p(o^A,o^B |s^A,s^B, s^C, o^C, \kappa(A,B), [C]^{\RN{1}})$ guarantee that $p^{[C]^{\RN{1}}} ( \kappa(A,B,C), A\ncp B, A\ncp C, o^B, o^C |s^A,s^B,s^C) = p^{[C]^{\RN{1}}} ( \kappa(A,B,C), A\ncp B, A\ncp C, o^B, o^C |s^B,s^C)$. The necessary and sufficient conditions for a causal process compatible with ${[A]^{\RN{1}}}$ and ${[B]^{\RN{1}}}$ are analogous.

\begin{figure}[!htb]
 \centering
 \includegraphics[scale=.5]{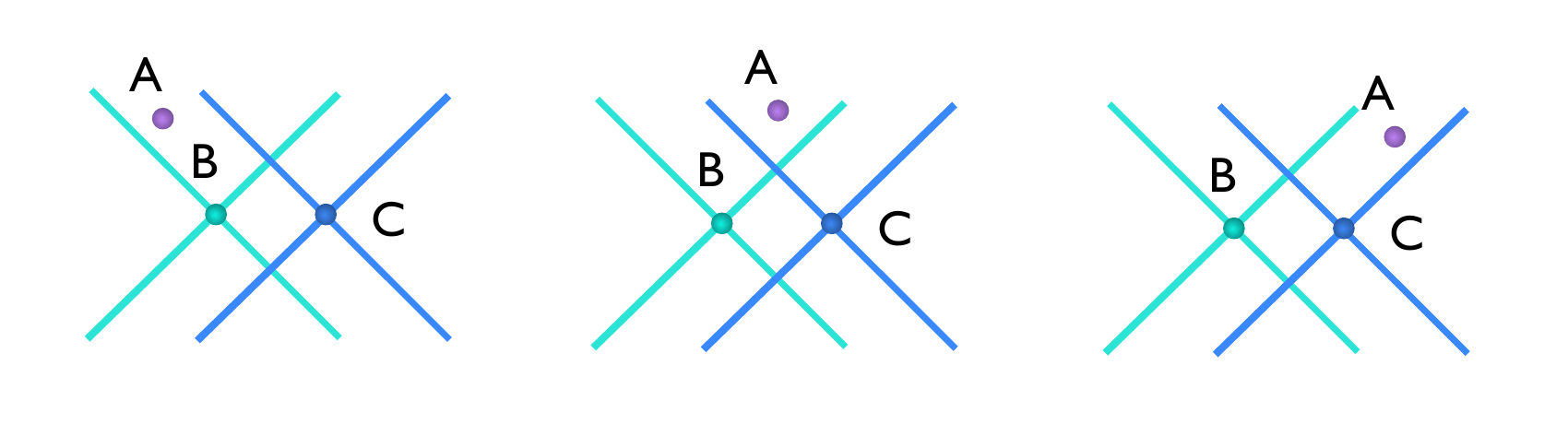}
 %\captionsetup{width=0.8\textwidth}
 \caption{The three possible tripartite causal configurations included in the group where B and C are first. From left to right: [$B\ind C$ and $B\cp A$ and $C\ind A$], [$B\ind C$ and $B\cp A$ and $C\cp A$], [$B\ind C$ and $B\ind A$ and $C\cp A$].}
 \label{fig:3configs}
 \end{figure}

Let us now consider the case where two parties are first, say $[B,C]^{\RN{1}}$. The possible causal configurations in this case (Table~\ref{tbl:the groups}) are depicted in Fig.~\ref{fig:3configs}. Similarly to the previous case, using the assumption of causality, we can expand the probabilities $p(o^A,...|s^A,..., [B,C]^{\RN{1}}) $ conditionally on the different configurations as follows:
\begin{gather}
 p(o^A, o^B, o^C|s^A, s^B, s^C, [B,C]^{\RN{1}}) =p(o^B,o^C| s^B,s^C,  [B,C]^{\RN{1}}) \times \notag\\ 
 [p(B\cp A, C\ind A|  s^B,o^B,s^C,  o^C   , [B,C]^{\RN{1}}) 
  p(o^A |s^A,  s^B,o^B,s^C,  o^C  ,B\cp A, C\ind A, [B,C]^{\RN{1}}) \notag\\
 +\ p(B\ind A, C\cp A|   s^B,o^B,s^C,  o^C  ,  [B,C]^{\RN{1}}) 
 p(o^A |s^A,  s^B,o^B,s^C,  o^C  , B\ind A, C\cp A, [B,C]^{\RN{1}}) \notag\\
 +\ p(B\cp A, C\cp A|   s^B,o^B,s^C,  o^C  ,   [B,C]^{\RN{1}}) 
 p(o^A |s^A,  s^B,o^B,s^C,  o^C,B\cp A, C\cp A, [B,C]^{\RN{1}})],\label{tripartBC}
\end{gather}
with
\begin{gather}
  p(B\cp A, C\ind A|  s^B,o^B,s^C,  o^C  , [B,C]^{\RN{1}})  +\ 
 (B\ind A, C\cp A|   s^B,o^B,s^C,  o^C  , [B,C]^{\RN{1}}) +\notag\\
 p(B\cp A, C\cp A|  s^B,o^B,s^C,  o^C  , [B,C]^{\RN{1}}) =1,
\end{gather}
where the probabilities $p(o^B,o^C| s^B,s^C,  [B,C]^{\RN{1}}) $ in Eq.~(\ref{tripartBC}) define a reduced bipartite process that involves no signaling between $B$ and $C$, and the probabilities in the square brackets describe a conditional process for $A$. The fact that there is no signaling between $B$ and $C$ in the first process follows easily from Proposition \ref{prop:2.3}.

It turns out that the decomposition over different causal configurations does not yield any nontrivial conditions on the probabilities of the conditional process enclosed in the square brackets, i.e., the simpler form
 \begin{gather}
  p(o^A, o^B, o^C|s^A, s^B, s^C, [B,C]^{\RN{1}}) =p(o^B,o^C| s^B,s^C,  [B,C]^{\RN{1}}) \ p(o^A |s^A, s^B, o^B, s^C, o^C, [B,C]^{\RN{1}}) \label{tripartBC2}
 \end{gather}
 is both necessary and sufficient for a valid $\mathcal{W}^{[B,C]^{\RN{1}}}$. Necessity is obvious since Eq.~(\ref{tripartBC}) implies Eq.~(\ref{tripartBC2}). Sufficiency follows from the fact that the right-hand side of Eq.~(\ref{tripartBC2}) is compatible with the particular case $p(B\cp A, C\cp A|  s^B,o^B,s^C,  o^C  , [B,C]^{\RN{1}}) =1$, where the only non-trivial constraints on the probabilities $p(o^A, o^B, o^C|s^A, s^B, s^C, [B,C]^{\RN{1}}) $ imposed by $\kappa(A,B, C)$ are that there is no signaling from Alice to Bob and Charlie, and no signaling between Bob and Charlie in their reduced bipartite process. These are clearly guaranteed by Eq.~(\ref{tripartBC2}) when the reduced process $\{ p(o^B,o^C| s^B,s^C,  [B,C]^{\RN{1}}) \} $ involves no signaling between Bob and Charlie. Therefore, similarly to Eq.~(\ref{tripartCshort}), we can write Eq.~(\ref{tripartBC2}) in the compact form
\begin{equation}
\mathcal{W}^{[B,C]^{\RN{1}}} = \mathcal{W}^{A|B,C}\circ  \mathcal{W}_{ns}^{B,C}, \label{tripartBCshort}
\end{equation}
 where $\mathcal{W}_{ns}^{B,C}$ is a non-signaling bipartite process for Bob and Charlie, and $\mathcal{W}^{A|BC}$ is a monopartite process for Alice conditional on the events in the laboratories of Bob and Charlie.

Finally, in the case where all of the parties are first, we only have the constraint that
\begin{equation}
\mathcal{W}^{[A,B,C]^{\RN{1}}} = \mathcal{W}_{ns}^{A,B,C} \label{tripartABCshort}
\end{equation}
is a tripartite non-signaling process. Again, this follows from Proposition \ref{prop:2.3}.

Therefore, we have obtained that a tripartite causal process $\mathcal{W}_{c}^{A,B,C} $ must have the form 
\begin{gather}
 \mathcal{W}_{c}^{A,B,C} =  p([A]^{\RN{1}})\ \mathcal{W}_{c}^{B,C|A}\circ  \mathcal{W}^A+\ p([B]^{\RN{1}})\  \mathcal{W}_{c}^{A,C|B}\circ  \mathcal{W}^B \nonumber\\
 +p([C]^{\RN{1}})\ \mathcal{W}_{c}^{A,B|C}\circ  \mathcal{W}^C \ +\ p([A,B]^{\RN{1}})\ \mathcal{W}^{C|A,B}\circ  \mathcal{W}_{ns}^{A,B} \nonumber\\
 +p([A,C]^{\RN{1}})\ \mathcal{W}^{B|A,C}\circ  \mathcal{W}_{ns}^{A,C} \ +\ p([B,C]^{\RN{1}})\ \mathcal{W}^{A|B,C}\circ  \mathcal{W}_{ns}^{B,C}\nonumber\\
 +p([A,B,C]^{\RN{1}})\ \mathcal{W}_{ns}^{A,B,C},\label{tripart3}
\end{gather}
with suitable probability weights $p([A]^{\RN{1}}) $, $p([B]^{\RN{1}}) $, $p([C]^{\RN{1}}) $, $p([A,B]^{\RN{1}}) $, $p([A,C]^{\RN{1}}) $, $p([B,C]^{\RN{1}}) $,  and $p([A,B,C]^{\RN{1}}) $. 
This form is also sufficient for a tripartite process to be causal because it explicitly gives a probability distribution  $p(\kappa(A,B,C), o^A, o^B, o^C|s^A, s^B, s^C) = \sum_{[\cdots]^{\RN{1}}}  p([\cdots]^{\RN{1}})\ p(\kappa(A,B,C), o^A, o^B, o^C|s^A, s^B, s^C, [\cdots]^{\RN{1}})$ that satisfies Eq.~(\ref{causalorderDEF}). Indeed, we have seen that each of the distributions $p(\kappa(A,B,C), o^A, o^B, o^C|s^A, s^B, s^C, [\cdots]^{\RN{1}})$ in this convex mixture is an extension of a causal process $\{ p(o^A, o^B, o^C|s^A, s^B, s^C, [\cdots]^{\RN{1}})\} $, and hence it satisfies Eq.~(\ref{causalorderDEF}). Since the weights $p([\cdots]^{\RN{1}})$ in the mixture are independent of $s^A$, $s^B$, and $s^C$, and Eq.~(\ref{causalorderDEF}) is linear in $p(\kappa(A,B,C), o^A, o^B, o^C|s^A, s^B, s^C, [\cdots]^{\RN{1}})$, the equation is satisfied by the whole mixture too. 
 
Condition (\ref{tripart3}) can be further simplified by noticing that the processes corresponding to the cases in which two or three parties are first have forms compatible with cases in which only a single party is first. For instance, $\mathcal{W}^{[B,C]^{\RN{1}}} $ satisfies the necessary and sufficient conditions for a valid $ \mathcal{W}^{[B]^{\RN{1}}} $ or a valid $ \mathcal{W}^{[C]^{\RN{1}}}$, while $\mathcal{W}^{[A,B,C]^{\RN{1}}}$ satisfies the necessary and sufficient conditions for any of $ \mathcal{W}^{[A]^{\RN{1}}} $, $ \mathcal{W}^{[B]^{\RN{1}}} $, or $ \mathcal{W}^{[C]^{\RN{1}}}$. The compatibility of $\mathcal{W}^{[B,C]^{\RN{1}}} $ with $[C]^{\RN{1}}$, for example, can be seen from the fact that Eq.~(\ref{tripartBC2}) (or Eq.~(\ref{tripartBCshort})) is compatible with the case $[C]^{\RN{1}}$ in which $C\cp B \cp A$, since the only constraints in that case are that Alice cannot signal to Bob and Charlie, and that Bob cannot signal to Charlie, which are satisfied by the probabilities in Eq.~(\ref{tripartBC2}). Similarly, $\mathcal{W}^{[B,C]^{\RN{1}}} $ is compatible with $[C]^{\RN{1}}$. A process $\mathcal{W}^{[A,B,C]^{\RN{1}}}$ is compatible with any causal configuration since it does not involve signaling between any of the parties. These observations suggest that we can group (in a generally non-unique way) the terms in the probabilistic mixture (\ref{tripart2}) so as to obtain a mixture of three processes 
\begin{gather}
 \mathcal{W}_{c}^{A,B,C} =  p(w^{(B,C)\ncp A})\ \mathcal{W}^{(B,C)\ncp A}\nonumber\\
 +\ p(w^{(A,C)\ncp B})\ \mathcal{W}^{(A,C)\ncp B}+p(w^{(A,B)\ncp C})\ \mathcal{W}^{(A,B)\ncp C}
\label{tripart4},
\end{gather}
where $w^{(B,C)\ncp A}$, $w^{(A,C)\ncp B}$, and $w^{(A,B)\ncp C}$, are some mutually exclusive variables whose probabilities satisfy $p(w^{(B,C)\ncp A}) +\ p(w^{(A,C)\ncp B})+p(w^{(A,B)\ncp C})  =1$, such that conditionally on these variables, the causal configuration of the parties belongs to one of the groups compatible with  $(B,C)\ncp A$ (meaning $B\ncp A \land C\ncp A$), $(A,C)\ncp B$, and $(A,B)\ncp C$, respectively, while the processes $\mathcal{W}^{(B,C)\ncp A}$, $\mathcal{W}^{(A,C)\ncp B}$, and $\mathcal{W}^{(A,B)\ncp C}$, satisfy the most general causal constraints compatible with these groups. For instance, conditionally on $w^{(B,C)\ncp A}$, the causal configurations of the parties may belong to any of the groups defined by $[A]^{\RN{1}}$, $[A,B]^{\RN{1}}$, $[A,C]^{\RN{1}}$, and $[A,B,C]^{\RN{1}}$. The process $\mathcal{W}^{(B,C)\ncp A}$ would itself be a probabilistic mixture of processes compatible with these groups, which most generally satisfy the constraints satisfied by $\mathcal{W}^{[A]^{\RN{1}}}$. That is,
\begin{equation}
\mathcal{W}^{(B,C)\ncp A} = \mathcal{W}_{c}^{B,C|A}\circ  \mathcal{W}^A, \label{Ashort}
\end{equation}
\begin{equation}
\mathcal{W}^{(A,C)\ncp B} = \mathcal{W}_{c}^{A,C|B}\circ  \mathcal{W}^B, \label{Bshort}
\end{equation}
\begin{equation}
\mathcal{W}^{(A,B)\ncp C} = \mathcal{W}_{c}^{A,B|C}\circ  \mathcal{W}^C. \label{Cshort}
\end{equation}
Obviously, the existence of a convex decomposition (\ref{tripart4}) is both necessary and sufficient for a tripartite process to be causal, since any process of the form (\ref{tripart3}) can be written in the form (\ref{tripart4}), while Eq.~(\ref{tripart4}) is a special case of Eq.~(\ref{tripart3}).

As in the bipartite case, for any fixed number of settings and fixed number of outcomes for each party, the constraint (\ref{tripart4}) provides a means of testing whether the corresponding tripartite probabilities are compatible with causality. The set of probabilities that satisfy Eq.~(\ref{tripart4}) is the convex hull of the probabilities compatible with causal configurations in which $(B,C)\ncp A$, $(A,C)\ncp B$, and $(A,B)\ncp C$. One can see that the latter form polytopes, since the constraints imposed by causality in each of these cases are linear. For example, in the case of  $(A,B)\ncp C$, we have the constraint that $p(o^A,o^B,o^C|  s^A,s^B,s^C,  w^{(A,B)\ncp C})$ involve no signaling from Alice and Bob to Charlie, and that for every $(s^C,o^C)$, the resultant process between Alice and Bob is causal. The first requirement corresponds to a set of linear constraints. The second requirement corresponds to the condition that for every fixed $(s^C,o^C)$, the probabilities $p(o^A,o^B|  s^A,s^B, s^C, o^ C,   w^{(A,B)\ncp C}) \equiv p(o^A,o^B,o^C|  s^A,s^B,s^C,  w^{(A,B)\ncp C} )/ p(o^C| s^C,  w^{(A,B)\ncp C})$ are the probabilities describing a causal process for Alice and Bob, which themselves belong to a polytope and hence respect a set of linear inequalities. Plugging these probabilities in the respective inequalities and multiplying both sides by $p(o^C| s^C,  w^{(A,B)\ncp C})$ would yield a set of linear inequalities for $p(o^A,o^B,o^C|  s^A,s^B,s^C,  w^{(A,B)\ncp C} )$. Therefore, these probabilities also form a polytope, and so do the probabilities of the form (\ref{tripart4}). The nontrivial facets of the polytope of probabilities (\ref{tripart4}) would define tripartite causal inequalities, whose violation indicates incompatibility with causal order. Examples of tripartite causal inequalities for binary inputs and outputs can be found in Refs.~\cite{Baumeler1, Baumeler2} (we have not investigated whether these are facets of the respective causal polytope).\\

The extension of the conditions for causality of a process to the case of $n$ parties can be defined iteratively. The following theorem provides the generalization of condition (\ref{tripart3}): \\
\begin{thm}\label{thm:1}
A process for a set of parties $\mathcal{S}=\{1,\cdots,n\}$, $n\geq 1$, is causal if and only if it can be written in the form 
\end{thm}
\begin{gather}
 \mathcal{W}_{c}^{\mathcal{S}} = \sum_{\mathcal{X} \subset \mathcal S, \mathcal{X} \neq \{\null \}} p_{\mathcal{X}} \mathcal{W}_c^{\mathcal{S} \backslash \mathcal{X} |\mathcal{X} }\circ \mathcal{W}_{ns}^\mathcal{X} 
\label{Nparty},
\end{gather}
where the sum is over all nonempty subsets $\mathcal{X} $ of the local experiments ${\mathcal{S}}$, $p_{\mathcal{X}} $ are suitable probability weights (which can be interpreted as the probability for $\mathcal{X} $ to be first, $p_{\mathcal{X}} = p([\mathcal{X}]^{\RN{1} } )$), $\mathcal{S} \backslash \mathcal{X}$ denotes the relative complement of $\mathcal{X}$ in $\mathcal{S}$, $\mathcal{W}_{ns}^\mathcal{X}$ is a non-signaling reduced process for $\mathcal{X} $, and the conditional process $\mathcal{W}_c^{\mathcal{S} \backslash \mathcal{X}|\mathcal{X} }$ is either the trivial process (when $\mathcal{X} = \mathcal{S}$) or otherwise can be written in the same form (\ref{Nparty}) for every given value of the possible events in $\mathcal{X}$. The Proof \ref{pr:thm:1} is given in the Appendix.\\

As in the bipartite and tripartite cases, we can simplify the conditions for an $n$-partite process to be causal by noticing that the constraints on a process compatible with a given set of $k$ ($1\leq k \leq n$) parties being first are compatible with the constraints on a process compatible with the case in which only a single one of the $k$ parties is first. Therefore, by an argument analogous to the one in the tripartite case, we obtain the following alternative formulation of the conditions. \\
\begin{thm}\label{thm:1'}
\textbf{(Canonical causal decomposition):} A causal process for $n$ parties is one that can be written in the (generally non-unique) form
\end{thm}
\begin{gather}
 \mathcal{W}_{c}^{1,\cdots,n} = \sum_{i=1}^n q_i \mathcal{W}^{(1,\cdots,{i-1},{i+1},\cdots, n)\ncp {i}}, \hspace{0.2cm}q_i \geq 0, \forall i, \hspace{0.1cm}\sum_{i=1}^{n}q_i=1,
\label{Npart}
\end{gather}
with
\begin{gather}
 \mathcal{W}^{(1,\cdots, {i-1}, {i+1},\cdots,n)\ncp {i}} = \mathcal{W}_{c}^{1,\cdots, {i-1}, {i+1},\cdots,n| i}\circ  \mathcal{W}^{i}, \label{Nshort}
\end{gather}
where the $(n-1)$-partite conditional process $\mathcal{W}_{c}^{1,\cdots, {i-1}, {i+1},\cdots,n| i}$ is either trivial (when $n=1$) or has the form (\ref{Npart}) for every value of the event in $i$.\\

The weights $q_i$ in Eq.~(\ref{Nshort}) can be thought of as the probabilities $q_i\equiv p(w^{(1,\cdots,{i-1}, {i+1},\cdots, n)\ncp i}) $ for a mutually exclusive set of variables $w^{(1,\cdots, i-1, i+1,\cdots,n)\ncp i}$ for which the causal configurations of the parties belong to a group such that $(1,\cdots, {i-1}, {i+1},\cdots, n)\ncp i$.

Theorem \ref{thm:1'} (alternatively Theorem \ref{thm:1}) gives iteratively formulated necessary and sufficient conditions for a process to be causal in the general multipartite case. It can be understood as describing an `unraveling' of the different possible sequences of operations in steps: first, the party that is first and his/her monopartite process are selected at random based on some probability distribution; next, the party that is second and his/her monopartite process are selected at random from some probability distribution that most generally can depend on the first party's setting and outcome; next, the party that is third and his/her monopartite process are selected from some probability distribution that most generally can depend on the settings and outcomes of the first two parties, and so on. We refer to this intuitive decomposition as the \textit{canonical causal decomposition} of a causal process.  

By an argument analogous to the one in the tripartite case, one easily sees from Theorem \ref{thm:1'} that for any fixed number of settings and outcomes for each party, the causal probabilities for $n$ parties form a polytope, provided that the causal probabilities for $(n-1)$ parties form a polytope. By induction, this implies a polytope structure for the general multipartite case. The nontrivial facets of such a polytope define causal inequalites. Examples of $n$-partite causal inequalities, where $n=2k+1$, for binary inputs and outputs can been found in Refs.~\cite{Baumeler1, Baumeler2}. It would be interesting to check if these inequalities are facets of the respective causal polytope.

\section{The quantum process framework} \label{SectionII}

\subsection{General quantum processes}

The quantum process framework introduced in Ref.~\cite{OCB} is a particular theory within the general operational framework for pre-selected processes discussed in the previous section. It is based on a set of assumptions about the local operations of the parties and the joint probabilities for their outcomes, which we review next.

The first main assumption is that of \textit{local quantum mechanics} \cite{OCB}, which says that each local experiment is described as in standard quantum mechanics. Specifically, let $X_1$ and $X_2$ denote the input and output systems of a local experiment $X$. It is assumed that these systems are associated with Hilbert spaces $\mathcal{H}^{X_1}$ and $\mathcal{H}^{X_2}$ of dimensions $\textrm{dim} \mathcal{H}^{X_1} = d_{X_1}$ and $\textrm{dim} \mathcal{H}^{X_2} = d_{X_2}$, respectively. The set of operations that can be performed between the input and output systems is the set of standard quantum operations (or quantum instruments \cite{instrument}). A quantum operation has a set of outcomes labeled by $j=1,\dots,n$. Each outcome induces a specific transformation from the input to the output, which corresponds to a completely positive (CP) map ${\cal M}_j^X:{\cal L}({\cal H}^{X_1})\rightarrow {\cal L}({\cal H}^{X_2})$, where ${\cal L}({\cal H})$ is the space of linear operators over the (finite-dimensional) Hilbert space ${\cal H}$. The action of each ${\cal M}_j^X$ on an operator $\sigma\in {\cal L}({\cal H}^{X_1})$ can be written in the Kraus form \cite{Kraus} $\mathcal{M}_j^X(\sigma)=\sum_{k=1}^{m}E_{jk}\sigma E_{jk}^{\dagger}$, $m = d_{X_1}d_{X_2}$, where the Kraus operators $E_{jk}:{\cal H}^{X_1}\rightarrow {\cal H}^{X_2}$ satisfy $\sum_{k=1}^{m} E^{\dagger}_{jk}E_{jk}\leq \id^{X_1}$, $\forall j$. The set of CP maps $\left\{{\cal M}_j^X\right\}_{j=1}^n$ corresponding to all possible outcomes of a quantum operation has the property that $\sum_{j=1}^n{\cal M}_j^X$ is CP and trace-preserving (CPTP), which is equivalent to the condition $\sum_{j=1}^{n}\sum_{k=1}^{m}E^{\dagger}_{jk}E_{jk} = \id^{X_1}$.

The second main assumption is that the joint probabilities for the outcomes of the operations of a set of parties, Alice, Bob, Charlie, $\cdots$, is a non-contextual function of the local CP maps,
\begin{gather}
 p(i, j, k, \cdots|\{\mathcal{M}^A_{i}\}, \{\mathcal{M}^B_{j}\} , \{\mathcal{M}^C_{k}\} \cdots)=\omega(\mathcal{M}^A_{i}, \mathcal{M}^B_j, \mathcal{M}^C_k, \cdots).\label{joint1}
\end{gather}
The requirement that local procedures agree with standard quantum mechanics implies that the function $\omega$ should be linear in the local CP maps \cite{OCB}.

Such a linear function can be written in a convenient form by expressing each local CP map as a positive semidefinite operator using a version of the Choi-Jamio{\l}kowsky (CJ) isomorphism \cite{jam, choi}. In this isomorphism, the CJ operator $M^{A_1A_2}_i\in{\cal L}({\cal H}^{A_1}\otimes{\cal H}^{A_2})$ corresponding to a linear map ${\cal M}^A_i:{\cal L}({\cal H}^{A_1})\rightarrow {\cal L}({\cal H}^{A_2})$ is defined as $M^{A_1A_2}_i:=\left[{\cal I}\otimes{\cal M}_i\left(|\phi^+\rangle\langle \phi^+|\right)\right]^{\mathrm T}$, where $|\phi^+\ \rangle=\sum_{j=1}^{d_{A_1}}|jj\rangle \in {\cal H}^{A_1}\otimes{\cal H}^{A_1}$ is a (not normalized) maximally entangled state on two copies of $\mathcal{H}^{A_1}$, the set of states $\left\{|j\rangle\right\}_{j=1}^{d_{A_1}}$ is an orthonormal basis of ${\cal H}^{A_1}$, ${\cal I}$ is the identity map, and ${\mathrm T}$ denotes matrix transposition in the basis $\left\{|j\rangle\right\}_{j=1}^{d_{A_1}}$ of ${\cal H}^{A_1}$ and some specific basis of ${\cal H}^{A_2}$. The CJ operator defined in this way does not depend on the choice of basis of $\mathcal{H}^{A_1}$, but does depend on the choice of basis of $\mathcal{H}^{A_2}$ \cite{footnote3}. For the purposes of the present paper, the latter basis can be an arbitrary fixed basis. We note, however, that within the time-symmetric generalization of the framework developed in Ref.~\cite{OC}, this basis has a nontrivial physical significance related to the transformation of time reversal. Specifically, in that formulation, the Hilbert space $\mathcal{H}^{A_2}$ on which the CJ operator is defined is not interpreted as the original output Hilbert space of the CP map, but a time-reversed copy of it. In this paper, we will not be concerned with that formulation, but will simply regard the CJ representation of CP maps, defined for an arbitrary choice of basis, as a mathematical convenience. Using the CJ representation, the joint probabilities (\ref{joint1}) can be written in the form
\begin{gather}
 p(i, j,k, \cdots|\{\mathcal{M}^A_{i}\}, \{\mathcal{M}^B_{j}\} ,\{\mathcal{M}^C_{k}\},\cdots)\nonumber\\= \textrm{Tr} \left[W^{A_1A_2B_1B_2C_1C_2\cdots}\left(M^{A_1A_2}_i\otimes M_j^{B_1B_2}\otimes M_k^{C_1C_2}\otimes \cdots\right)\right], \label{Wmain}
\end{gather}
where $W^{A_1A_2B_1B_2C_1C_2\cdots} \in \mathcal{L}({\cal H}^{A_1}\otimes{\cal H}^{A_2}\otimes{\cal H}^{B_1}\otimes{\cal H}^{B_2}\otimes{\cal H}^{C_1}\otimes{\cal H}^{C_2}\otimes \cdots)$.

The last main assumption behind the quantum process framework is that the local operations of the parties can be extended to act on input ancillas $A_1'$, $B_1'$, $C_1'$, $\cdots$, that are allowed to be prepared in an arbitrary quantum state $\rho^{A_1'B_1'C_1'\cdots}$, $\rho^{A_1'B_1'C_1'\cdots}\geq 0$, $\Tr \rho^{A_1'B_1'C_1'\cdots} = 1$. Upon such an extension, the original operator $W^{A_1A_2B_1B_2C_1C_2\cdots}$ becomes $W^{A_1A_2B_1B_2C_1C_2\cdots}\otimes \rho^{A_1'B_1'C_1'\cdots}$ \cite{OCB}. The requirement that the probabilities are non-negative for any combination of local CP maps $\mathcal{M}^A_i$, $\mathcal{M}^B_j$, $\mathcal{M}^C_k$, $\cdots$, on the extended systems $A=A_1A_1'A_2$, $B=B_1B_1'B_2$, $C=C_1C_1'C_2$, $\cdots$, implies that \cite{OCB}
\begin{gather}
W^{A_1A_2B_1B_2C_1C_2\cdots}\geq 0.\label{W1}
\end{gather}
In addition, since the probabilities should sum up to $1$ for a complete set of local outcomes, we have the condition that
\begin{gather}
\textrm{Tr} \left[W^{A_1A_2B_1B_2C_1C_2\cdots}\left(M^{A_1A_2}\otimes M^{B_1B_2}\otimes M^{C_1C_2}\otimes\cdots\right)\right] = 1, \label{W2} \\
\forall M^{A_1A_2},M^{B_1B_2}, M^{C_1C_2}, \cdots \geq 0,\nonumber\\
\textrm{Tr}_{A_2}M^{A_1A_2}=\id^{A_1}, \textrm{Tr}_{B_2}M^{B_1B_2}=\id^{B_1},   \textrm{Tr}_{C_2}M^{C_1C_2}=\id^{C_1}, \cdots ,\nonumber
\end{gather}
where $\textrm{Tr}_{X_2}$ denotes partial trace over $X_2$. Here, we have used the fact that a linear map $\mathcal{M}^X$ is CPTP if and only if its CJ operator satisfies $M^{X_1X_2}\geq 0$ and $\textrm{Tr}_{X_2}M^{X_1X_2}=\id^{X_1}$. An operator $W^{A_1A_2B_1B_2C_1C_2\cdots}$ that satisfies conditions (\ref{W1}) and (\ref{W2}) is called a \textit{process matrix} \cite{OCB}. Knowing the process matrix, by Eq.~(\ref{Wmain}) we have the probabilities for the outcomes of any combination of local operations of the parties, i.e., the process matrix provides a complete description of a process. (Here, the set $S^X$ of possible settings of a given party is the set of quantum operations with the respective input and output systems.)

The process matrix can be expanded in a Hilbert-Schmidt basis of orthogonal matrices on the Hilbert spaces of the input and output systems of the parties, which is helpful in analyzing different properties of the correlations that the process allows. A Hilbert-Schmidt basis of $\mathcal{L}(\mathcal{H}^X)$ is given by a set of Hermitian operators $\{\sigma_\mu^X\}_{\mu = 0}^{d_X^2 - 1}$, with $\sigma_0^X = \id^X$, $\Tr\sigma_\mu^X\sigma_\nu^X = d_X \delta_{\mu\nu}$, and $\Tr\sigma_j^X = 0$ for $j=1,...,d_X^2 - 1$. In such a basis, a process matrix can be written
\begin{gather}\label{eq:Wexpansion}
 W^{A_1A_2B_1B_2C_1C_2\cdots} = \sum_{i,j,k,l,m,n\cdots} w_{ijklmn\cdots} \sigma^{A_1}_i\otimes\sigma^{A_2}_j\otimes\sigma^{B_1}_k\otimes\sigma^{B_2}_l\otimes\sigma^{C_1}_m\otimes\sigma^{C_2}_n\otimes \cdots,\\
w_{ijklmn\cdots}\in \mathbb{R}, \hspace{0.2cm}\forall i,j,k,l,m,n,\cdots.\nonumber
\end{gather}
It turns out that many properties of process matrices can be formulated entirely as statements about the nonzero terms in the above expansion \cite{OCB}. For this purpose, it is convenient to introduce the following terminology. Non-zero terms proportional to $\sigma^{A_1}_{i}\otimes\id^{rest}$ ($i\geq1$) will be called terms of type $A_1$, non-zero terms proportional to $\sigma^{A_2}_{i}\otimes\sigma^{B_1}_{j}\otimes\id^{rest}$ ($i$, $j\geq1$) will be called terms of type $A_2B_1$, etc. Every process matrix also contains a non-zero term proportional to the identity operator on all systems. This term will be referred to as of type $\id$, or as the \textit{identity term}.

In Ref.~\cite{OCB}, it was shown that, in the bipartite case, an operator $W^{A_1A_2B_1B_2}$ satisfies condition (\ref{W2}) if and only if it contains at most terms from the following types: $\id$, $A_1$, $B_1$, $A_2B_1$, $A_1B_2$, $A_1A_2B_1$, $A_1B_1B_2$. This rule also includes the monopartite case, which is obtained when the input and output systems of one of the parties is trivial (the one-dimensional Hilbert space $\mathbb{C}^1$). Specifically, a monopartite operator $W^{A_1A_2}$ satisfies condition (\ref{W2}) if and only if it contains at most terms of type $\id$ and $A_1$. The types of allowed terms can be generalized to the $n$-partite case as follows.\\

\begin{prop}\label{prop:3.1}
An operator of the form \eqref{eq:Wexpansion} satisfies condition (\ref{W2}) if and only if in addition to the identity term it contains at most terms in which there is a nontrivial $\sigma$ operator on $X_1$ and a trivial one (the identity operator) on $X_2$ for some party $X\in\{A,B,C, \cdots\}$.\\
\end{prop}

\begin{table}
\caption{\label{tbl:validsigmas}The types of terms allowed in a tripartite process matrix $W^{A_1A_2B_1B_2C_1C_2}$}
\footnotesize\rm
\begin{tabular*}{\textwidth}{@{}l*{15}{@{\extracolsep{0pt plus12pt}}l}}
\hline
     $C_1$&$B_2C_1$&$B_1$&$B_1C_2$&$B_1C_1$  \\ \hline
     $B_1C_1C_2$&$B_1B_2C_1$&$A_2C_1$&$A_2B_2C_1$&$A_2B_1$ \\ \hline
     $A_2B_1C_2$ & $A_2B_1C_1$ & $A_2B_1C_1C_2$ & $A_2B_1B_2C_1$ & $A_1$  \\ \hline
     $A_1C_2$ & $A_1C_1$ & $A_1C_1C_2$ & $A_1B_2$ & $A_1B_2C_2$ \\ \hline
     $A_1B_2C_1$ & $A_1B_2C_1C_2$ & $A_1B_1$ & $A_1B_1C_2$ & $A_1B_1C_1$ \\ \hline
      $A_1B_1C_1C_2$ & $A_1B_1B_2$ & $A_1B_1B_2C_2$ & $A_1B_1B_2C_1$ & $A_1B_1B_2C_1C_2$ \\ \hline
      $A_1A_2C_1$ & $A_1A_2B_2C_1$ & $A_1A_2B_1$ & $A_1A_2B_1C_2$ & $A_1A_2B_1C_1$  \\ \hline
      $A_1A_2B_1C_1C_2$&  $A_1A_2B_1B_2C_1$ & $\id$  \\
\hline
\end{tabular*}
\end{table}
In the Appendix, we present Proof \ref{pr:prop:3.1} of the above proposition for the case of three parties and the general case follows accordingly. From the analysis in Proof~\ref{pr:prop:3.1} we see that a general operator $W^{A_1A_2B_1B_2C_1C_2}$  can contain up to 64 types of terms. The condition for normalization of probabilities (\ref{W2}) narrows the types of terms to the 38 types listed in Table~\ref{tbl:validsigmas}. The positive semidefiniteness condition (\ref{W1}) does not limit any further the allowed types of terms, because one can conceive of a positive semidefinite matrix containing nonzero terms of any chosen type (this can be ensured by taking the nontrivial $\sigma$ terms with non-zero coefficients of sufficiently small magnitude relative to the weight of the identity term which is always fixed). Thus, an operator $W^{A_1A_2B_1B_2C_1C_2}$ is a valid tripartite process matrix, i.e., it satisfies conditions  (\ref{W1}) and (\ref{W2}), if and only if it satisfies condition (\ref{W1}) and contains only terms of the types listed in Table~\ref{tbl:validsigmas}, where the identity term comes with the weight $w_{000000} = \frac{1}{d_{A_1}d_{B_1}d_{C_1}}$. In a similar way, one proves the allowed types of terms in the general $n$-partite case. (For an alternative formulation of the conditions for an  operator to be a valid process matrix, see Ref.~\cite{Araujo3}.)\\

The types of terms that appear in the expansion of a process matrix are closely related to the signaling between the parties that the process allows. For example, a bipartite process involves signaling from Bob to Alice if and only if the process matrix contains terms of type $A_1B_2$ or $A_1B_1B_2$ \cite{OCB}. To state the condition for (no) signaling in the multipartite case, it is convenient to introduce the following terminology (see also Ref.~\cite{Araujo3}).  Consider a Hilbert-Schmidt term $\sigma^{A_1}_i\otimes\sigma^{A_2}_j\otimes\sigma^{B_1}_k\otimes\sigma^{B_2}_l\otimes\sigma^{C_1}_m\otimes\sigma^{C_2}_n\otimes \cdots$ as in Eq.~(\ref{eq:Wexpansion}). The \textbf{\textit{restriction}} of this term onto, say, subsystems $B_2C_1C_2\cdots$ is defined as $\sigma^{B_2}_l\otimes\sigma^{C_1}_m\otimes\sigma^{C_2}_n\otimes \cdots$.\\

\begin{prop}\label{prop:3.2}
An $n$-partite process matrix for a set of parties $\{1, \cdots, n\}$ does not permit signaling from, say, ($1$ and $2$ and $\cdots$ and $k$) to ($k+1$ and $k+2$ and $\cdots$ and $n$) if an only if it contains only terms whose restriction onto $1_11_2\cdots k_1k_2$ are of the allowed types for a process matrix on $\{1,\cdots, k\}$ as described in Proposition \ref{prop:3.1}. The Proof \ref{pr:prop:3.2} is given in the Appendix.\\
 \end{prop}

As an example, a tripartite quantum process that is causal and compatible with a situation in which Charlie is first (Fig.~\ref{fig:3Causal config}) should involve no signaling from Alice and Bob to Charlie, and hence it can only contain the types of terms listed in Table~\ref{tbl:Cfirst}. These constraints on the allowed types of terms imposed by causal order will turn out to play an important role in the characterization of the so-called causally separable quantum processes, which we define in the next subsection.

\begin{table}
\caption{The types of terms allowed in a causal process matrix $W^{A_1A_2B_1B_2C_1C_2}_{(A,B)\ncp C}$ compatible with $(A,B)\ncp C$.}
\footnotesize\rm
\begin{tabular*}{\textwidth}{@{}l*{15}{@{\extracolsep{0pt plus12pt}}l}}
\hline
     $C_1$ &  $B_1$ & $B_1C_2$ & $B_1C_1$  &
     $B_1C_1C_2$ \\ \hline $A_2B_1$ & $A_2B_1C_2$ & $A_2B_1C_1C_2$ &
     $A_1$&  $A_1C_2$ \\ \hline $A_1C_1$ & $A_1C_1C_2$ &
     $A_1B_2$ & $A_1B_2C_2$ & $A_1B_2C_1C_2$ \\ \hline $A_1B_1$ &
      $A_1B_1C_2$ & $A_1B_1C_1$ & $A_1B_1C_1C_2$ & $A_1B_1B_2$ \\ \hline
     $A_1B_1B_2C_2$ & $A_1B_1B_2C_1C_2$ & $A_1A_2B_1$ & $A_1A_2B_1C_2$ &
     $A_1A_2B_1C_1C_2$ \\ \hline $A_1B_2C_1$ & $A_2B_1C_1$& $A_1A_2B_1C_1$ & $A_1B_1B_2C_1$& $\id$ \\
\hline
\label{tbl:Cfirst}
  \end{tabular*}
\end{table}

\subsection{Causally separable quantum processes}

Given that quantum processes have a simple description in terms of process matrices, it is natural to ask whether the property of causality can also be expressed in terms of simple conditions on these matrices. Consider a bipartite quantum process for Alice and Bob, and assume that it is a fixed-order process compatible with the causal configuration $A\cp B$. In that case, as argued earlier, the only constraint imposed by causal order is that the process should involve no signaling from Bob to Alice. As pointed out in the previous subsection, there can be signaling from Bob to Alice if and only if the process matrix $W^{A_1A_2B_1B_2}$ contains terms of type $A_1B_2$ or $A_1B_1B_2$. Therefore, a process matrix is compatible with $A\cp B$ if and only if none of these types of terms appear in its expansion. This means that such a process matrix has the form
\begin{equation}
W^{A\cp B} = W^{A_1A_2B_1}\otimes \id^{B_2},\label{ApB}
\end{equation}
where $ W^{A_1A_2B_1}\geq 0$ (with $\textrm{Tr} W^{A_1A_2B_1} = d_{A_2}$) contains at most terms of type $\id$, $A_1$, $B_1$, $A_1B_1$, $A_2B_1$, $A_1A_2B_1$. (This is equivalent to saying that $W^{A_1A_2B_1}$ is a valid process matrix for the case where Bob has a trivial output system, $\mathcal{H}^{B_1}= \mathbb{C}^1$.)

Similarly, in the case where $A\ncp\ncf B$, the process matrix has the form
\begin{equation}
W^{A\ncp\ncf B} = W^{A_1B_1}\otimes \id^{A_2B_2},
\end{equation}
where $W^{A_1B_1}\geq 0$, $\textrm{Tr}W^{A_1B_1}=1$. Such a process is realized in a situation in which Alice and Bob receive input systems in a joint quantum state with a density matrix $W^{A_1B_1}$, and their output systems are discarded.

We can unify these two conditions to write down the form of a process matrix compatible with $B\ncp A$, which is identical to (\ref{ApB}),
\begin{equation}
W^{B\ncp A} = W^{A_1A_2B_1}\otimes \id^{B_2},\label{BnpA}
\end{equation}
where $W^{A_1A_2B_1}$ is a valid process matrix for the case where $\mathcal{H}^{B_1}= \mathbb{C}^1$.

As shown in Ref.~\cite{networks} within a different framework, all process matrices of the type (\ref{BnpA}) can be realized by embedding the experiments of Alice and Bob in a quantum circuit, so that Bob's experiment does not precede Alice's experiment in the order of the circuit composition. Most generally, this corresponds to providing Alice with an input system that is entangled with an ancilla, then sending Alice's output together with the ancilla through a quantum channel into Bob's input, and then discarding Bob's output. Such a process is referred to as quantum `channel with memory'.

As we have seen earlier, a bipartite causal process is one that can be written in the form (\ref{eq:biconstraintW2}), where $\mathcal{W}^{A\ncp B}$ and $\mathcal{W}^{B\ncp A}$ are two processes compatible with ${A\ncp B}$ and ${B\ncp A}$, respectively. It is then tempting to conjecture that the class of causal quantum processes might be those whose process matrices can be written in the form
\begin{equation}
W^{A_1A_2B_1B_2} = q\ W^{A\ncp B}\ +\ (1-q)\ W^{B\ncp A}, \hspace{0.1cm} 0\leq q \leq 1, \label{Wcs}
\end{equation}
where $W^{A\ncp B}$ and $W^{B\ncp A}$ have the form defined in Eq.~(\ref{BnpA}). Certainly, since the probabilities for the outcomes in the quantum process framework are linear functions of the process matrix, a process matrix of the form (\ref{Wcs}) describes a causal process.

However, the condition for a process to be causal (Eq.~(\ref{eq:biconstraintW2})) does not imply that $\mathcal{W}^{A\ncp B}$ and $\mathcal{W}^{B\ncp A}$ in the convex decomposition of the process should themselves be quantum process; only their convex mixture needs to be. While it is conceivable that the structure of quantum processes might imply the form (\ref{Wcs}) (indeed, this has been shown to hold for a limited class of bipartite quantum processes \cite{brukner2}), there is no obvious reason to expect this to hold in the general case. In fact, we will see that the natural generalization of condition (\ref{Wcs}) to the multipartite case is not equivalent to the condition that a process is causal (the same holds also for other possible generalizations that we will discuss later). Very recently, the same was shown to hold also in the bipartite case, by Feix, Ara\'{u}jo, and Brukner \cite{Feix}.

A bipartite quantum process that admits the decomposition  (\ref{Wcs}) was called \textit{causally separable} \cite{OCB}. One way to think of the relation between causal and causally separable quantum processes is in analogy with the relation between Bell-local and separable (non-entangled) quantum states. Given an arbitrary multipartite quantum state with a density matrix $\rho^{AB\cdots}$, the probabilities for the outcomes of a set of local POVM measurements $\{M^A_{i}\}_{i\in O^A}$, $\{M^B_{j}\}_{j\in O^B}$, $\cdots$ ( $\sum_{i\in O^A} M_i^A = \id^A$, $\sum_{j\in O^B} M^B_j = \id^A$, $\cdots$) are given by
\begin{equation}
 p(i,j, \cdots | \{M^A_{i}\}_{i\in O^A}, \{M^B_{j}\}_{j\in O^B}, \cdots)= \textrm{Tr}(\rho^{AB\cdots } M_i^A\otimes M_j^B\otimes\cdots ).\label{Born}
\end{equation}
A Bell-local state is one for which the joint probabilities for the outcomes of any combination of local measurements admits a local hidden variable description (and hence such a state cannot be used to violate any Bell inequality \cite{Bell}), i.e., the joint distribution can be written as a probabilistic mixture of factorizing local distributions,
\begin{equation}
 p(o^A, o^B, \cdots|s^A, s^B, \cdots , \rho^{AB})= \sum_{\lambda}\ p(\lambda)\ p(o^A|s^A, \lambda)\ p(o^B|s^B, \lambda)\cdots,
\end{equation}
where $\lambda$ is some variable with a probability distribution $p(\lambda)$, $s^A$, $s^B$, $\cdots$ are the local measurement settings (each corresponding to a specific local POVM measurement $\{M^A_{i}\}_{i\in O^A}$, $\{M^B_{j}\}_{j\in O^B}$, $\cdots$), and $o^A$, $o^B$, $\cdots$ are their outcomes (corresponding to $i$, $j$, $\cdots$ in the expression (\ref{Born})). A separable quantum state is one for which each of the local distributions $p(o^A|s^A, \lambda)$, $p(o^B|s^B, \lambda)$, $\cdots$ in Eq.~(\ref{Born}) itself can be thought of as arising from the respective local measurement being applied on a local quantum state, which means that the density matrix of the state can be written
\begin{equation}
\rho^{AB\cdots} = \sum_{\lambda} p(\lambda)\ \rho^A(\lambda)\otimes \rho^B(\lambda)\otimes \cdots. \label{separablestate}
\end{equation}
A separable quantum state is clearly Bell local, but the reverse is known not to be true \cite{Werner}. The relation between causal (\ref{eq:biconstraintW2}) and causally separable (\ref{Wcs}) bipartite quantum processes can be seen in an analogous way -- a causally separable process is one for which the processes into which we decompose the process are themselves valid quantum processes.

Here, we propose to extend the notion of causal separability to the multipartite case based on this analogy. \\

\begin{defn}
\textbf{(Causally separable quantum process):} A quantum process is called causally separable if and only if it can be decomposed in the canonical form given by Theorem \ref{thm:1'}, with the additional condition that each process on the right-hand side of Eq.~(\ref{Npart}) is a quantum process. (Note that since the canonical form is defined iteratively, the latter is understood to hold for all conditional processes in this definition.)
\end{defn}

By a direct analogy, causally separable processes can be defined for any theory formulated in the process framework, but here we will be interested specifically in quantum processes. The process matrix of a causally separable quantum process will be called a causally separable process matrix.\\

%Our goal will be to reduce the definition of causal separability to an equivalent set of simpler conditions on the process matrix, similarly to the bipartite case.

\subsection{Non-equivalence between causal and causally separable multipartite processes: a tripartite example}

We now give an example of a tripartite quantum process that is causal but causally non-separable, which demonstrates that these two concepts are not equivalent, at least in the case of more that two parties. A similar conclusion based on the same example has been obtained independently by Costa and is presented in Ref.~\cite{Araujo3}.

The example is inspired by the idea of superposition of causally ordered quantum circuits by means of the so-called \textit{quantum switch} technique \cite{Chiribella12}, where the order of two black-box quantum operations is made to depend on the value of a quantum control bit prepared in superposition of the two possible logical values. Each of the input and output systems of Alice and Bob in our example will be assumed to be a two-dimensional (qubit) system. We can imagine that this is the spin degree of freedom of a spin-$\frac{1}{2}$ particle, which enters each laboratory, interacts with the devices inside, and leaves. The particle could be prepared so as to go in superposition along two different possible paths -- along one path, it goes first through Alice's laboratory and then through Bob's, whereas along the other path it goes first through Bob's laboratory and then through Alice's. For simplicity, we can imagine that the experiment is arranged in such a way that the particle would always go through Bob's laboratory at a fixed time, but depending on the value of the control bit, it would go through Alice's laboratory before or after that. It is assumed that independently of the time at which the system may go through Alice' laboratory in a given run, Alice would apply the same operation on it. To understand the effect of such a setup, consider first the case in which Alice and Bob each apply a unitary operation on the system, $U_A$ and $U_B$, respectively. Let us denote the Hilbert space of the control qubit (path degree of freedom) by $\mathcal{H}^c$, and that of the system (spin degree of freedom) by $\mathcal{H}^s$. If $|0\rangle^c$ corresponds to the path in which Alice is before Bob and $|1\rangle^c$ to the path in which Bob is before Alice, if we initially prepare the particle in the state, say, $\rho_{in}^{cs} =|\Psi\rangle \langle \Psi|_{in}^{cs} $, where $|\Psi\rangle_{in}^{cs}= (\alpha |0\rangle^c +\ \beta |1\rangle^c)|\psi \rangle^s$, at the end it will be in the state  $\rho_{fi}^{cs} = |\Psi\rangle \langle \Psi|_{fi}^{cs} $, where  $ |\Psi\rangle_{fi}^{cs} = \alpha |0\rangle^c U^s_BU^s_A |\psi\rangle^s +\ \beta |1\rangle^c U^s_AU^s_B |\psi\rangle^s $. Now, if a third party, Charlie, performs an operation on the joint system $\mathcal{H}^c\otimes \mathcal{H}^s$ subsequently, he can distinguish this situation from a situation in which the order between the operations of Alice and Bob is conditioned on a classical bit (e.g., modeled by the initial state of the control qubit being in a `classical' mixture of the two possible values, $|\alpha|^2 |0\rangle\langle 0|^s+\ |\beta|^2 |1\rangle\langle 1|^s$, instead of a coherent superposition) by performing a suitable measurement. In fact, it was shown in Ref.~\cite{Chiribella12b} that by exploiting such a coherent strategy, Charlie can perfectly distinguish between a pair of unitaries $U^A$ and $U^B$ that commute or anticommute by using each of the unitaries only once, which is impossible if the order of the unitaries is conditioned on a classical bit. An experimental demonstration of this effect was recently reported in Ref.~\cite{Procopio}. 

In the general case, the operations of Alice and Bob need not be unitary and may have different possible outcomes. Every such operation, however, can be seen as the result of a joint unitary on the input system and a local ancilla, such that the outcome remains stored on the local ancilla in a particular basis. Similarly, any local `choice' of operation can be modeled by a larger unitary on all systems involved plus a local ancilla that carries the `choice' variable. Thus, we can have Alice and Bob perform general operations in this setup by purifying their local operations to unitaries and deferring the reading of their outcomes to the end of the whole experiment. (Note that in order not to destroy the superposition, the whole experiments needs to be performed coherently, which may be unrealistic for local operations performed by macroscopic devices, but is in principle compatible with standard quantum mechanics.)

\begin{figure}[!htb]
\centering
\includegraphics[scale=.25]{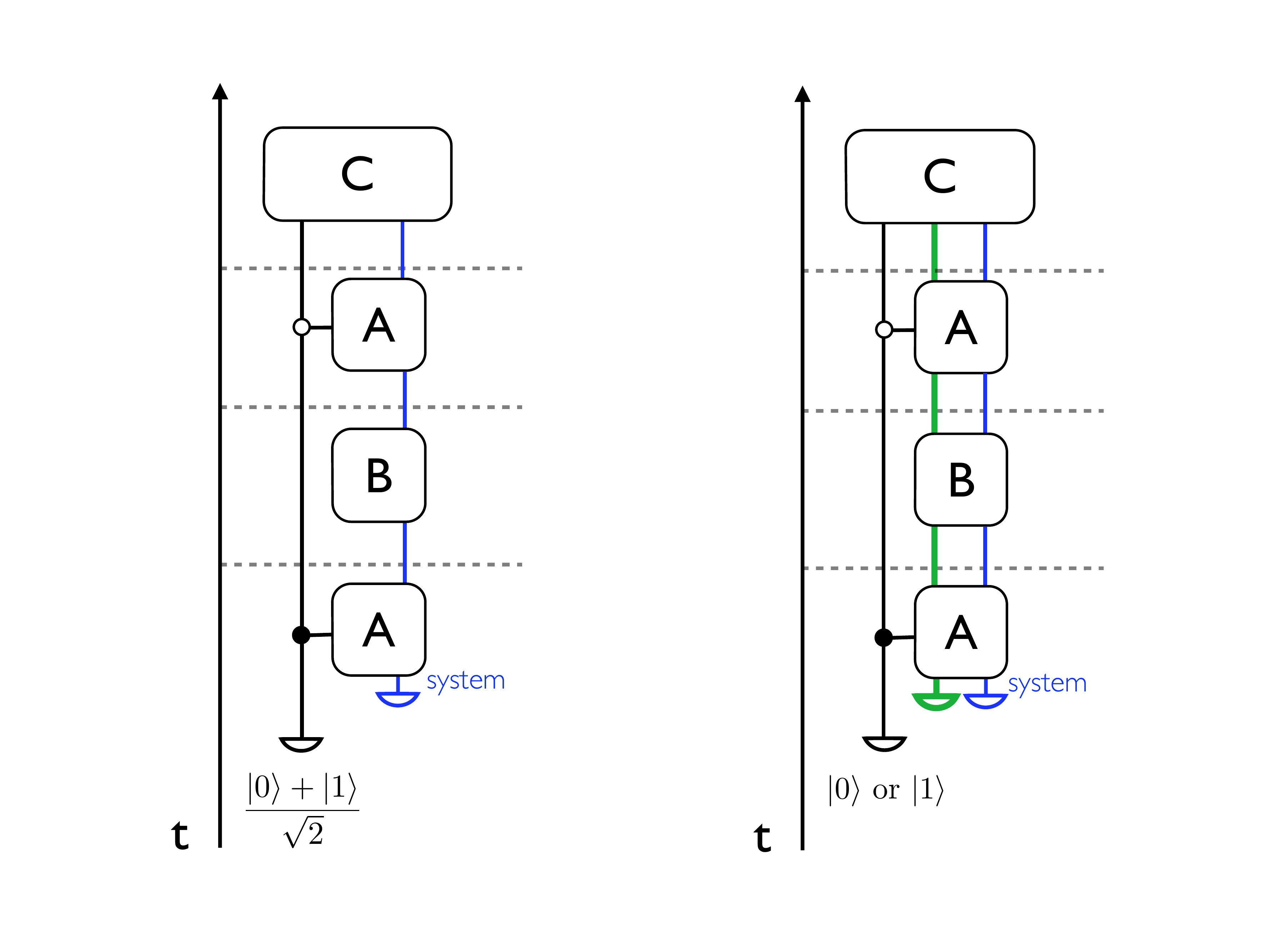}
\caption{The left diagram illustrates the circuit with quantum control. The right diagram illustrates a simulation of the same correlations with a classically controlled circuit using input and output systems of larger dimensions.}
\label{fig:simulation}
\end{figure}

In our example, we will take $\alpha = \beta = \frac{1}{\sqrt{2}}$, and we will assume, as described above, that Charlie can operate on both the path and spin degrees of freedom of the particle after it has interacted with Alice and Bob. In other words, Charlie's input system will be four dimensional, and we will formally decompose it into two qubit subsystems, $\mathcal{H}^{C_1} = \mathcal{H}^{C^c_1} \otimes \mathcal{H}^{C^s_1} $, where $\mathcal{H}^{C^c_1}$ and $\mathcal{H}^{C^s_1}$ correspond to the path and spin degrees of freedom, respectively. Since Charlie operates last, we do not need to introduce a non-trivial output system for him, i.e., his output system will be assumed one-dimensional. The process matrix relating the local experiment of Alice, Bob, and Charlie in this setup can easily be obtained by describing the experiment in the form of a circuit in which Alice's operation is represented by two controlled operations at two possible times, such that one of them would act nontrivially depending on the state of the control qubit (left diagram on Fig.~\ref{fig:simulation}). Using the CJ representation of the channels connecting the different boxes, we obtain
\begin{gather}
W^{A_1A_2B_1B_2C_1C_2} = |W\rangle\langle W|^{A_1A_2B_1B_2C_1C_2},\label{nonsep1}
\end{gather}
where
\begin{gather}
|W\rangle^{A_1A_2B_1B_2C_1C_2} =     (|0\rangle^{C_1^c}  |\psi\rangle^{A_1}|\Phi^+\rangle ^{A_2B_1} |\Phi^+\rangle^{B_2C^s_1}  +\   |1\rangle^{C_1^c}|\psi\rangle^{B_1} |\Phi^+\rangle^{B_2A_1}|\Phi^+\rangle^{A_2C^s_1})/\sqrt{2},\label{nonsep2}
\end{gather}
with $ |\Phi^+\rangle  =|00\rangle+\ |11\rangle$. It can be verified that $W^{A_1A_2B_1B_2C_1C_2} $ contains only allowed terms. This process matrix is a rank-one projector, and hence it cannot be written as a convex mixture of different process matrices. Therefore, if it is causally separable, it must be of one of the types ${W}^{(A,B)\ncp C}$, ${W}^{(B,C)\ncp A}$, or ${W}^{(A,C)\ncp B}$. But each of these types of process matrices should permit no signaling from two of the parties to the third one (e.g., in the first case there can be no signaling from Alice and Bob to Charlie). However, the above process matrix permits signaling to any of the parties from some of the other parties. Indeed, to see that there can be signaling from Alice and Bob to Charlie, imagine that Alice and Bob choose to perform the unitary operations $U_A$ and  $U_B$. In this case, Charlie will receive the state $[|0\rangle^{C_1^c} (U_BU_ A|\psi\rangle)^{C_1^s}+\ |1\rangle^{C_1^c} (U_AU_ B|\psi\rangle)^{C_1^s} ]/\sqrt{2}$, which can be different for different choices of the unitaries of Alice and Bob, and can therefore yield different probabilities for the outcomes of some measurement of Charlie. To see that we can have signaling from Alice to Bob or vice versa, notice first that there can be no signaling from Charlie to Alice and Bob (Charlie has a trivial output system). This means that we have a well-defined reduced process for Alice and Bob, whose process matrix is  
\begin{gather}
W^{A_1A_2B_1B_2}= \frac{1}{2} (|\psi\rangle\langle \psi|^{A_1} \otimes |\Phi^+\ \rangle\langle \Phi^+|^{A_2B_1} \otimes \id^{B_2} +\ |\psi\rangle\langle \psi|^{B_1} \otimes |\Phi^+\ \rangle\langle \Phi^+|^{B_2A_1} \otimes \id^{A_2} ).\label{sep3}
\end{gather}
This is a causally separable bipartite process matrix that can be interpreted as describing an equally weighted probabilistic mixture of two fixed-order processes -- the first one describes a situation in which the input state $|\psi\rangle $ is sent into Alice's input, her output is sent into Bob's input through the identity channel, and Bob's output is discarded; the second one describes the analogous situation with the roles of Alice and Bob interchanged. Clearly, since in the first situation there is an ideal channel from Alice to Bob, there can be signaling from Alice to Bob in this process (even if imperfect on average), and similarly from Bob to Alice. Therefore, the process matrix given by Eqs.~(\ref{nonsep1}) and (\ref{nonsep2}) is not causally separable.

The fact that the process is causal follows immediately from the fact that the reduced process for Alice and Bob is causally separable (and hence also causal). Specifically, we have $\mathcal{W}^{AB} = \frac{1}{2} \mathcal{W}^{B\ncp A} +\ \frac{1}{2} \mathcal{W}^{A\ncp B} = \frac{1}{2} \mathcal{W}^{B|A}_{[A]^{\RN{1}}}\circ \mathcal{W}_{[A]^{\RN{1}}}^A +\ \frac{1}{2} \mathcal{W}^{A|B}_{[B]^{\RN{1}}}\circ \mathcal{W}^B_{[B]^{\RN{1}}}$. But the tripartite process is simply $\mathcal{W}^{ABC} = \mathcal{W}^{C|AB} \circ  \mathcal{W}^{AB} = \frac{1}{2}\mathcal{W}^{C|AB} \circ \mathcal{W}^{B|A}_{[A]^{\RN{1}}}\circ \mathcal{W}^A_{[A]^{\RN{1}}} +\ \frac{1}{2}\mathcal{W}^{C|AB} \circ \mathcal{W}^{A|B}_{[B]^{\RN{1}}}\circ \mathcal{W}^B_{[B]^{\RN{1}}}  $, which is the form of a causal process. This observation suggests how the probabilities of Alice, Bob, and Charlie can be simulated without using a quantum switch, if we allow the parties to have larger input and output systems. Since the reduced probabilities of Alice and Bob can be realized by conditioning their order on a classical random bit, all that is needed in order for the tripartite process to be reproduced in this way is for Charlie to receive the information about the settings and outcomes of Alice and Bob so as to produce the necessary $p(o^C|s^A,o^A,s^B,o^B,s^C)$. Therefore, if in addition to the qubit system that goes between Alice and Bob there is another (possibly infinite-dimensional) system on which each party writes down his/her setting and outcome (right diagram on Fig.~\ref{fig:simulation}), and this system at the end enters Charlie's laboratory (or, alternatively, the state on Charlie's original input system is prepared based on this information), the process can be simulated using classically random causal configurations. 

By a similar argument we can construct a large class of multipartite processes that are causal but not causally separable. Consider a situation in which the order of all but one of the parties is conditioned on the state of a control system prepared in superposition, and subsequently all systems on which these parties have operated together with the control system are sent into the input of the last party. If all systems were initially prepared in a pure state and all channels are unitary ones, the process matrix will have rank $1$, and unless the process is fixed-order causal, it cannot be causally separable. Yet, it will be causal because the reduced process for all parties except for the last one will be causally separable (and hence causal) due to the fact that when we trace out the control system, the process for these parties would be a classical probabilistic mixture of fixed-order processes. Since the full process is obtained by multiplying the conditional process of the last party with the reduced process of the previous ones, the full process is causal. It can be simulated using classical control of the order of the parties by allowing larger input and output systems by which the settings and outcomes of all other parties are made available to the last one.

\subsection{Non-causality can be activated by shared entanglement}

We now show another peculiar property of the concepts of causality and causal separability of quantum processes. One of the key assumptions in the derivation of the quantum process matrix framework is that every process can be extended by supplying the parties with ancillary input systems in an arbitrary quantum state, yielding another valid process. Intuitively, since a joint input state is a non-signaling process that is compatible with any causal configuration, one may expect that by adding such a state to a causal quantum process would yield again a causal process. We now show that this is not the case. We refer to this effect as \textit{activation} of non-causality. 

We give a particular example of a tripartite causal quantum process matrix, constructed on the basis of the bipartite process matrix presented in Ref.~\cite{OCB}, 
\begin{gather}
W^{A_1A_2B_1B_2} = \frac{1}{4}  ( \id^{A_1A_2B_1B_2} +\ \frac{1}{\sqrt{2}}\sigma_z^{A_1}\sigma_x^{B_1} \sigma_z^{B_2} +\   \frac{1}{\sqrt{2}}\sigma_z^{A_2} \sigma_z^{B_1}) ,\label{WOCB}
\end{gather}
which itself can violate a causal inequality and is hence non-causal (see Ref.~\cite{OCB}). Here, the input and output systems of Alice and Bob are two-level systems. In our tripartite construction, the input and output systems of Alice and Bob are also two-level systems, and we add Charlie, who has a trivial input system and a two-level output system. In terms of the Pauli matrices $\sigma_x$, $\sigma_y$, $\sigma_z$, the process matrix we consider has the form  

\begin{gather}
W^{A_1A_2B_1B_2C_2} = \frac{1}{4} ( \id^{A_1A_2B_1B_2C_2} +\ \frac{1}{\sqrt{2}}\sigma_z^{A_1} \sigma_z^{B_1} \sigma_z^{B_2} \sigma_x^{C_2}+\   \frac{1}{\sqrt{2}}\sigma_z^{A_2} \sigma_z^{B_1} \sigma_z^{C_2}).
\end{gather} 
 
 The fact that this is a valid process matrix follows from the fact that it has the right normalization, contains only allowed $\sigma$ terms, and is positive semidefinite. The latter is easy to see by noticing that relative to the $\{|0\rangle, |1\rangle \}$ basis of system $B_1$ (this is the eigenbasis of $\sigma_z$ corresponding to eigenvalues $+1$ and $-1$, respectively), the process matrix can be written 
 \begin{gather}
 W^{A_1A_2B_1B_2C_2} = |0\rangle\langle 0|^{B_1}  \otimes  \frac{1}{4}  ( \id^{A_1A_2B_2C_2} +\ \frac{1}{\sqrt{2}}\sigma_z^{A_1} \sigma_z^{B_2} \sigma_x^{C_2}+\   \frac{1}{\sqrt{2}}\sigma_z^{A_2} \sigma_z^{C_2}) +\   |1\rangle\langle 1|^{B_1}  \otimes  \frac{1}{4}  ( \id^{A_1A_2B_2C_2} - \frac{1}{\sqrt{2}}\sigma_z^{A_1} \sigma_z^{B_2} \sigma_x^{C_2}-   \frac{1}{\sqrt{2}}\sigma_z^{A_2} \sigma_z^{C_2}) .
 \end{gather} 
Now, the operator $\frac{1}{4}  ( \id^{A_1A_2B_2C_2} +\ \frac{1}{\sqrt{2}}\sigma_z^{A_1} \sigma_z^{B_2} \sigma_x^{C_2}+\   \frac{1}{\sqrt{2}}\sigma_z^{A_2} \sigma_z^{C_2}) $ is identical to that in Eq.~\eqref{WOCB} except that we have the system $C_2$ in the place of $B_1$, and this operator has been shown to be positive semidefinite. The operator $\frac{1}{4}  ( \id^{A_1A_2B_2C_2} - \frac{1}{\sqrt{2}}\sigma_z^{A_1} \sigma_z^{B_2} \sigma_x^{C_2}-   \frac{1}{\sqrt{2}}\sigma_z^{A_2} \sigma_z^{C_2})$ differs only by the fact that the nontrivial $\sigma$ terms come with a minus sign, and can be obtained from the first operator by a unitary transformation (e.g., one that takes $\sigma_x^{C_2}$ to $-\sigma_x^{C_2}$  and  $\sigma_z^{C_2}$ to $-\sigma_z^{C_2}$, such as $\sigma_y^{C_2}$).

To see that this process matrix describes a causally separable process, note that it permits no signaling from Alice and Bob to Charlie, i.e., it can be formally written as $\mathcal{W}^{A,B,C} = \mathcal{W}^{A,B|C}\circ\mathcal{W}^{C}$. But conditionally on any event in Charlie's laboratory, which is most generally described by some CP map with CJ operator $M^{C_2}\geq 0$, Alice and Bob are left with a bipartite process with process matrix
\begin{gather}
W^{A_1A_2B_1B_2}_{M^{C_2}} = \Tr_{C_2} [(M^{C_2}\otimes \id^{A_1A_2B_1B_2})\ W^{A_1A_2B_1B_2C_2}]/ \Tr [M^{C_2}].
\end{gather}    
This process matrix is obviously a linear combination of the identity and terms containing only $\sigma_z$ operators on different subsystems, i.e., it is diagonal in a given local basis (the $\{|0\rangle, |1\rangle \}$ basis for each subsystem). It was shown in Ref.~\cite{OCB} that all such bipartite process matrices are causally separable (though we remark that the same was shown not to hold for multipartite processes \cite{Baumeler2}).

Imagine now that we supply Bob and Charlie with the entangled input state $\frac{1}{2}|\Phi^+\rangle \langle \Phi^+|^{C_1'B_1'} $, which yields the new process
\begin{gather}
W^{A_1A_2B_1B_1'B_2C_1'C_2} = W^{A_1A_2B_1B_2C_2} \otimes \frac{|\Phi^+\rangle \langle \Phi^+|^{C_1'B_1'}}{2}.\label{tripartOCB2}
\end{gather} 
If Charlie performs the identity unitary channel from $C_1'$ to $C_2$ in his laboratory, which is described by $M^{C_1'C_2} =|\Phi^+\rangle \langle \Phi^+|^{C_1'C_2}$, Alice and Bob are left with the bipartite process
\begin{gather}
W^{A_1A_2B_1B_1'B_2} = \frac{1}{4}  ( \id^{A_1A_2B_1B_1'B_2} +\ \frac{1}{\sqrt{2}}\sigma_z^{A_1}\sigma_z^{B_1}\sigma_x^{B_1'} \sigma_z^{B_2} +\   \frac{1}{\sqrt{2}}\sigma_z^{A_2} \sigma_z^{B_1} \sigma_z^{B_1'}).\label{WOCB2}
\end{gather}
This can be easily seen from the fact that taking the partial trace of $W^{A_1A_2B_1B_1'B_2C_1'C_2}$ with the operator $|\Phi^+\rangle \langle \Phi^+|^{C_1'C_2}$ is formally identical (up to a normalization) to a local projection in a quantum-state teleportation protocol \cite{teleportation}, which amounts to `teleporting' the part of the matrix on $C_2$ onto $B_1'$. (Note that the standard notion of teleportation is defined for quantum states and not process matrices, and the protocol requires a correcting operation on the receiver's side since a projection of the kind above, which does not require correction, cannot be accomplished deterministically \cite{teleportation}). The process matrix \eqref{WOCB2} is similar to \eqref{WOCB}, except that the local operators on $B_1$ in the non-trivial sigma terms in Eq.~\eqref{WOCB} are now on $B_1'$, and there is a $\sigma_z$ operator on $B_1$ in each such term. This process matrix is non-causal, because it allows Alice and Bob to obtain any correlations that they could obtain using the non-causal process matrix \eqref{WOCB}. This can be done as follows. Alice always performs the same operations that she would perform with the process matrix \eqref{WOCB}. Bob performs a measurement on system $B_1$ in the $\{|0\rangle,|1 \rangle\}$ basis. If he obtains the outcome $|0\rangle$, then it is as if Alice and Bob share the process matrix \eqref{WOCB} with $B_1'$ in the place of $B_1$. He will then apply any operation from $B_1'$ to $B_2$ that he would apply from $B_1$ to $B_2$ with the process matrix \eqref{WOCB}, which yields the same joint probabilities for Alice and Bob as those with the process matrix \eqref{WOCB}. If Bob obtains the outcome $|1\rangle$ for his measurement on $B_1$, then it is as if Alice and Bob share the same process matrix as \eqref{WOCB} with $B_1'$ in the place of $B_1$ but with a minus sign in front of each of the two nontrivial $\sigma$ terms. This process matrix is equivalent to the previous one under a change of basis by the unitary $\sigma_y^{B_1'}$. Therefore, Bob can simply apply from $B_1'$ to $B_2$ the same operations he would apply from $B_ 1$ to $B_2$ with the process matrix \eqref{WOCB} but transformed by the unitary transformation $\sigma_y^{B_1'}$. Again, this yields the same joint probabilities for Alice and Bob as with the process matrix \eqref{WOCB}. In particular, Alice and Bob can use this strategy to violate the causal inequality described in Ref.~\cite{OCB}. The process matrix \eqref{WOCB2} is thus non-causal, and so is the tripartite process matrix \eqref{tripartOCB2}. 

It is not known at present whether non-causal processes can be realized in agreement with the known laws of quantum mechanics without resorting to post-selection. We have seen in the previous subsection that we can realize causally non-separable processes, which are nevertheless causal. Here, we see that certain causal processes can become non-causal when supplied with shared entanglement. The ability to extend a process with shared entanglement seems natural to expect for any experimentally realizable process. From this perspective, this result suggests that either non-causal processes may be possible, or that there may exist causally separable processes, as defined above, that cannot be realized in practice. 

\subsection{Extensibly causal and extensibly causally separable quantum processes}

The fact that according to our definition of causal separability there exist causal processes that may be activated to non-causal ones by shared entanglement naturally suggests the definition of the following classes of processes that do not have this counterintuitive property.\\

\begin{defn}
\textbf{(Extensibly causal quantum process):} A quantum process that is causal and remains causal under extension with input systems in an arbitrary joint quantum state is called extensibly causal.\\\
\end{defn}

\begin{defn}
\textbf{(Extensibly causally separable (ECS) quantum process):} A quantum process that is causally separable and remains causally separable under extension with input systems in an arbitrary joint quantum state is called extensibly causally separable (ECS). 
\end{defn}

The process matrices of these types of processes will also be referred to as extensibly causal and ECS process matrices, respectively. \\

\textit{Note.} These definitions can be formulated analogously for more general process theories that permit composite local systems. \\

Do these classes of processes correspond to something easy to describe in practice, and are they different at all? It is immediate to see the following facts. \\ 

\textbf{Observation 1:}  All bipartite causally separable processes are ECS. This is because, if we add an arbitrary joint input ancilla to a process matrix of the form \eqref{Wcs}, we again obtain a process matrix of the same form. Therefore, the notion of extensible causal separability can be seen as another possible multipartite extension of the bipartite notion of causal separability, which, however, is linked in a less direct way to the theory-independent notion of causality.\\

\textbf{Observation 2:}  Extensibly causal and ECS processes are not equivalent in general. Indeed, the causally non-separable tripartite process \eqref{nonsep1} based on the quantum switch is also extensibly causal (our proof that it is causal applies also if the parties share entangled input ancillas).\\ 

\textbf{Comment:} Recently, Feix, Ara\'{u}jo, and Brukner gave an example of a bipartite quantum process that is causal but not extensibly causal \cite{Feix}, proving that causality and extensible causality are different in the bipartite case too. While in the tripartite case we have seen that extensible causality is also different from causal separability, it is currently an open problem whether the same holds in the bipartite case. \\

In the next subsection, we derive a characterization o  f the tripartite ECS processes in terms of conditions on the form of the process matrix which generalize the conditions in the bipartite case (Eqs. \eqref{BnpA}, \eqref{Wcs}).

\subsection{Structure of tripartite ECS process matrices}

Recalling the definition of causally separable process, let us first state an obvious consequence of this definition for the structure of causally separable (though not necessarily ECS) process matrices. Since the probabilities of a quantum process are linear in the process matrix, the requirement that a causally separable process decomposes as in Theorem \ref{thm:1'} where all processes on the right-hand side of Eq.~(\ref{Npart}) are valid quantum processes means that a causally separable process matrix is one that can be written in the form
\begin{gather}
 {W}_{cs}^{1_11_2\cdots n_1n_2} = \sum_{i=1}^n q_i {W}^{(1,\cdots,{i-1},{i+1},\cdots, n)\ncp {i}}, \hspace{0.2cm}0\leq q_i, \forall i, \hspace{0.1cm}\sum_{i=1}^{n}q_i=1, \label{cslong}
\end{gather}
where ${W}^{(1,\cdots, {i-1}, {i+1},\cdots,n)\ncp {i}}$ is a process matrix which describes a process $\mathcal{W}^{(1,\cdots, {i-1}, {i+1},\cdots,n)\ncp {i}}$ with the property
\begin{gather}
 \mathcal{W}^{(1,\cdots, {i-1}, {i+1},\cdots,n)\ncp {i}} = \mathcal{W}_{cs}^{1,\cdots, {i-1}, {i+1},\cdots,n| i}\circ  \mathcal{W}^{i},\label{cslong2}
\end{gather}
where for $n>1$ the conditional process $\mathcal{W}_{cs}^{1,\cdots, {i-1}, {i+1},\cdots,n| i}$ is a causally separable process for every value of the event in $i$, and for $n=1$ it is the trivial process. Note that the requirement that $\mathcal{W}^{(1,\cdots, {i-1}, {i+1},\cdots,n)\ncp {i}} $ is a quantum process that permits no signaling from the rest of the parties to $i$ guarantees that both the reduced and the conditional process on the right-hand side of Eq.~\eqref{cslong2} are valid quantum processes (this can be seen from the (no) signaling condition in Proposition \ref{prop:3.2}). 

In the case of two parties, we have seen that the process matrices $W^{A\ncp B}$, whose processes obey $\mathcal{W}^{A\ncp B} = \mathcal{W}_{cs}^{A|B}\circ \mathcal{W}^{B}$ (note that any monopartite process is trivially causally separable and ECS), are those that can be written in the form $W^{A\ncp B}= W^{B_1B_2A_1}\otimes \id^{A_2}$, and the general form of bipartite causally separable process matrices is \eqref{Wcs}. As noted already, this is also the general form of the bipartite ECS process matrices. Our goal is to obtain a similar conditionfor triparite ECS processes. 

First, let us consider a process of the form $\mathcal{W}^{(A,B )\ncp C } = \mathcal{W}_{cs}^{A,B|C}\circ  \mathcal{W}^{C}$, where $\mathcal{W}^{C}$ is a monopartite quantum process  and $\mathcal{W}_{cs}^{A,B|C}$ is a bipartite conditional process which is causally separable for each possible event in $C$. Since in particular there should be no signaling from Alice and Bob to Charlie in such a process, its process matrix, which we will denote $W_{(A,B )\ncp C}^{A_1A_2B_1B_2C_1C_2} $, can at most contain the types of terms listed in Table~\ref{tbl:Cfirst}. These are the terms that do not permit signaling from Alice and Bob to Charlie according to Proposition \ref{prop:3.2}.

We will first obtain necessary and sufficient conditions for such a process to be ECS. Note that we have not proven yet that a general ECS process matrix should have the form \eqref{cslong} where each of the terms ${W}^{(1,\cdots, {i-1}, {i+1},\cdots,n)\ncp {i}}$ is itself ECS. This will be shown later.

Every event in Charlie's laboratory is described by some CP map with CJ operator $M^{C_1C_2}\geq 0$, $\Tr M^{C_1C_2} \leq d_{C_1}$. Conditionally on such an event, Alice and Bob are left with the process matrix 
\begin{gather}
 W^{A_1A_2B_1B_2}_{ M^{C_1C_2} } = {\Tr_{C_1C_2} [W_{(A,B)\ncp C}^{A_1A_2B_1B_2C_1C_2} (\id^{A_1A_2B_1B_2} \otimes   M^{C_1C_2} )]} /{p( M^{C_1C_2} )},
\end{gather}
where $ {p( M^{C_1C_2} )} $ is the probability for the event $M^{C_1C_2}$ to occur in Carlie's laboratory (given the appropriate setting), which is independent of the operations performed by Alice and Bob since the process involves no signaling from Alice and Bob to Charlie. More specifically,   
\begin{gather}
 {p( M^{C_1C_2} )} = { \Tr [W^{C_1C_2}   M^{C_1C_2} ]},
 \end{gather}
 where 
 \begin{gather}
 W^{C_1C_2}  = \Tr_{A_1A_2B_1B_2} [W_{(A,B)\ncp C}^{A_1A_2B_1B_2C_1C_2} ( \frac{ \id^{A_1A_2B_1B_2} } {d_{A_2} d_{B_2} } \otimes \id^{C_1C_2}) ]
 \end{gather}
 is the reduced process of Charlie. The requirement that the conditional process for Alice and Bob is causally separable means that for all $M^{C_1C_2}$,
\begin{gather}
 W^{A_1A_2B_1B_2}_{ M^{C_1C_2} } =    q_{M^{C_1C_2}} W_{M^{C_1C_2}} ^{A\ncp B}+\ (1- q_{M^{C_1C_2}} ) W_{M^{C_1C_2}} ^{B\ncp A},
\end{gather}
where $W_{M^{C_1C_2}} ^{A\ncp B}$ and $W_{M^{C_1C_2}} ^{B\ncp A}$ are valid quantum processes compatible with ${A\ncp B}$ and ${B\ncp A}$, respectively, and $ q_{M^{C_1C_2}} \in [0,1]$ (all objects generally depend on $M^{C_1C_2}$). For convenience, we will write this simply in the form 
\begin{gather}
 W^{A_1A_2B_1B_2}_{ M^{C_1C_2} } =    \id^{A_2}\otimes \tilde{W}_{M^{C_1C_2}}^{A_1B_1B_2}+\ \id^{B_2}\otimes \tilde{W}_{M^{C_1C_2}}^{A_1A_2B_1},
\end{gather}
where $\tilde{W}_{M^{C_1C_2}}^{A_1B_1B_2}\geq 0$ and $\tilde{W}_{M^{C_1C_2}}^{A_1A_2B_1} \geq 0$, and the whole operator is a valid process matrix, i.e., it contains only allowed terms and is properly normalized. 

A sufficient condition for this to hold is that 
\begin{gather}
W_{(A,B)\ncp C}^{A_1A_2B_1B_2C_1C_2}  =  \id^{A_2}\otimes \tilde{W}^{A_1B_1B_2C_1C_2}+\ \id^{B_2}\otimes \tilde{W}^{A_1A_2B_1C_1C_2}, \label{firstcond1}
\end{gather}
where $\tilde{W}^{A_1B_1B_2C_1C_2}\geq 0$ and $\tilde{W}^{A_1A_2B_1C_1C_2}\geq 0$ are some positive semidefinite operators, whose sum gives a properly normalized quantum process matrix containing only the types of terms listed in Table \ref{tbl:Cfirst}. (We remark that each of $\tilde{W}^{A_1B_1B_2C_1C_2}\geq 0$ and $\tilde{W}^{A_1A_2B_1C_1C_2}\geq 0$ may contain terms that are forbidden in a process matrix, such as terms of type $C_2$, but these terms have to cancel in the sum.) Indeed, we have
\begin{gather}
 {\Tr_{C_1C_2} [W_{(A,B)\ncp C}^{A_1A_2B_1B_2C_1C_2} (\id^{A_1A_2B_1B_2} \otimes   M^{C_1C_2} )]} /p (M^{C_1C_2} )=   W^{A_1A_2B_1B_2}_{ M^{C_1C_2} }  = \id^{A_2}\otimes \tilde{W}_{M^{C_1C_2}}^{A_1B_1B_2}+\ \id^{B_2}\otimes \tilde{W}_{M^{C_1C_2}}^{A_1A_2B_1},\nonumber\\
\forall M^{C_1C_2}\geq 0,
\end{gather}
where 
\begin{gather}
\tilde{W}_{M^{C_1C_2}}^{A_1B_1B_2} = {\Tr_{C_1C_2} [\tilde{W}^{A_1B_1B_2C_1C_2}  (\id^{A_1B_1B_2} \otimes   M^{C_1C_2} )]} /p (M^{C_1C_2} ) \geq 0,
\end{gather}
\begin{gather}
\tilde{W}_{M^{C_1C_2}}^{A_1A_2B_1} = {\Tr_{C_1C_2} [\tilde{W}^{A_1A_2B_1C_1C_2}  (\id^{A_1A_2B_1} \otimes   M^{C_1C_2} )]} /p (M^{C_1C_2} ) \geq 0,
\end{gather}
and it is easy to see that since $W_{(A,B)\ncp C}^{A_1A_2B_1B_2C_1C_2}$ contains only the types of terms listed in Table \ref{tbl:Cfirst}, $W^{A_1A_2B_1B_2}_{ M^{C_1C_2} }  $ can only contain allowed terms.

It is immediate to see that this condition is sufficient also for the process matrix $W_{(A,B)\ncp C}^{A_1A_2B_1B_2C_1C_2}  $ to be ECS. This is because if $W^{A_1A_2B_1B_2C_1C_2}  $ has the above properties, any extension $W^{A_1A_2B_1B_2C_1C_2}  \otimes \rho^{A_1'B_1'C_1'}$, where $\rho^{A_1'B_1'C_1'}$ is a density matrix, also has these properties.  
 
 We now show that the form \eqref{firstcond1} is also a necessary condition for an ECS process matrix compatible with $(A,B)\ncp C$, which we will denote by $W_{ecs; (A,B)\ncp C}^{A_1A_2B_1B_2C_1C_2}  $. The proof makes use of the `teleportation' technique that we used in showing the activation of non-causality. Imagine that we supply Alice and Charlie respectively with ancillary systems $A_1'$ and $C_1'$ of dimension $d_{C_{1}} d_{C_{2}}$ each, which are prepared in the maximally entangled state $|\phi^+\rangle \langle\phi^+|^{A_1'C_1'}/(d_{C_{1}} d_{C_{2}})$, where $|\phi^+\rangle = \sum_{i=1}^{d_{C_{1}} d_{C_{2}}} |i\rangle^{A_1'} |i\rangle^{C_1'}$. Conditionally on Charlie performing a suitable operation and obtaining an outcome with CP map $M^{C_1C_2C_1' } \propto |\phi^+\rangle \langle\phi^+|^{(C_1C_2)C_1'}$, Alice and Bob will be left sharing a process matrix which, up to a normalization factor, has an identical form to that of $W_{ecs; (A,B)\ncp C}^{A_1A_2B_1B_2C_1C_2}  $ but with $A_1'$ in the place of $C_1C_2$. The requirement that this is a causally separable bipartite process matrix means that $W_{ecs; (A,B)\ncp C}^{A_1A_2B_1B_2C_1C_2}  $ must be of the form \eqref{firstcond1}.

So far, we have only obtained necessary and sufficient conditions for an ECS process matrix $W_{ecs; (A,B)\ncp C}^{A_1A_2B_1B_2C_1C_2}  $ compatible with $(A,B)\ncp C$ (and similarly for permutations of $A$, $B$, $C$). We next prove the general case.\\

\begin{prop}\label{prop:3.3}
Every tripartite ECS process matrix can be written in the form 
\begin{gather} \label{tripartecs}
{W}_{ecs}^{A_1A_2B_1B_2C_1C_2} =  q_1 W_{ecs; (A,B)\ncp C}^{A_1A_2B_1B_2C_1C_2}  +q_2 W_{ecs; (A,C)\ncp B}^{A_1A_2B_1B_2C_1C_2}   +\ q_3 W_{ecs; (B,C)\ncp A}^{A_1A_2B_1B_2C_1C_2}  , \hspace{0.3cm} q_i\geq 0, \hspace{0.1cm} \forall i = 1,2,3, \hspace{0.2cm} \sum_{i=1}^{3} q_i =1,
\end{gather}
where $W_{ecs; (A,B)\ncp C}^{A_1A_2B_1B_2C_1C_2} $ contains only terms from table \ref{tbl:Cfirst} and has the form \eqref{firstcond1}, and analogously for $W_{ecs; (A,C)\ncp B}^{A_1A_2B_1B_2C_1C_2}   $ and $W_{ecs; (B,C)\ncp A}^{A_1A_2B_1B_2C_1C_2}  $ by permutation. The Proof \ref{pr:prop:3.3} is given in the Appendix.\\
\end{prop}

The extension of this form to an arbitrary number of parties is left for future investigation.

\subsection{Processes realizable by classically controlled quantum circuits}

Bipartite ECS processes have a clear experimental realization. This raises the question of whether multipartite ECS processes can also be realized in practice, and if so, whether they correspond to a natural class of experimental procedures. (Note that in the bipartite case, ECS processes are equivalent to causally separable processes, but we have already seen that there are multipartite causally separable processes that can become non-causal under extension with entangled ancillas, and these do not have a known experimental realization.) Here, we will show that a particular class of processes which can be realized in practice, referred to as classically controlled quantum circuits, belong to the class of ECS processes, which is the smallest class of causal quantum processes that we have considered so far. Based on certain considerations, we furthermore conjecture that all ECS processes can be realized in this way (this is certainly true in the bipartite case).   

The idea of a classically controlled circuit can be thought of as falling within the paradigm of quantum lambda calculus with classical control \cite{Knill, SelingerValiron}. If we regard the local experiments of the parties as black-box operations, we may think that they are called, only once each, as part of a computation where at every time step a quantum operation is applied on some part of a quantum register depending on a classical protocol that may use as a variable the outcomes of past operations.  If black-box operations are involved in such a computation, their outcomes cannot be directly used (they remain `inside the box' until the end), but the order of subsequent operations of the circuit may nevertheless depend indirectly on the event inside such a black box, since it can be decided based on a measurement on the output system.

More concretely, we define such a process to have the following general realization. We begin with some sufficiently large quantum system (or `register') in a given quantum state. We perform a quantum operation on it and conditionally on the outcome of that operation we determine which party will be first, which subsystem of the register will be his/her input system, and what operation will be applied after the black box of that party, all according to some specified rule. We apply the black-box operation of the first party on the decided subsystem, perform the decided operation after it, and depending on its outcome and the outcome of our first operation decide which party will be second, and so on. This continues until all parties are called (by definition, the protocol is such that each party is called exactly once). This model can be formalized in different equivalent ways, which may be suitable for different purposes, and we will consider some simplifications below when we discuss a tripartite example. The fact that this model gives rise to valid quantum processes can be seen from the fact that if we formally write the operation inside each box and calculate the joint probabilities for the outcomes of all boxes using the standard rules of quantum mechanics for all possible outcomes of the protocol, we see that they are linear and non-contextual functions of the respective CP maps of the parties. The same holds if we introduce ancillary systems prepared in an arbitrary state and consider extended operations of the parties that act on parts of them.   

In the case of only two parties, we know that any (extensibly) causally separable process can be implemented in this way, since it most generally corresponds to embedding at random the local experiments of Alice and Bob into one of two possible fixed circuits, which can be chosen conditionally on the outcome of a measurement on some state at the very beginning. Since after the first party is chosen there is only one possible choice for the second party, no measurement after the first party is needed. Reversely, any bipartite process that we may obtain via this model has the form of an ECS process. Fist notice that the process is independent of the operation applied after the last party. Also, the outcome of any operation after the first party can be ignored since there is only one choice for the last party, i.e., that operation can be assumed deterministic. Finally, the outcomes of the operation before the first party can be grouped into two coarse-grained outcomes such that conditionally on one of them the first party is Alice and on the other one it is Bob. But since after the outcome of that operation and before the input of the first party the quantum register is in some particular quantum state, the rest of the experiment simply corresponds to a deterministic circuit in which Alice and Bob are embedded in a particular order. Therefore, the process realized by such a procedure is just a probabilistic mixture of the processes of two fixed-order circuits, which is the claimed form. 

In the case of more than two parties, the equivalence between the two concepts is less obvious, but we can easily argue that all processes obtained by classically controlled circuits are ECS. First, it is clear that depending on the outcome of the first measurement (which has a probability independent of any future operations and therefore of the settings of the parties), there will be one party that is first and hence the subsequent process that results from the protocol can involve no signaling from the rest of the parties to that first party. Therefore, the subsequent process has a well-defined reduced process for the first party. Taking into account all possible outcomes of the first measurement, the whole process will be just a probabilistic mixture of processes of this kind where one party is first, which is Eq.~\eqref{cslong}. But conditionally on the outcome of the first party, the procedure for the rest of the parties looks analogously, so Eq.~\eqref{cslong2} holds too, i.e., the process is causally separable. Including ancillas onto which the operations of the parties can be extended does not change anything in this argument. Therefore, every process realizable with a classically controlled quantum circuit is ECS. 

We conjecture that the reverse also holds. We provide some partial considerations that support this conjecture, based on analysis of the restrictions on the allowed terms in processes realized by classically controlled quantum circuits in the tripartite case. We will focus on the question of implementing by a classically controlled quantum circuit an ECS process matrix of the type $W_{ecs}^{(A,B)\ncp C}$, which has the form \eqref{firstcond1}. Implementability of a matrix of this kind is both necessary and sufficient for the implementability of a general tripartite ECS process matrix as described in Proposition \ref{prop:3.3}, since by using a suitable measurement at the beginning we can select with the right probability which of the three process matrices in the mixture on the right-hand side of Eq. \eqref{tripartecs} to realize subsequently.

\begin{figure}[!htb]
\centering
\includegraphics[scale=.3]{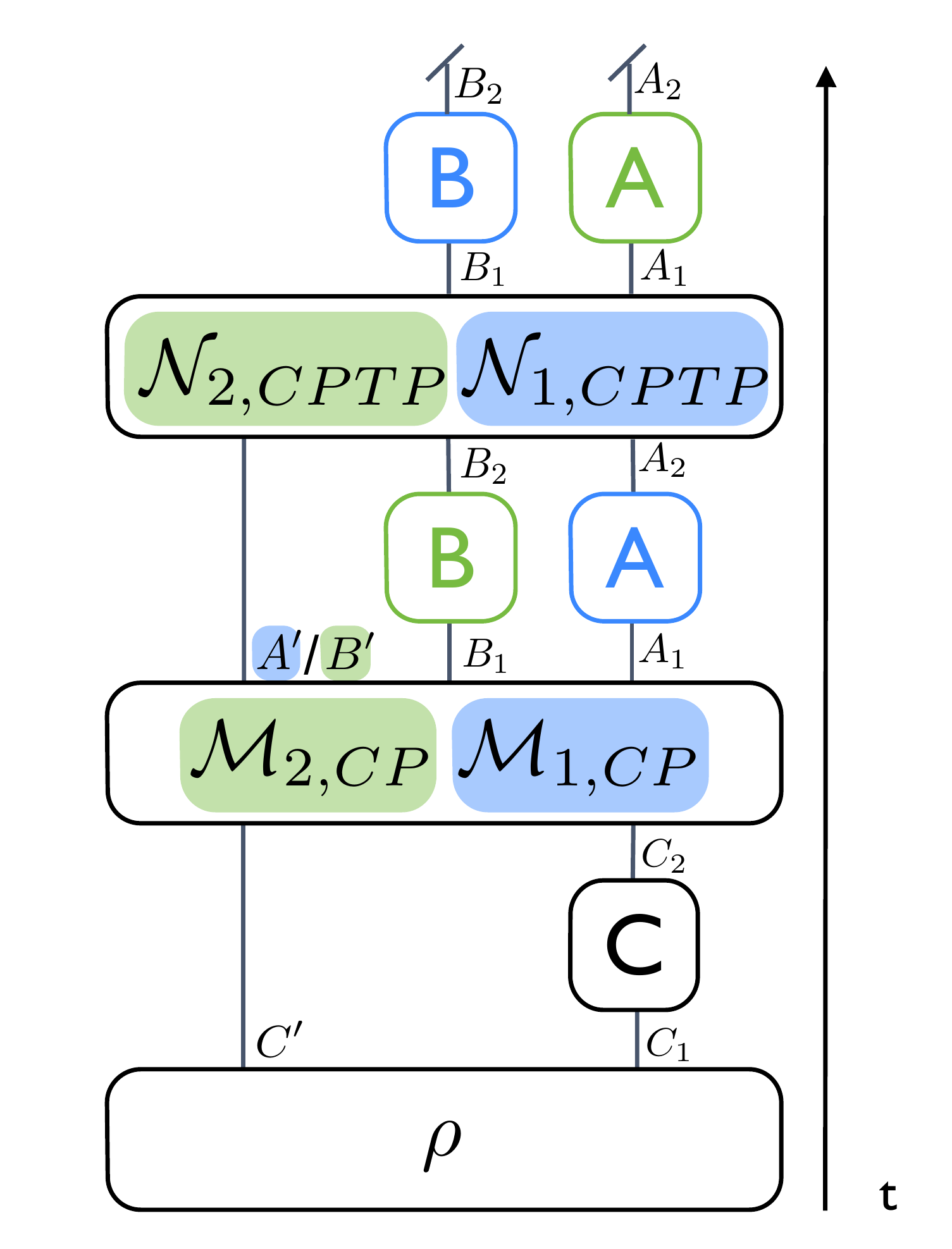}
%\captionsetup{width=0.8\textwidth}
\caption{Realization of an ECS process compatible with $(A,B)\ncp C$ by a classically controlled quantum circuit.}
\label{fig:realization}
\end{figure}

%Without loss of generality, we may assume that the protocol begins with a quantum system in a pure state, since any state is the reduced state of some pure state on a larger system. 
The protocol begins most generally with some quantum system prepared in a state $\rho$. After Charlie operates on some subsystem, we apply some operation based on whose outcome we determine who is second, on what subsystem he/she would act, and what operation will be applied after that. Note that without loss of generality we may assume that there is a pre-specified subsystem on which the second party will operate since any subsystem of the same dimension can be mapped onto the designated subsystem by a unitary transformation that can be absorbed as part of the definition of the present operation. Also, without loss of generality we may assume that this operation has only two outcomes, since we can group the outcomes into those for which Alice will be next, and those for which Bob will be next, and any conditioning of the operation following the next party on the fine-grained outcome within each group can be equivalently done by a single future operation acting on a larger system that includes some subsystem on which the classical information about the outcome at this step is copied (still something that we can include as part of the definition of the operation at this step). Since there is only a single possibility for the last party, the operation after the second party can be regarded as a deterministic operation (or a CPTP map) from all systems to the input of the last party. We leave the possibility that this last operation may be defined conditionally on the first outcome rather than absorb the conditioning on that outcome into a larger operation, in order to avoid complications arising from the fact that the different parties may have input and output systems of different dimensions. The outlined procedure is sketched in Fig.~\ref{fig:realization}, where the two possible sequences of transformations arising from the two possible outcomes of our first operation are depicted in blue and green, respectively. The two CP maps corresponding to the outcomes of the operation after Charlie must sum up to a CPTP map, since they correspond to the two possible outcomes of a standard quantum operation. 

Each of the two possible developments (blue and green) of this protocol is a non-deterministic linear supermap \cite{supermaps} from the local CP maps of the parties into the real numbers, the result of which equals the probability for the particular sequence of events. This can be written in a similar form as the formula for the probabilities of the outcomes of the parties in a valid process, except that in the place of the process matrix we would have an operator $\tilde{W}_{i}^{A_1A_2B_1B_2C_1C_2}\geq 0$, where $i=1,2$, labels the particular development, which generally would not be a valid process matrix. However, $\tilde{W}_{1}^{A_1A_2B_1B_2C_1C_2} +\ \tilde{W}_{2}^{A_1A_2B_1B_2C_1C_2} = {W}_{cs;(A,B)\ncp C}^{A_1A_2B_1B_2C_1C_2}$ would be a valid process matrix realized through this classically controlled quantum circuit. 

Consider now just one of the two possible developments, say, the blue one, in which Alice is second and Bob is last (labeled by $1$). One can see that since Bob is last and his output system is discarded, we have $\tilde{W}_{1}^{A_1A_2B_1B_2C_1C_2}= \id^{B_2}\otimes  \tilde{W}_{1}^{A_1A_2B_1C_1C_2}$ (similarly, in the other case we have $\tilde{W}_{2}^{A_1A_2B_1B_2C_1C_2}= \id^{A_2}\otimes  \tilde{W}_{2}^{A_1B_1B_2C_1C_2}$). Notice that if the transformation $\mathcal{N}_{1,CPTP} $ after Alice was not required to be CPTP but could be any CP map $\mathcal{N}_{1,CP}$, for a suitable choice of the initial state $\rho$ and of the CP maps $\mathcal{M}_{1,CP}$ and $\mathcal{N}_{1,CP}$ we could realize any $\tilde{W}_{1}^{A_1A_2B_1C_1C_2}\geq 0$. This is simply because we can choose the density operator $\rho^{C_1C'}$ proportional to $\tilde{W}_{1}^{A_1A_2B_1C_1C_2}$, where the part of $\tilde{W}_{1}^{A_1A_2B_1C_1C_2}$ on $A_1A_2B_1C_2 $ is stored on $C'$, and we can `teleport' this part of the operator onto its desired subsystem by using CP maps  $\mathcal{M}_{1,CP}$ and $\mathcal{N}_{1,CP}$ that have CJ operators proportional to projectors on maximally entangled states as needed to realize the `teleportation' (the traces of these CP maps can be chosen to ensure the overall trace of the resultant operator $\tilde{W}_{1}^{A_1A_2B_1C_1C_2}$). However, the restriction that the transformation after Alice is trace-preserving, $\mathcal{N}_{1,CPTP} $, places constraints on what kind of $\tilde{W}_{1}^{A_1A_2B_1C_1C_2}$ can be obtained. Indeed, the CJ operator of $\mathcal{N}_{1,CPTP} $ cannot contain terms of type $A_2$, $A'A_2$, and $A'$. Considering the calculation of $\tilde{W}_{1}^{A_1A_2B_1C_1C_2}$  based on the CJ operators of $\rho$, $\mathcal{M}_{1,CP}$ and $\mathcal{N}_{1,CP}$, we see that the lack of these types of terms in $\mathcal{N}_{1,CPTP} $ implies the lack of any term with a nontrivial $\sigma$ on $A_2$ in $\tilde{W}_{1}^{A_1A_2B_1C_1C_2}$. This is the only constraint on the possible types of terms in $\tilde{W}_{1}^{A_1A_2B_1C_1C_2}$. The possible types of terms are exactly those allowed in the operator $\tilde{W}^{A_1A_2B_1C_1C_2}$ in Eq.~\eqref{firstcond1}. Similarly, we see that the allowed terms in  $\tilde{W}_{2}^{A_1B_1B_2C_1C_2}$  (Bob second, Alice last) are the same as those in $\tilde{W}^{A_1B_1B_2C_1C_2}$ in Eq.~\eqref{firstcond1}.  These are the terms allowed in a process matrix compatible with Charlie being first, except that both $\tilde{W}_{1}^{A_1A_2B_1C_1C_2}$ and $\tilde{W}_{2}^{A_1B_1B_2C_1C_2}$ may contain terms of type $C_2$ and $C_1C_2$ . The fact that these terms should cancel in the sum $\id^{B_2}\otimes  \tilde{W}_{1}^{A_1A_2B_1C_1C_2}+\ \id^{A_2}\otimes  \tilde{W}_{2}^{A_1B_1B_2C_1C_2}= {W}_{cs;(A,B)\ncp C}^{A_1A_2B_1B_2C_1C_2} $ follows from the fact that this is a valid ECS process, and can be seen to be ensured by the requirement that $\mathcal{M}_{1,CP}+\ \mathcal{M}_{2,CP}$ is CPTP. 

The only restriction on the operators $\id^{B_2}\otimes  \tilde{W}_{1}^{A_1A_2B_1C_1C_2}$ and $\id^{A_2}\otimes  \tilde{W}_{2}^{A_1B_1B_2C_1C_2}$ imposed by this model, apart from their positive-semidefiniteness and the normalization of their sum,  seems to be the absence of the forbidden terms in each of them, as well as of the forbidden terms in their sum. If this is indeed the case, then any ECS process could be realized by a suitable classically controlled quantum circuit. 
A strictly rigorous proof requires showing that apart from the lack of these forbidden terms, there can be no other hidden constraints on the pair of operators $\id^{B_2}\otimes  \tilde{W}_{1}^{A_1A_2B_1C_1C_2}$ and $\id^{A_2}\otimes  \tilde{W}_{2}^{A_1B_1B_2C_1C_2}$ (which, of course, are guaranteed to be properly normalized). One way of doing it could be by exhibiting an explicit constructive procedure for implementing any given ECS process, which would be of additional interest on its own right. We leave this question, and the multipartite case, for future investigation.

\section{Conclusion}

\begin{figure}
\centering
\subfloat[a) Multipartite case.]{\includegraphics[scale=.36]{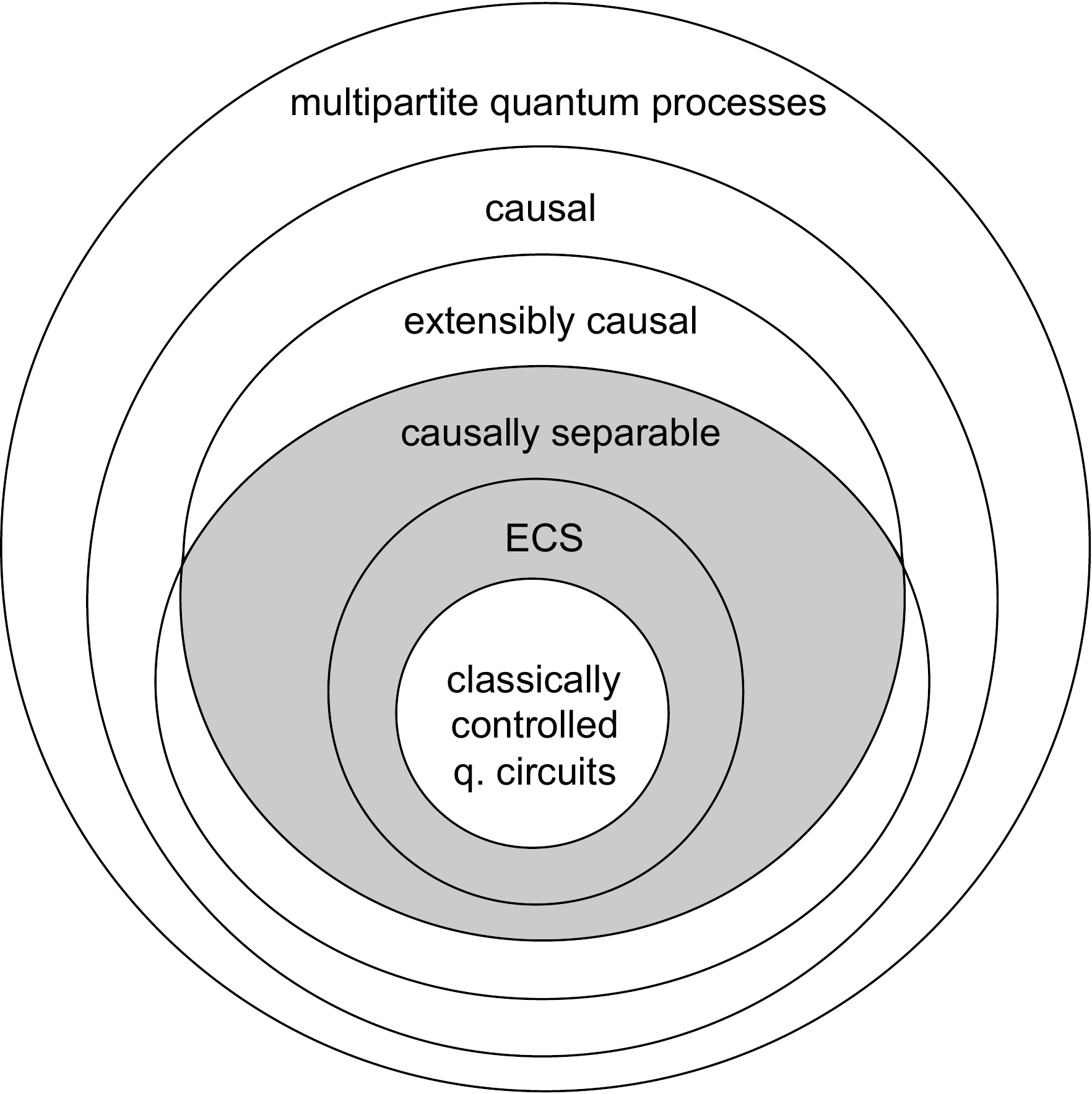}{\label{subfig:multipartitesets}}}\hspace{0.6cm}
\subfloat[b) Bipartite case.]{\includegraphics[scale=.36]{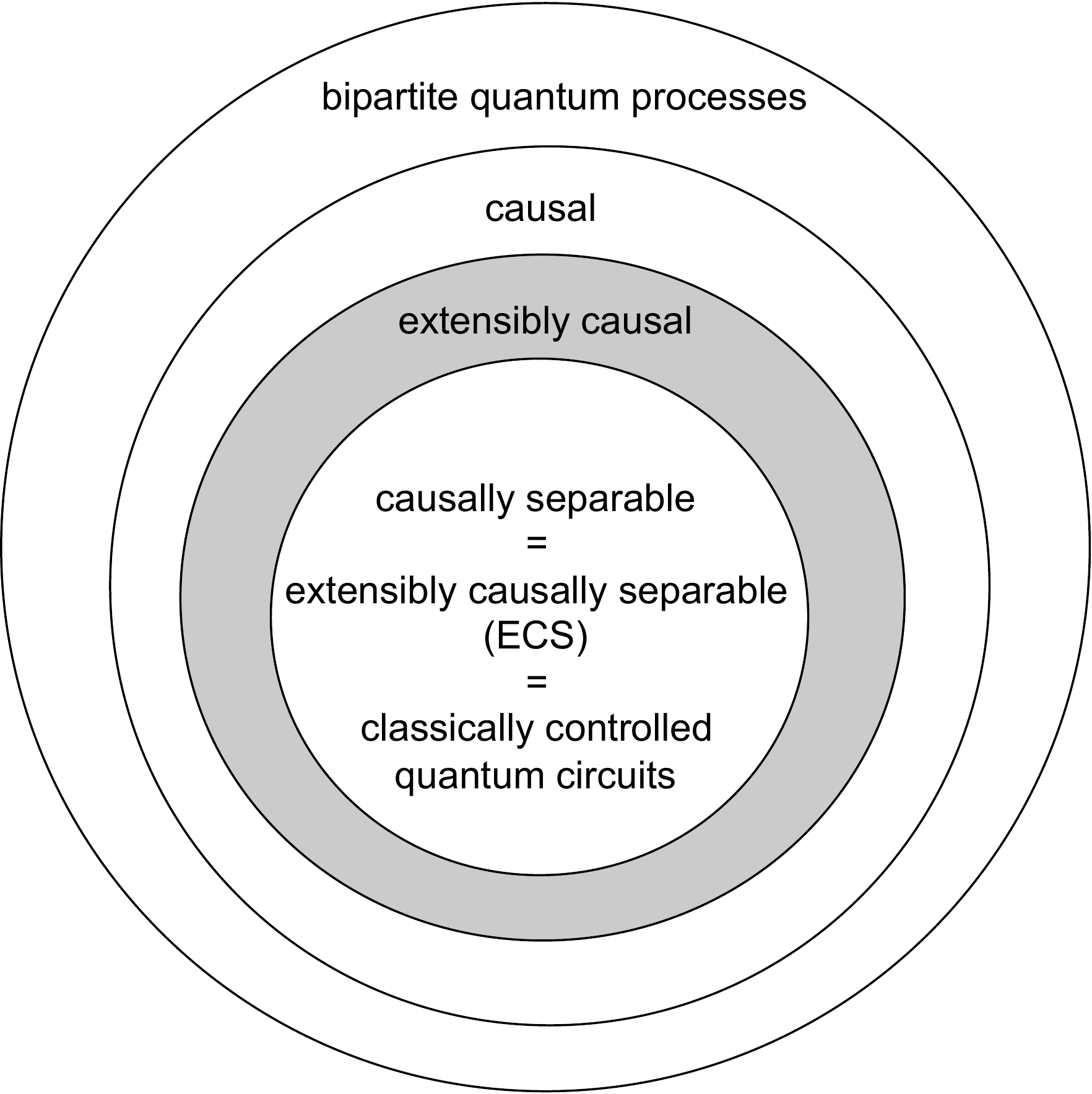}{\label{subfig:bipartitesets}}}
\caption{A Venn diagrammatic sketch of our present knowledge of the different sets of quantum processes that we have introduced, in the general multipartite case and in the bipartite case. The white segments are non-empty. The gray segments are sets for which at present we do not know if they are empty or not.}
\end{figure}

In this paper, we proposed a rigorous definition of causality in the process framework \cite{OCB}, which takes into account the fact that the causal order between a set of local experiments may in general be random and correlated with the settings of some of them. We derived the structure of causal processes permitting such `dynamical' causal order in the general multipartite case, which is captured by an iteratively formulated canonical form expressed in terms of reduced and conditional processes. The  canonical form can be interpreted as an unraveling of the process into a sequence of local experiments, which agrees with the condition that the order and outcomes of the experiments prior to a given step is independent of the settings of future experiments. We showed that for any fixed number of settings and outcomes for each party, the probabilities of a causal processes form a polytope, referred to as the causal polytope. The facets of this polytope define causal inequalities, whose violation by a given process can be interpreted as demonstrating the non-existence of causal order between the local experiments. 

We investigated this concept and the related concept of causal separability in the quantum process theory introduced in Ref.~\cite{OCB}, whose properties were detailed here in the multipartite case. We proposed a definition of causal separability, which reduces to the one for the case of two parties \cite{OCB}, based on the canonical form of causal processes. Specifically, a causally separable quantum process was defined as a causal quantum process that has a causal decomposition such that the different processes appearing in this decomposition are themselves valid quantum processes. We showed that the set of causally separable quantum processes is strictly within the set of causal quantum processes, by exhibiting an example of a tripartite process that is causal but not causally separable. Very recently, the same was shown to hold also in the bipartite case \cite{Feix}. We also gave an example of a causally separable (and hence also causal) process that becomes non-causal when extended by supplying the parties with an entangled ancillary state. Based on this observation, we proposed two extended notions of causality and causal separability called extensible causality and extensible causal separability, which require preservation of the respective property under extending the process with entangled input ancillas. Although they are different in the general case, the sets of causally separable and ECS processes are equivalent in the bipartite case. We showed that the sets of extensibly causal and causally separable processes are different in general via the same tripartite example that we used to show that causal and causally separable processes are different. At present we do not know if the same separation holds in the bipartite case. However, it was recently shown that causal and extensibly causal processes are different in the bipartite case, similarly to the multipartite case \cite{Feix}. 

Finally, we derived a simple characterization of the ECS quantum processes in the tripartite case in terms of conditions on the form of their process matrices, which extends the conditions for (extensibly) causally separable process matrices in the bipartite case. We conjectured that the set of ECS processes is equivalent to the processes that can be obtained within the paradigm of classically controlled quantum circuits and provided evidence for this based on analysis of the restrictions that this paradigm imposes on the tripartite process matrices it can create. The ECS processes and the processes obtainable by classically controlled quantum circuits are equivalent in the bipartite case. 

Our present understanding of the relation between all these different classes of quantum processes is illustrated for the general multipartite case and for the bipartite case in Fig.~\ref{subfig:multipartitesets} and Fig.~\ref{subfig:bipartitesets}, respectively. An obvious open problem is whether the gray segments in these figures are empty or not. 

Another problem of fundamental importance is to understand the class of quantum processes that are physically admissible in agreement with the known laws of quantum mechanics, and where this class stands with respect to all of the above classes. Are the processes that can be realized by classically controlled quantum circuits all the physically admissible causally separable processes? Where does the class of quantum-controlled quantum circuits stand? At present, this is the most general operationally feasible paradigm that we are aware of and all known processes realizable through it seem to be extensibly causal. Could the class of extensibly causal processes be equivalent to quantum-controlled quantum circuits? And most intriguingly, are there physically admissible non-causal processes? 

The implications of our results are not limited to the subject of indefinite causal order in quantum mechanics. They can be useful also for the problem of inferring causal structure \cite{Pearl}, both in classical and quantum theory \cite{Ried}. The subject of causal inference concerns many disciplines, from philosophy and machine learning to sociology and medicine. Our formulation of a background-independent operational notion of causality that admits dynamical causal relations opens the road to a more general paradigm for causal inference than the one assuming deterministic underlying variables and static causal relations \cite{Pearl}. The decomposition of causal processes derived here implies constraints on the possible causal orders compatible with given setting-outcome correlations, which can serve as a basis for developing more sophisticated causal inference tools.

%When adding new references: Add the references of the "endnote" form to file :'thereferences.bib using bibdesk and cite them here. Then the two lines below need to be uncommented, run bibtex, then latex on this file. Then the "multipartite causally separable processes.bbl" file will contain the \bibitems to be added below. Add the bibitems. Finally, comment again the following two lines.
%%
%\bibliographystyle{ieeetr}
%\bibliography{thereferences}

\begin{thebibliography}{10}
\bibitem{footnote3} Note that both the entangled state $|\phi^+\rangle$ and the overall transposition ${\mathrm T}$ that appear in our definition of the CJ operator depend on the choice of basis $\left\{|j\rangle\right\}_{j=1}^{d_{A_1}}$, which altogether makes the CJ operator independent of that basis. In contrast, in the original version of the Choi isomorphism \cite{choi}, the Choi operator is defined without the transposition ${\mathrm T}$, which is why it depends on  $\left\{|j\rangle\right\}_{j=1}^{d_{A_1}}$. However, the original Choi operator does not depend on any choice of basis for ${\cal H}^{A_2}$, whereas our CJ operator depends on such a choice through the transposition ${\mathrm T}$.
\end{thebibliography}

\section*{Acknowledgments}

We thank M. Ara\'{u}jo, \"{A}. Baumeler, C. Branciard, \v{C}. Brukner, and A. Feix for discussions. We are grateful to C. Branciard for pointing out several errors in previous versions of the manuscript. %A similar conclusion to the one in Sec. III B, based on the same example, has been proven independently by F. Costa and is published in Ref.~\cite{Araujo3}. %During the final stages of the preparation of this manuscript, after we had obtained and presented our proof of the nonequivalence between causal and causally separable quantum processes \cite{Christina}, we were informed that an analogous conclusion based on the same example has been reached independently by coauthors of one of us in a concurrent work ~\cite{Araujo3}. The analysis of Ref.~\cite{Araujo3} is based on a more restricted notion of causal separability that does not consider the possibility for dynamical causal order. 

This work was supported in part by the F.R.S.-FNRS under Project T.0199.13 and by HIPERCOM (ERA-Net). O.O. acknowledges support by the European Commission under the Marie Curie Intra-European Fellowship Programme (PIEF-GA-2010-273119) and by the F.R.S.--FNRS under the Charg\'{e} de recherches (CR) Fellowship Programme. C.G. was supported in part by the ARC Center of Excellence for Engineered Quantum Systems (Grant No. CE110001013) and the ARC Center of Excellence for Quantum Computation and Communication Technology (Grant No. CE110001027).

%%%%%%%%%% Merge with supplemental materials %%%%%%%%%%
\pagebreak
\widetext

%%%%%%%%%% Prefix a "S" to all equations, figures, tables and reset the counter %%%%%%%%%%
\setcounter{equation}{0}
\setcounter{figure}{0}
\setcounter{table}{0}
\setcounter{section}{0}
\makeatletter
\renewcommand{\theequation}{S\arabic{equation}}
\renewcommand{\thefigure}{S\arabic{figure}}
\renewcommand{\thetable}{S\arabic{table}}
\renewcommand{\thepr}{S\arabic{pr}}
\renewcommand{\thelem}{S\arabic{lem}}
\renewcommand{\bibnumfmt}[1]{[S#1]}
\renewcommand{\citenumfont}[1]{S#1}

\begin{center}
\textbf{\large Appendix: Causal and causally separable processes}
\end{center}

\vspace{1cm}
%\section{The process framework}\label{sec:proofs_sec_II_B}

\begin{pr}\label{pr:prop2.2}
\textbf{Proposition \ref{prop:2.2}}. The `only if' part is contained in the very Proposition \ref{prop:2.1}. To prove the `if' part, take an arbitrary experiment, say, $1$. Let $\{2, \cdots, k\}$, up to relabeling, be the set of local experiments that are in the causal past or causal elsewhere of $1$, and $\{k+1, \cdots, n\}$ be the set of local experiments that are in the causal future of $1$. Since the causal configuration of the local experiments is assumed fixed, the condition for the process to be causal reduces to the requirement that for every such $1$, we have $p(o^{2},\cdots,o^{k}|s^{1}, s^{2}, \cdots, s^{n}) = p(o^{2},\cdots,o^{n}| s^{2}, \cdots, s^{n})$. But from the transitivity and anti-symmetry of causal order it follows that none of the experiments $\{1, \cdots, k\}$ is in the causal future of any of the experiments $\{k+1, \cdots, n\}$. This implies that we have a reduced $k$-partite process for $\{1, \cdots, k\}$, i.e., $p(o^{1},\cdots,o^{k}|s^{1}, s^{2}, \cdots, s^{n}) = p(o^{1},\cdots,o^{k}| s^{1}, \cdots, s^{k})$. The desired condition then follows from Proposition \ref{prop:2.1} applied to the $k$-partite process. \\
\end{pr}

\begin{pr}\label{pr:prop:2.3}
\textbf{Proposition \ref{prop:2.3}}. First, observe that the property (\ref{inductive}) holds for the case where the specified $\textrm{K}$ consecutive sets exhaust all local experiments $\{1,\cdots, n\}$. This is because, in this case, each of the local experiments in the ${\textrm{K}}^{th}$ consecutive set is causally preceded by or causally independent from every other local experiment. Hence, the definition of causality (\ref{causalorderDEF}) directly implies the desired relation. The general case follows by induction from this special case and the following Lemma. \\
%\end{pr}

\begin{lem}\label{lem:1}
Let the property (\ref{inductive}) hold for $\textrm{K}=\textrm{K}'+\RN{1}$, where $\textrm{K}'\geq 1$. Then it also holds for $\textrm{K}=\textrm{K}'$. \\
\end{lem}

\textbf{Proof.} Observe that  
\begin{eqnarray}
 &p(\kappa({1}_{\RN{1}}, \cdots, n_{\textrm{K}'}) , [1_{\RN{1}}, \cdots, n_{\RN{1}}]^{\RN{1}},\cdots, [1_{\textrm{K}'}, \cdots, n_{\textrm{K}'}]^{\textrm{K}'}, o^{1_{\RN{1}}}, \cdots,  o^{g_{\textrm{K}'} }|s^{{1}}, \cdots, s^{{n}}) \\
 &= \sum_{[1_{\textrm{K}'+\RN{1}}, \cdots, n_{\textrm{K}'+\RN{1}}]^{\textrm{K}'+\RN{1}}} {
 p(\kappa(1_{\RN{1}}, \cdots, n_{\textrm{K}'}), [1_{\RN{1}}, \cdots, n_{\RN{1}}]^{\RN{1}}, \cdots, [1_{\textrm{K}'}, \cdots, n_{\textrm{K}'}]^{\textrm{K}'}, [1_{\textrm{K}'+\RN{1}}, \cdots, n_{\textrm{K}'+\RN{1}}]^{\textrm{K}'+\RN{1}}, o^{1_{\RN{1}}}, \cdots,  o^{g_{\textrm{K}'}}|s^{{1}}, \cdots, s^{{n}})},\nonumber 
\end{eqnarray}
where the sum on the right-hand side is over all sets of local experiments that can be the $({\textrm{K}'+\RN{1}})^{th}$ set when the first $\textrm{K}'$ consecutive sets are the specified ones. If Eq.~(\ref{inductive}) holds for $\textrm{K}=\textrm{K}'+\RN{1}$, all terms in the sum can depend non-trivially only on the settings of the parties in the first $\textrm{K}'$ consecutive sets, and hence the same must hold for the quantity on the left-hand side:
 \begin{eqnarray}
 p( \kappa( 1_{\RN{1}}, \cdots, n_{\textrm{K}'}), [1_{\RN{1}}, \cdots, n_{\RN{1}}]^{\RN{1}},\cdots, [1_{\textrm{K}'}, \cdots, n_{\textrm{K}'}]^{{\textrm{K}'}}, o^{1_{\RN{1}}}, \cdots, o^{g_{\textrm{K}'} }|s^{{1}}, \cdots, s^{{n}})\label{intemediate}\\
 =p(\kappa( 1_{\RN{1}}, \cdots, n_{\textrm{K}'}), [1_{\RN{1}}, \cdots, n_{\RN{1}}]^{\RN{1}},\cdots, [1_{\textrm{K}'}, \cdots, n_{\textrm{K}'}]^{{\textrm{K}'}}, o^{1_{\RN{1}}}, \cdots, o^{g_{\textrm{K}'} }|s^{1_{\RN{1}}}, \cdots, s^{n_{\textrm{K}'}}).\nonumber
\end{eqnarray}
What remains to be shown is that this probability cannot depend on the settings $s^{(g+1)_{\textrm{K}'}}, \cdots, s^{n_{\textrm{K}'}}$. 

Note that here we cannot apply straightforwardly the causality condition (\ref{causalorderDEF}) as we did in the case when the first $\textrm{K}'$ consecutive sets were assumed to contain all local experiments. This is because for a particular causal configuration $\kappa_*(1_{\RN{1}}, \cdots, n_{\textrm{K}'})$ compatible with  $[1_{\RN{1}}, \cdots, n_{\RN{1}}]^{\RN{1}}$, $\cdots$, $[1_{\textrm{K}'}, \cdots, n_{\textrm{K}'}]^{{\textrm{K}'}}$, it is generally \textit{not} the case that
\begin{gather}
p(\kappa_*( 1_{\RN{1}}, \cdots, n_{\textrm{K}'}), [1_{\RN{1}}, \cdots, n_{\RN{1}}]^{\RN{1}},\cdots, [1_{\textrm{K}'}, \cdots, n_{\textrm{K}'}]^{{\textrm{K}'}}, o^{1_{\RN{1}}}, \cdots, o^{g_{\textrm{K}'}}|s^{{1}}, \cdots, s^{{n}}) \notag\\
= p(\kappa_*( 1_{\RN{1}}, \cdots, n_{\textrm{K}'}), o^{1_{\RN{1}}}, \cdots, o^{g_{\textrm{K}'}}|s^{{1}}, \cdots, s^{{n}}). \label{possible}
\end{gather}
Indeed, in order for the first $\textrm{K}'$ consecutive sets to be the specified ones, it is necessary and sufficient that: 1) the local experiments in the specified $\textrm{K}'$ consecutive sets have a causal configuration compatible with these sets, {and} 2) each of the local experiments that are not in the specified $\textrm{K}'$ consecutive sets is in the causal future of at least one of the local experiments in the ${\textrm{K}'}^{th}$ consecutive set. [In the case where the ${\textrm{K}'}$ sets were assumed to contain all local experiments, only condition 1) was relevant and hence the equality (\ref{possible}) held.] Consider a particular causal configuration $\kappa_*(1_{\RN{1}}, \cdots, n_{\textrm{K}'})$ compatible with  $[1_{\RN{1}}, \cdots, n_{\RN{1}}]^{\RN{1}}$, $\cdots$, $[1_{\textrm{K}'}, \cdots, n_{\textrm{K}'}]^{{\textrm{K}'}}$ (when the causal configuration $\kappa(1_{\RN{1}}, \cdots, n_{\textrm{K}'})$ in the probability on the left-hand side of Eq.~(\ref{inductive}) is not compatible with the specified consecutive sets, that probability is trivially zero). Let us denote by  $1_{rest}, \cdots, l_{rest}$, $l=n-\sum_{\textrm{K}=\RN{1}}^{\textrm{K}'}\sum_{m=1}^{n_{\textrm{K}}} m$, the rest of the local experiments, i.e, those that do not belong to the assumed first $\textrm{K}'$ consecutive sets. We have 
\begin{gather}
 p(\kappa_*( 1_{\RN{1}}, \cdots, n_{\textrm{K}'}) , [1_{\RN{1}}, \cdots, n_{\RN{1}}]^{\RN{1}},\cdots, [1_{\textrm{K}'}, \cdots, n_{\textrm{K}'}]^{{\textrm{K}'}}, o^{1_{\RN{1}}}, \cdots, o^{g_{\textrm{K}'}}|s^{{1}}, \cdots, s^{{n}      })  \nonumber\\
  =  p(\kappa_*(1_{\RN{1}}, \cdots, n_{\textrm{K}'}), (1_{\textrm{K}'} \cp 1_{rest} \lor  \cdots \hspace{0.1cm} \lor \hspace{0.1cm} n_{\textrm{K}'} \cp 1_{rest}) ,
\cdots, (1_{\textrm{K}'} \cp l_{rest} \lor \cdots \lor  n_{\textrm{K}'}\cp l_{rest} ), o^{1_{\RN{1}}}, \cdots, o^{g_{\textrm{K}'}}|s^{{1}}, \cdots, s^{{n}}).
\end{gather}
We will show that the probability on the right-hand side can be written as a linear combination of probabilities for which the condition of causality (\ref{causalorderDEF}) straightforwardly implies independence of $s^{(g+1)_{\textrm{K}'}}, \cdots, s^{n_{\textrm{K}'}}$.

To this end, we write 
\begin{gather}
 p(\kappa_*(1_{\RN{1}}, \cdots, n_{\textrm{K}'}), (1_{\textrm{K}'} \cp 1_{rest} \lor  \cdots \hspace{0.1cm} \lor \hspace{0.1cm} n_{\textrm{K}'} \cp 1_{rest}) ,
\cdots, (1_{\textrm{K}'} \cp l_{rest} \lor \cdots \lor  n_{\textrm{K}'}\cp l_{rest} ), o^{1_{\RN{1}}}, \cdots, o^{g_{\textrm{K}'}}|s^{{1}}, \cdots, s^{{n}      })\nonumber\\
 = p(\kappa_*(1_{\RN{1}}, \cdots, n_{\textrm{K}'}), o^{1_{\RN{1}}}, \cdots, o^{g_{\textrm{K}'}}|s^{{1}}, \cdots, s^{{n}      })
 - p(\kappa_*(1_{\RN{1}}, \cdots, n_{\textrm{K}'}),  (\mathcal{K}' \ncp  1_{rest}) \lor 
 \cdots \lor (\mathcal{K}'  \ncp l_{rest}) , o^{1_{\RN{1}}}, \cdots, o^{g_{\textrm{K}'}}|s^{{1}}, \cdots, s^{{n}      }), \nonumber
\end{gather}
where $\mathcal{K}'=\{1_{\textrm{K}'}, \cdots,  n_{\textrm{K}'}\}$, and $\mathcal{K}' \ncp l$ means $ 1_{\textrm{K}'} \ncp l \land \cdots \land n_{\textrm{K}'} \ncp l$.
 
By the definition of causality, the term $p(\kappa_*(1_{\RN{1}}, \cdots, n_{\textrm{K}'}),o^{1_{\RN{1}}}, \cdots, o^{g_{\textrm{K}'}}|s^{{1}}, \cdots, s^{{n}      })$ on the right-hand side is independent of $s^{(g+1)_{\textrm{K}'}}, \cdots, s^{n_{\textrm{K}'}}$. We need to prove that the second term is also independent of $s^{(g+1)_{\textrm{K}'}}, \cdots, s^{n_{\textrm{K}'}}$. Observe that the proposition ``$      (\mathcal{K}' \ncp 1_{rest}) \lor \cdots \lor (\mathcal{K}' \ncp l_{rest}) $'' is true when a proposition of the following kind is true: for some nonempty subset of $\{1_{rest}, \cdots, l_{rest}\}$, say, $\{1_{rest}, \cdots, r_{rest}\} $, $1\leq r\leq l$, define the proposition ``$(\mathcal{K}' \ncp 1_{rest}) \land  \cdots \land  (\mathcal{K}' \ncp r_{rest}) \land \lnot [ (\mathcal{K}' \ncp (r+1)_{rest})\lor \cdots \lor (\mathcal{K}' \ncp l_{rest})]$''. The different nonempty subsets of $\{1_{rest}, \cdots, l_{rest}\}$ yield different such propositions that describe a complete set of mutually exclusive scenarios for which ``$  (\mathcal{K}' \ncp 1_{rest}) \lor \cdots \lor (\mathcal{K}' \ncp l_{rest}) $'' is true. Therefore, the probability $p(\kappa_*(1_{\RN{1}}, \cdots, n_{\textrm{K}'}),  (\mathcal{K}'\ncp 1_{rest}) \lor \cdots \lor (\mathcal{K}' \ncp l_{rest})  , o^{1_{\RN{1}}}, \cdots, o^{g_{\textrm{K}'}}|s^{{1}}, \cdots, s^{{n}      })$ is a sum of probabilities of the form $p(\kappa_*(1_{\RN{1}}, \cdots, n_{\textrm{K}'}),  (\mathcal{K}' \ncp 1_{rest}), \cdots ,  (\mathcal{K}' \ncp r_{rest}), \lnot [(\mathcal{K}' \ncp (r+1)_{rest}) \lor  \cdots \lor (\mathcal{K}' \ncp l_{rest}) ], o^{1_{\RN{1}}}, \cdots, o^{g_{\textrm{K}'}} |s^{{1}}, \cdots, s^{{n}      })$, up to relabeling of $1_{rest}, \cdots, l_{rest}$, where $1\leq r \leq l$. But every such probability can be further written as $p(\kappa_*(1_{\RN{1}}, \cdots, n_{\textrm{K}'}),  (\mathcal{K}' \ncp 1_{rest}), \cdots ,  (\mathcal{K}' \ncp r_{rest}), o^{1_{\RN{1}}}, \cdots, o^{g_{\textrm{K}'}}|s^{{1}}, \cdots, s^{{n}      }) - p(\kappa_*(1_{\RN{1}}, \cdots, n_{\textrm{K}'}),  (\mathcal{K}' \ncp 1_{rest}), \cdots ,  (\mathcal{K}' \ncp r_{rest}), (\mathcal{K}' \ncp (r+1)_{rest}) \lor  \cdots \lor (\mathcal{K}' \ncp l_{rest}), o^{1_{\RN{1}}}, \cdots, o^{g_{\textrm{K}'}}|s^{{1}}, \cdots, s^{{n}      })$. By the definition of causality, the first of these terms is independent of $s^{(g+1)_{\textrm{K}'}}, \cdots, s^{n_{\textrm{K}'}}$. Considering again the different realizations of ``$ (\mathcal{K}' \ncp (r+1)_{rest}) \lor  \cdots \lor (\mathcal{K}' \ncp l_{rest})$'' by propositions of the form ``$(\mathcal{K}' \ncp (r+1)_{rest}) \land  \cdots \land  (\mathcal{K}' \ncp (r+q)_{rest}) \land \lnot [ (\mathcal{K}' \ncp (r+q+1)_{rest})\lor \cdots \lor (\mathcal{K}' \ncp l_{rest})]$'' for $1\leq q \leq l-r $ (up to relabeling of the local experiments), the probability in the second term can again be written as a sum of probabilities of the form $p(\kappa_*(1_{\RN{1}}, \cdots, n_{\textrm{K}'}),  (\mathcal{K}' \ncp 1_{rest}), \cdots ,  (\mathcal{K}' \ncp r_{rest}), \lnot [(\mathcal{K}' \ncp (r+1)_{rest}) \lor  \cdots \lor (\mathcal{K}' \ncp l_{rest}) ], o^{1_{\RN{1}}}, \cdots, o^{g_{\textrm{K}'}}|s^{{1}}, \cdots, s^{{n}      })$, where now $r$ is strictly larger than the one in the previous step. We can continue this for every new term until we reach $r=l$. In this way, the probability $p(\kappa_*(1_{\RN{1}}, \cdots, n_{\textrm{K}'}), [1_{\RN{1}}, \cdots, n_{\RN{1}}]^{\RN{1}},\cdots, [1_{\textrm{K}'}, \cdots, n_{\textrm{K}'}]^{{\textrm{K}'}}, o^{1_{\RN{1}}}, \cdots, o^{g_{\textrm{K}'}}|s^{{1}}, \cdots, s^{{n}      })$ is decomposed entirely into a linear combination of probabilities of the form $p(\kappa_*(1_{\RN{1}}, \cdots, n_{\textrm{K}'}),  (\mathcal{K}' \ncp 1_{rest}), \cdots ,  (\mathcal{K}' \ncp r_{rest}), o^{1_{\RN{1}}}, \cdots, o^{g_{\textrm{K}'}}|
 s^{{1}}, \cdots, s^{{n}      })$, $1\leq r \leq l$, which by the definition of causality are independent of  $s^{(g+1)_{\textrm{K}'}}, \cdots, s^{n_{\textrm{K}'}}$. This completes the proof of Lemma \ref{lem:1}. \\
\end{pr}

\begin{pr}\label{pr:thm:1}
\textbf{Theorem \ref{thm:1}}. The necessity of the form (\ref{Nparty}) follows from Proposition \ref{prop:2.3}. Indeed, let $s^{\mathcal{M}}$ and $o^{\mathcal{M}}$ denote the collection of settings and outcomes, respectively, of the local experiments in a subset $\mathcal{M}\subset \mathcal{S}$. In terms of the set of parties $\mathcal{X} $ that are first, the probabilities of a causal process  $\mathcal{W}_{c}^{\mathcal{S}}$ can most generally be expanded  
\begin{gather}
 p_c^{\mathcal{S}} (o^{\mathcal{S}}| s^{\mathcal{S}}) = \sum_{\mathcal{X} \subset \mathcal S, \mathcal{X} \neq \{\null \}}  p([\mathcal{X}]^{\RN{1} } |s^{\mathcal{S}}) p(o^{\mathcal{X}}| s^{\mathcal{S}}, [\mathcal{X}]^{\RN{1} } ) p(o^{\mathcal{S}\backslash \mathcal{X} }|s^{\mathcal{S}\backslash \mathcal{X} }, s^{\mathcal{X}}, o^{\mathcal{X}}, [\mathcal{X}]^{\RN{1} }  ).
\end{gather}
But, as noted earlier, Proposition \ref{prop:2.3} implies that $p([\mathcal{X}]^{\RN{1} } |s^{\mathcal{S}}) = p([\mathcal{X}]^{\RN{1} } )$, and that $p(o^{\mathcal{X}}| s^{\mathcal{S}}, [\mathcal{X}]^{\RN{1} } ) =  p([\mathcal{X}]^{\RN{1} } , o^{\mathcal{X}}| s^{\mathcal{S}})/ p([\mathcal{X}]^{\RN{1} }|s^{\mathcal{S}} )  = p([\mathcal{X}]^{\RN{1} } , o^{\mathcal{X}}| s^{\mathcal{X}})/ p([\mathcal{X}]^{\RN{1} } )   =  p_{ns}^{\mathcal{X}} (o^{\mathcal{X}}| s^{\mathcal{X}}, [\mathcal{X}]^{\RN{1} } ) $ are the probabilities of a non-signaling process for ${\mathcal{X}} $. We therefore have
\begin{gather}
 \mathcal{W}_{c}^{\mathcal{S}} = \sum_{\mathcal{X} \subset \mathcal S, \mathcal{X} \neq \{\null \}} p_{\mathcal{X}} \mathcal{W}^{\mathcal{S} \backslash \mathcal{X} |\mathcal{X} }\circ \mathcal{W}_{ns}^\mathcal{X} 
\label{Nparty1},
\end{gather}
where $p_{\mathcal{X}} = p([\mathcal{X}]^{\RN{1} } )$. Next, if $\mathcal{X} \neq \mathcal{S}$, we can similarly expand the probabilities of the process $\mathcal{W}^{\mathcal{S} \backslash \mathcal{X} |\mathcal{X} }$ in terms of the set of parties $\mathcal{Y} $ that are second:
\begin{gather}
 p(o^{\mathcal{S}\backslash \mathcal{X} }|  s^{\mathcal{S}\backslash \mathcal{X}} , s^{\mathcal{X}}, o^{\mathcal{X}}, [\mathcal{X}]^{\RN{1} } ) = \sum_{\mathcal{Y} \subset \mathcal S, \mathcal{Y} \neq \{\null \}}  p([\mathcal{Y}]^{\RN{2} } |s^{\mathcal{S}}, o^{\mathcal{X}}, [\mathcal{X}]^{\RN{1} }) p(o^{\mathcal{Y}}| s^{\mathcal{S}}, o^{\mathcal{X}}, [\mathcal{X}]^{\RN{1} }, [\mathcal{Y}]^{\RN{2} })      \nonumber\\
\times p(o^{\mathcal{S\backslash(X\cup Y)}}| s^{\mathcal{S}}, o^{\mathcal{X}}, o^{\mathcal{Y}}, [\mathcal{X}]^{\RN{1} }, [\mathcal{Y}]^{\RN{2} } ).
\end{gather}
Again, from Proposition \ref{prop:2.3} we have that $p([\mathcal{Y}]^{\RN{2} } |s^{\mathcal{S}}, o^{\mathcal{X}}, [\mathcal{X}]^{\RN{1} }) =    p( [\mathcal{X}]^{\RN{1} }, [\mathcal{Y}]^{\RN{2} }, o^{\mathcal{X}} |s^{\mathcal{S}})/$ $p( [\mathcal{X}]^{\RN{1} }, o^{\mathcal{X}}|s^{\mathcal{S}}) =     p( [\mathcal{X}]^{\RN{1} }, [\mathcal{Y}]^{\RN{2} } , o^{\mathcal{X}}|s^{\mathcal{X}})/p( [\mathcal{X}]^{\RN{1} },o^{\mathcal{X}}|s^{\mathcal{X}}) = p([\mathcal{Y}]^{\RN{2} } |s^{\mathcal{X}}, o^{\mathcal{X}}, [\mathcal{X}]^{\RN{1} })$. Similarly, $ p(o^{\mathcal{Y}}| s^{\mathcal{S}}, o^{\mathcal{X}}, [\mathcal{X}]^{\RN{1} }, [\mathcal{Y}]^{\RN{2} }) = p( [\mathcal{X}]^{\RN{1} }, [\mathcal{Y}]^{\RN{2} }, o^{\mathcal{X}} , o^{\mathcal{Y}} | s^{\mathcal{S}})/ p ( [\mathcal{X}]^{\RN{1} }, [\mathcal{Y}]^{\RN{2} } ,o^{\mathcal{X}}  | s^{\mathcal{S}}) =  p( [\mathcal{X}]^{\RN{1} }, [\mathcal{Y}]^{\RN{2} }, o^{\mathcal{X}} , o^{\mathcal{Y}} | s^{\mathcal{X}}, s^{\mathcal{Y}})/ p ( [\mathcal{X}]^{\RN{1} }, [\mathcal{Y}]^{\RN{2} } ,o^{\mathcal{X}}  | s^{\mathcal{X}}) = p_{ns}^{\mathcal{Y}} (o^{\mathcal{Y}}| s^{\mathcal{Y}}, s^{\mathcal{X}}, o^{\mathcal{X}}, [\mathcal{X}]^{\RN{1} }, [\mathcal{Y}]^{\RN{2} }) $ are the probabilities of a non-signaling process for $\mathcal{Y}$ for each value of $(s^{\mathcal{X}}, o^{\mathcal{X}})$. In other words, we obtain that for each value of the events $(s^{\mathcal{X}}, o^{\mathcal{X}})$ in ${\mathcal{X}}$, $\mathcal{W}^{\mathcal{S} \backslash \mathcal{X} |\mathcal{X} }$ has the form (\ref{Nparty1}). The argument is completely analogous for the next conditional process that appears, $ \mathcal{W}^{\mathcal{S} \backslash (\mathcal{X}\cup  \mathcal{Y}) |\mathcal{X}\cup  \mathcal{Y} }  $, which, if nontrivial, can be expanded in terms of the different possibilities for the third consecutive set, and so on. This can be continued until we reach the last consecutive set in every possible grouping of the parties into consecutive sets, which proves the necessity of the form (\ref{Nparty}). 

To prove sufficiency, we will show that if every process of the form (\ref{Nparty}) is causal for $1\leq n\leq n'$, then the same must hold for $n=n'+1$. The general case then follows by induction from this and the fact that a monopartite ($n=1$) process, which has the form  (\ref{Nparty}), is causal. Let an $n'$-partite process have the form (\ref{Nparty}), i.e., its probabilities can be written
\begin{gather}
p( \kappa(\mathcal{S}), o^{   \mathcal{S}}|s^{\mathcal{S} }) = \sum_{\mathcal{X} \subset \mathcal S, \mathcal{X} \neq \{\null \}} p_{\mathcal{X} } p_{ns} (o^{\mathcal{X}}|s^{\mathcal{X}}) p( o^{   \mathcal{S} \backslash \mathcal{X}}|s^{\mathcal{S} \backslash \mathcal{X} } , s^{\mathcal{X} }, o^{\mathcal{X} } ),
\end{gather}
where the probabilities $p( o^{   \mathcal{S} \backslash \mathcal{X}}|s^{\mathcal{S} \backslash \mathcal{X} } , s^{\mathcal{X} }, o^{\mathcal{X} } )$ describe a conditional process $\mathcal{W}_c^{\mathcal{S} \backslash \mathcal{X} |\mathcal{X} }$, which, if non-trivial, has an analogous form for every possible value of $(s^{\mathcal{X} }, o^{\mathcal{X} } )$. Such a conditional process is therefore causal for every possible value of $(s^{\mathcal{X} }, o^{\mathcal{X} } )$ according to our assumption. This means that there exists a probability distribution $p( \kappa(\mathcal{S} \backslash \mathcal{X} ), o^{   \mathcal{S} \backslash \mathcal{X}}|s^{\mathcal{S} \backslash \mathcal{X} } , s^{\mathcal{X} }, o^{\mathcal{X} } )$, where $\kappa(\mathcal{S} \backslash \mathcal{X} )$ is the causal configurations of the experiments in $\mathcal{S} \backslash \mathcal{X}$, such that $\sum_{ \kappa (\mathcal{S} \backslash \mathcal{X} )}  p( \kappa(\mathcal{S} \backslash \mathcal{X} ), o^{   \mathcal{S} \backslash \mathcal{X}}|s^{\mathcal{S} \backslash \mathcal{X} } , s^{\mathcal{X} }, o^{\mathcal{X} } ) =  p(o^{   \mathcal{S} \backslash \mathcal{X}}|s^{\mathcal{S} \backslash \mathcal{X} } , s^{\mathcal{X} }, o^{\mathcal{X} } )$, which for every fixed $(s^{\mathcal{X} }, o^{\mathcal{X} })$ obeys the causality condition (\ref{causalorderDEF}). We want to show that there exists a distribution $p( \kappa(\mathcal{S}), o^{   \mathcal{S}}|s^{\mathcal{S} })$, where $ \kappa(\mathcal{S})$ is the causal configuration of all experiments $\mathcal{S}$, such that $\sum_{\kappa(\mathcal{S})} p( \kappa(\mathcal{S}), o^{ \mathcal{S}}|s^{\mathcal{S} }) = p( o^{   \mathcal{S}}|s^{\mathcal{S} }) $, which also obeys the causality condition (\ref{causalorderDEF}). The following distribution will be shown to satisfy these desiderata: 
\begin{gather}
p( \kappa(\mathcal{S}), o^{   \mathcal{S}}|s^{\mathcal{S} }) = p_{\mathcal{X} } p_{ns} (o^{\mathcal{X}}|s^{\mathcal{X}}) p( \kappa(\mathcal{S} \backslash \mathcal{X} ), o^{   \mathcal{S} \backslash \mathcal{X}}|s^{\mathcal{S} \backslash \mathcal{X} } , s^{\mathcal{X} }, o^{\mathcal{X} } )\nonumber
\end{gather}

 for $
\kappa(\mathcal{S}) = [\kappa(\mathcal{S} \backslash \mathcal{X} ); i\cp j, \forall i\in \mathcal{X},  \forall j\in \mathcal{S}\backslash \mathcal{X}; i\ind j, \forall i,j\in \mathcal{X} ]$,
\begin{gather}
p( \kappa(\mathcal{S}), o^{   \mathcal{S}}|s^{\mathcal{S} }) =0
\end{gather}

 for $\kappa(\mathcal{S}) \neq [\kappa(\mathcal{S} \backslash \mathcal{X} ); i\cp j, \forall i\in \mathcal{X},  \forall j\in \mathcal{S}\backslash \mathcal{X}; i\ind j, \forall i,j\in \mathcal{X} ]$.\\
According to this distribution, $p_{ \mathcal{X}} = p([\mathcal{X}]^{\RN{1}})$, and the causal configuration of all local experiments for $[ \mathcal{X}]^{\RN{1}}$ is always such that each of the local experiments in $\mathcal{X}$ is in the causal past of all local experiments in $ \mathcal{S}\backslash \mathcal{X}$, while the probability for the causal configuration and outcomes of $\mathcal{S}\backslash \mathcal{X}$ given the events in $\mathcal{X}$ and the settings in $\mathcal{S}\backslash \mathcal{X}$ is $p( \kappa(\mathcal{S} \backslash \mathcal{X} ), o^{   \mathcal{S} \backslash \mathcal{X}}|s^{\mathcal{S} \backslash \mathcal{X} } , s^{\mathcal{X} }, o^{\mathcal{X} } )$. The distribution $p( \kappa(\mathcal{S}), o^{   \mathcal{S}}|s^{\mathcal{S} }) $ has the correct marginal $p( o^{   \mathcal{S}}|s^{\mathcal{S} })$ by construction. To show that it satisfies condition (\ref{causalorderDEF}), we will show that $p( \kappa(\mathcal{S}), o^{   \mathcal{S}}|s^{\mathcal{S} }, [ \mathcal{X}]^{\RN{1}} )$, which equals $ p_{ns} (o^{\mathcal{X}}|s^{\mathcal{X}}) p( \kappa(\mathcal{S} \backslash \mathcal{X} ), o^{   \mathcal{S} \backslash \mathcal{X}}|s^{\mathcal{S} \backslash \mathcal{X} } , s^{\mathcal{X} }, o^{\mathcal{X} } )$ for $\kappa(\mathcal{S}) = [\kappa(\mathcal{S} \backslash \mathcal{X} ); i\cp j, \forall i\in \mathcal{X},  \forall j\in \mathcal{S}\backslash \mathcal{X}; i\ind j, \forall i,j\in \mathcal{X} ]$ and vanishes otherwise, satisfies this condition. The fact that the whole mixture $ p( \kappa(\mathcal{S}), o^{   \mathcal{S}}|s^{\mathcal{S} }) = {\sum_{\mathcal{X} \subset \mathcal S, \mathcal{X} \neq \{\null \}} }  p([ \mathcal{X}]^{\RN{1}}) p( \kappa(\mathcal{S}), o^{   \mathcal{S}}|s^{\mathcal{S} }, [ \mathcal{X}]^{\RN{1}})  $ satisfies it then follows from the linearity of the condition. Consider a given local experiment $l\in \mathcal{S}\backslash \mathcal{X}$. Let $\mathcal{X}'\subset \mathcal{X}$ and $\mathcal{Y}'\subset (\mathcal{S}\backslash \mathcal{X} )\backslash l$. We have $ p (\kappa(\mathcal{X}', \mathcal{Y}', l), l\ncp \mathcal{X}',  l\ncp \mathcal{Y}'  , o^{ \mathcal{X}'}, o^{ \mathcal{Y}'}| s^{\mathcal{S}})  = p (\kappa(\mathcal{X}', \mathcal{Y}', l),  l\ncp \mathcal{Y}'  , o^{ \mathcal{X}'}, o^{ \mathcal{Y}'}| s^{\mathcal{S}}) =  p_{ns} (o^{\mathcal{X}'}|s^{\mathcal{X}'}) p( \kappa(\mathcal{Y}', l ), l\ncp \mathcal{Y}' ,  o^{   \mathcal{Y}' }|s^l, s^{\mathcal{Y}'} , s^{\mathcal{X} }, o^{\mathcal{X} } ) $ for $\kappa(\mathcal{X}', \mathcal{Y}', l) = [\kappa(\mathcal{Y}', l ); i\cp j, \forall i\in \mathcal{X}',  \forall j\in \mathcal{Y}'\cup l; i\ind j, \forall i,j\in \mathcal{X}']$, and $p (\kappa(\mathcal{X}', \mathcal{Y}', l), l\ncp \mathcal{X}',  l\ncp \mathcal{Y}'  , o^{ \mathcal{X}'}, o^{ \mathcal{Y}'}| s^{\mathcal{S}}) =0$ otherwise. But from the fact that $p( \kappa(\mathcal{S} \backslash \mathcal{X} ), o^{   \mathcal{S} \backslash \mathcal{X}}|s^{\mathcal{S} \backslash \mathcal{X} } , s^{\mathcal{X} }, o^{\mathcal{X} } )$ satisfies condition (\ref{causalorderDEF}), it follows that $ p( \kappa(\mathcal{Y}', l ), l\ncp \mathcal{Y}' ,  o^{   \mathcal{Y}' }|s^l, s^{\mathcal{Y}'} , s^{\mathcal{X} }, o^{\mathcal{X} } ) =  p( \kappa(\mathcal{Y}', l ), l\ncp \mathcal{Y}' ,  o^{   \mathcal{Y}' }|s^{\mathcal{Y}'} , s^{\mathcal{X} }, o^{\mathcal{X} } ) $. This proves that $p (\kappa(\mathcal{X}', \mathcal{Y}', l), l\ncp \mathcal{X}',  l\ncp \mathcal{Y}'  , o^{ \mathcal{X}'}, o^{ \mathcal{Y}'}| s^{\mathcal{S}}) = p (\kappa(\mathcal{X}', \mathcal{Y}', l), l\ncp \mathcal{X}',  l\ncp \mathcal{Y}' , o^{ \mathcal{X}'}, o^{ \mathcal{Y}'}| s^{\mathcal{S}\backslash l}) $, which is condition (\ref{causalorderDEF}). Similarly, if we take $l\in \mathcal{X}$, consider two arbitrary subsets $\mathcal{X}'\subset \mathcal{X}\backslash l$, $\mathcal{Y}'\subset \mathcal{S}\backslash \mathcal{X}$. When $\mathcal{Y}' \neq \{ \}$, we have $p (\kappa(\mathcal{X}', \mathcal{Y}', l), l\ncp \mathcal{X}',  l\ncp \mathcal{Y}'  , o^{ \mathcal{X}'}, o^{ \mathcal{Y}'}| s^{\mathcal{S}}) =0$. When  $\mathcal{Y}' = \{ \}$, we have $p (\kappa(\mathcal{X}', \mathcal{Y}', l), l\ncp \mathcal{X}',  l\ncp \mathcal{Y}'  , o^{ \mathcal{X}'}, o^{ \mathcal{Y}'}| s^{\mathcal{S}}) =  p (\kappa(\mathcal{X}', l), l\ncp \mathcal{X}', o^{ \mathcal{X}'}| s^{\mathcal{S}}) = p_{ns} (o^{ \mathcal{X}'}| s^{\mathcal{X}})= p_{ns} (o^{ \mathcal{X}'}|s^{ \mathcal{X}'} ) $, which again proves that $p (\kappa(\mathcal{X}', \mathcal{Y}', l), l\ncp \mathcal{X}',  l\ncp \mathcal{Y}'  , o^{ \mathcal{X}'}, o^{ \mathcal{Y}'}| s^{\mathcal{S}}) = p (\kappa(\mathcal{X}', \mathcal{Y}', l), l\ncp \mathcal{X}',  l\ncp \mathcal{Y}' , o^{ \mathcal{X}'}, o^{ \mathcal{Y}'}| s^{\mathcal{S}\backslash l}) $, i.e., we have seen that condition (\ref{causalorderDEF}) is satisfied for every $l$. This completes the proof of Theorem \ref{thm:1}.\\  
\end{pr}

\begin{pr}\label{pr:prop:3.1}
\textbf{Proposition \ref{prop:3.1}}. The proof follows the idea of the proof for the bipartite case in Ref.~\cite{OCB}. Here, we detail it for the case of three parties. The $n$-partite follows analogously.\\

Expanding the CJ operator of a local CP map in the Hilbert-Schmidt basis, $M^{X_1X_2}= \sum_{\mu \nu} r_{\mu \nu} \sigma^{X_1}_{\mu}\sigma^{X_2}_{\nu}$, $r_{\mu \nu} \in \mathbb {R}$, we observe that the trace-preserving condition $\textrm{Tr}_{X_2} M^{X_1X_2} = \id^{X_1}$ is equivalent to the requirement $r_{00} = \frac{1}{d_{X_2}}$, $r_{i0}=0$ for $i>0$. Thus, CJ operators corresponding to CPTP maps are positive semidefinite operators of the form
\begin{gather}
\label{CPTP}
M^{X_1X_2} = \frac{1}{d_{X_2}}\left(\id+\sum_{i>0}a_{i}\sigma^{X_2}_{i}+\ \sum_{i,j>0}t_{ij}\sigma^{X_1}_{i}\sigma^{X_2}_{j}\right), \hspace{0.1cm}
a_{i}, t_{ij} \in \mathbb {R}.
\end{gather}

It turns out that condition (\ref{W2}) can be equivalently imposed only for operators $M^{X_1X_2}$ of the form (\ref{CPTP}) without the constraint $M^{X_1X_2}\geq 0$. Clearly, an operator $W^{A_1A_2B_1B_2C_1C_2\cdots}$ that satisfies Eq.~(\ref{W2}) for all operators $M^{X_1X_2}$ of the form (\ref{CPTP}) satisfies Eq.~(\ref{W2}) for positive semidefinite operators $M^{X_1X_2}$ of this form in particular. The converse follows from the fact that any operator $M^{X_1X_2}$ of the form (\ref{CPTP}) can be written as a real linear combination of positive semidefinite operators of the form (\ref{CPTP}): $M^{X_1X_2} = \sum_i \alpha_i M_i^{X_1X_2}$, where  $M_i^{X_1X_2}\geq 0$ satisfy (\ref{CPTP}) for all $i$, and $\sum_i \alpha_i =1$, $\alpha_i\in \mathbb{R}$, $\forall i$. We will use this fact to recast condition Eq.~(\ref{W2}) as a statement about the types of non-zero terms in the Hilbert-Schmidt expansion of  $W^{A_1A_2B_1B_2C_1C_2\cdots}$.

In the case of three parties, the expansion of $W^{A_1A_2B_1B_2C_1C_2}$ reads
\begin{gather}
W^{A_1A_2B_1B_2C_1C_2} = \sum_{i,j,k,l,m,n} w_{ijklmn} \sigma^{A_1}_i\sigma^{A_2}_j\sigma^{B_1}_k\sigma^{B_2}_l\sigma^{C_1}_m\sigma^{C_2}_n,\\
w_{ijklmn}\in \mathbb{R}, \forall i,j,k,l,m,n.
\end{gather}
Let us fix $M^{A_1A_2} = \frac{\id^{A_1A_2}}{d_{A_2}}$ and $M^{B_1B_2} = \frac{\id^{B_1B_2}}{d_{B_2}}$, and consider an arbitrary $M^{C_1C_2}$ of the form (\ref{CPTP}). Condition (\ref{W2}) becomes
\begin{gather}
 \frac{1}{d_{A_2}d_{B_2}d_{C_2}}\Tr[W^{A_1A_2B_1B_2C_1C_2}(\id^{A_1A_2}\otimes\id^{B_1B_2}
\otimes (\id^{C_1C_2} +\nonumber\\ \sum_{n>0}c_{n}\sigma^{C_2} +\sum_{mn>0}p_{mn}\sigma^{C_1}_m\sigma^{C_2}_n))] = 1,
\end{gather}
which, using the expansion of the process matrix, becomes
\begin{gather}\label{eq:}
d_{A_1}d_{B_1}d_{C_1}(w_{000000} +\ \sum_{n>0}w_{00000n}c_n +\ \sum_{mn>0}w_{0000mn}p_{mn}) = 1,\\
\forall c_n,\ p_{mn} \in \mathbb{R}. \nonumber
\end{gather}
This implies $w_{000000} = \frac{1}{d_{A_1}d_{B_1}d_{C_1}}$ and $w_{00000n} = w_{0000mn} = 0$, $\forall m,n>0$.

Likewise, by fixing $M^{A_1A_2} = \frac{\id^{A_1A_2}}{d_{A_2}}$ and $M^{C_1C_2} = \frac{\id^{C_1C_2}}{d_{C_2}}$, and considering an arbitrary $M^{B_1B_2}$ of the form (\ref{CPTP}), we obtain $w_{000l00} = w_{00kl00} = 0$ for all $k,l>0$, while by fixing $M^{B_1B_2} = \frac{\id^{B_1B_2}}{d_{B_2}}$ and $M^{C_1C_2} = \frac{\id^{C_1C_2}}{d_{C_2}}$, and considering an arbitrary $M^{A_1A_2}$ of the form (\ref{CPTP}), we obtain $w_{0j0000} = w_{ij0000} = 0$ for all $i,j>0$.

Now, if we fix only $M^{A_1A_2} = \frac{\id^{A_1A_2}}{d_{A_2}}$, and we use the previously obtained constraints, we obtain $w_{000l0n} = w_{00kl0n} = w_{000lmn} = w_{00klmn} = 0$ (each of these coefficients can be shown to vanish by suitably choosing the parameters in $M^{B_1B_2} $ and $M^{C_1C_2} $ in order to select only the term with that coefficient). Then, if we fix $M^{B_1B_2} = \frac{\id^{B_1B_2}}{d_{B_2}}$, we obtain $w_{0j000n} = w_{0j00mn} = w_{ij000n} = w_{ij00mn} = 0$. Similarly, if we fix $M^{C_1C_2} = \frac{\id^{C_1C_2}}{d_{C_2}}$, we obtain $w_{0j0l00} = w_{0jkl00} = w_{ij0l00} = w_{ijkl00} = 0$.

\begin{table}
\caption{\label{tbl:forbidden}The types of terms that are forbidden in a tripartite process matrix $W^{A_1A_2B_1B_2C_1C_2}$.}
\footnotesize\rm
\begin{tabular*}{\textwidth}{@{}l*{15}{@{\extracolsep{0pt plus12pt}}l}}
\hline
$C_2$&$C_1C_2$&$B_2$&$B_2C_2$  \\ \hline
$B_2C_1C_2$&$B_1B_2$&$B_1B_2C_2$&$B_1B_2C_1C_2$\\ \hline
$A_2$ & $A_2C_2$ & $A_2C_1C_2$ & $A_2B_2$  \\ \hline
$A_2B_2C_2$ &$A_2B_2C_1C_2$ & $A_2B_1B_2$ & $A_2B_1B_2C_2$  \\ \hline
 $A_2B_1B_2C_1C_2$&$A_1A_2$&$A_1A_2C_2$ & $A_1A_2C_1C_2$   \\ \hline
 $A_1A_2B_2$ & $A_1A_2B_2C_2$&$A_1A_2B_2C_1C_2$ & $A_1A_2B_1B_2$ \\ \hline
 $A_1A_2B_1B_2C_2$  & $A_1A_2B_1B_2C_1C_2$ \\
\hline
  \end{tabular*}
\end{table}

Finally, we impose condition (\ref{W2}) for arbitrary $M^{A_1A_2}$, $M^{B_1B_2}$, and $M^{C_1C_2}$, of the form (\ref{CPTP}). Using the constraints obtained from the special cases above, we obtain $w_{0j0l0n} = w_{0j0lmn} = w_{0jkl0n} = w_{0jklmn} =  w_{ij0l0n} =  w_{ij0lmn} = w_{ijkl0n} = w_{ijklmn} = 0$. Thus, we have shown that all coefficients $w_{ijklmn}$, except for $w_{000000}$, that may appear in the result of taking the trace of $W^{A_1A_2B_1B_2C_1C_2}$  with a general combination of $M^{A_1A_2}$, $M^{B_1B_2}$, $M^{C_1C_2}$ of the form (\ref{CPTP}), must vanish. This is also a sufficient condition for the normalization condition (\ref{W2}) to hold. All these forbidden terms for a process matrix are listed in Table (\ref{tbl:forbidden}).\\
\end{pr}

\begin{pr}\label{pr:prop:3.2}
\textbf{Proposition \ref{prop:3.2}}.

%An $n$-partite process matrix permits signaling from, say, ($1$ or $2$ or $\cdots$ or $k$) to ($k+1$ or $k+2$ or $\cdots$ or $n$) if and only if it contains at least one term with: {(i)} a nontrivial $\sigma$ operator on $i_2$ for some $i=1,\cdots, k$; (ii) no nontrivial $\sigma$ operator on $j_1$ together with $\id$ on $j_2$ for any $j=1,\cdots,k$. 

Explicitly, by the definition of (no) signaling \eqref{Defsignaling} and the expression for the probabilities of a process in terms of the process matrix \eqref{Wmain}, there is no signaling from ($1$ and $2$ and $\cdots$ and $k$) to ($k+1$ and ${k+2}$ and $\cdots$ and $n$) if and only if 
\begin{gather}
p(o^1,\cdots, o^{k+1}|\{\mathcal{M}^{1}_{o^{1}} \}, \cdots,  \{\mathcal{M}^{n}_{o^{n}} \} ) \equiv \sum_{ o^{1},\cdots, o^{k} }  \Tr\left[W^{1_11_2 \cdots n_1n_2}\left(M^{1_11_2}_{o^{1}}\otimes \cdots \otimes M^{n_1n_2}_{o^{n}} \right)\right]\notag\\
\equiv \Tr\left[W^{1_11_2 \cdots n_1n_2}\left(\overline{M}^{1_11_2}\otimes \cdots \otimes \overline{M}^{k_1k_2} \otimes M^{(k+1)_1(k+1)_2}_{o^{k+1}} \otimes \cdots\otimes   M^{n_1n_2}_{o^{n}} \right)\right]\notag\\
= \Tr\left[W^{(k+1)_1(k+1)_2 \cdots n_1n_2}\left(M^{(k+1)_1(k+1)_2}_{o^{k+1}} \otimes \cdots\otimes   M^{n_1n_2}_{o^{n}} \right)\right]\notag \equiv  p(o^{k+1},\cdots, o^{k+1}|\{\mathcal{M}^{k+1}_{o^{k+1}} \}, \cdots,  \{\mathcal{M}^{n}_{o^{n}} \} ) , 
\end{gather}
for all local quantum operations $\{\mathcal{M}^{1}_{o^{1}} \}, \cdots,  \{\mathcal{M}^{n}_{o^{n}} \}$, where $\overline{M}^{X^i_1X^i_2} = \sum_{o^{X^i}} {M}^{X^i_1X^i_2}_ {o^{X^i}}$, $\forall i$. Here, the operator $W^{(k+1)_1(k+1)_2 \cdots n_1n_2}$ is given by  
\begin{gather}
W^{(k+1)_1(k+1)_2 \cdots n_1n_2} =\frac{ \Tr_{1_11_2 \cdots k_1k_2} W^{1_11_2 \cdots n_1n_2}} { d_{1_1} \cdots d_{k_1}  },
\end{gather}
which is obtained for the case where $\overline{M}^{i_1i_2} = \id^{i_1i_2}/ d_{i_1}$, $\forall i=1,\cdots,k$. This condition is equivalent to the condition that 
\begin{gather}
\Tr_{1_11_2 \cdots k_1k_2} \left[ W^{1_11_2 \cdots n_1n_2} \left(\overline{M}^{1_11_2}\otimes \cdots \otimes \overline{M}^{k_1k_2} \otimes \id^{(k+1)_1(k+1)_2 \cdots}\right)   \right] =  W^{(k+1)_1(k+1)_2 \cdots n_1n_2}  , \hspace{0.4cm} \forall   \overline{M}^{1_11_2}, \cdots, \overline{M}^{k_1k_2}, \label{NSW1}
\end{gather}
where $\overline{M}^{1_11_2}, \cdots, \overline{M}^{k_1k_2}$ are the CJ operators of CPTP maps (this is because any linear operator $V^{(k+1)_1(k+1)_2 \cdots n_1n_2}  $ is fully determined by the values of $\Tr \left[ V^{(k+1)_1(k+1)_2 \cdots n_1n_2}  \left(M^{(k+1)_1(k+1)_2}\otimes \cdots\otimes   M^{n_1n_2}\right)   \right] $ for all possible $M^{(k+1)_1(k+1)_2} \geq 0, \cdots, M^{n_1n_2}\geq 0$). To analyze the role of the different types of terms in satisfying or violating condition \eqref{NSW1}, consider the representation of $W^{1_11_2 \cdots n_1n_2}$ as a linear combination of Hilbert-Schmidt terms of different types and the contribution that each such term makes to the quantity on the left-hand side of Eq.~\eqref{NSW1}. Assume that $W^{1_11_2 \cdots n_1n_2}$ contains only terms of the types stated in Proposition \ref{prop:3.2}. The identity term is such a term. When the identity term is partially traced with any combination of local CPTP maps $\overline{M}^{1_11_2}, \cdots, \overline{M}^{k_1k_2}$, it yields exactly the right-hand side of Eq.~\eqref{NSW1}. From the rest of the terms that satisfy the condition in Proposition \ref{prop:3.2}, we can distinguish two types. The first type are those that have a nontrivial $\sigma$ operator on $i_1$ and $\id$ on $i_2$ for some $i=1,\cdots, k$. They yield zero when partially traced with any combination of local CPTP maps $\overline{M}^{1_11_2}, \cdots, \overline{M}^{k_1k_2}$, since a CPTP map $\overline{M}^{i_1i_2}$ does not contain terms of type $i_1$ (which is necessary to get a non-trivial partial trace with the term in question). The second type of terms are those that do not have any nontrivial $\sigma$ operator on any of the systems $i_1$ and $i_2$, $i=1,\cdots, k$, and hence, when partially traced with any combination of local CPTP maps $\overline{M}^{1_11_2}, \cdots, \overline{M}^{k_1k_2}$, only the $\id$ components of those CPTP maps contribute to the result, which by definition yields the right-hand side of Eq.~\eqref{NSW1}. Therefore, if $W^{1_11_2 \cdots n_1n_2}$ contains only the types of terms stated in Proposition \ref{prop:3.2}.

To prove the reverse, assume that $W^{1_11_2 \cdots n_1n_2}$ contains at least one term whose restriction onto $1_11_2\cdots k_1k_2$ is not a valid term for a process matrix for $\{1,\cdots, k\}$. Every such term has the form $O^{\alpha^{1}_1} \otimes \sigma^{\alpha^{1}_2} \otimes \cdots \otimes O^{\alpha^{m}_1} \otimes \sigma^{\alpha^{m}_2} \otimes    \id^{{\alpha^{m+1}_1}\alpha^{m+1}_2} \otimes \cdots \otimes \id^{\alpha^k_1 \alpha^k_2}    \otimes  Q^{(k+1)_1 (k+1)_2\cdots {n}_1{n}_2}$, where $\alpha^i$, $i=1,\cdots, k$, are different numbers from $1$ to $k$, $1\leq m \leq k$, $O^{\alpha^{i}_1}$ is either the identity or some nontrivial $\sigma$ operator on $\alpha^{i}_1 $, $\sigma^{\alpha^{i}_2}$ is a nontrivial $\sigma$ operator on $\alpha^{i}_2$, and $Q^{(k+1)_1 (k+1)_2\cdots {n}_1{n}_2}$ is a non-zero operator on $(k+1)_1 (k+1)_2\cdots {n}_1{n}_2$, which is proportional to a tensor product of nontrivial $\sigma$ operators and $\id$ on the different subsystems, such that the whole term is an allowed term for a process matrix. We want to show that if such a term is present in the process matrix, Eq.~\eqref{NSW1} can be violated for a specific choice of the local CPTP maps $\overline{M}^{1_11_2}, \cdots, \overline{M}^{k_1k_2}$. Out of all such terms, consider one for which $m$ has the smallest value (there may be more than one of these). Consider the following choice of local CPTP maps constructed based on this term: for $i=m+1,\cdots,k$, choose $\overline{M}^{\alpha^i_1 \alpha^i_2} = \frac{1}{d_{\alpha^i_1}} \id^{\alpha^i_1 \alpha^i_2}$, and for $j=1,\cdots,m$, choose $\overline{M}^{\alpha^j_1 \alpha^j_2} = \frac{1}{d_{ \alpha^j_1}} (\id^{\alpha^j_1 \alpha^j_2} +\    {\epsilon_{\alpha^j}}  O^{\alpha^{j}_1} \otimes \sigma^{\alpha^{j}_2} ) $, where $\epsilon_{\alpha^j}> 0$ is such that $(\id^{\alpha^j_1 \alpha^j_2} +\ {\epsilon_{\alpha^j}} O^{\alpha^{j}_1} \otimes \sigma^{\alpha^{j}_2} )\geq 0$ (this can always be ensured for sufficiently small non-zero $\epsilon_{\alpha^j}$). Consider the Hilbert-Schmidt expansion of the tensor product $\overline{M}^{1_11_2}\otimes \cdots \otimes \overline{M}^{k_1k_2}$. From this expansion, only the identity term and the term proportional to $O^{\alpha^{1}_1 } \otimes \sigma^{\alpha^{1}_2 } \otimes \cdots \otimes O^{\alpha^{m}_1 } \otimes \sigma^{\alpha^{m}_2 } $ will survive when we plug  $\overline{M}^{1_11_2}\otimes \cdots \otimes \overline{M}^{k_1k_2}$ in the expression on the left-hand side of Eq.~\eqref{NSW1}. This is because in order for any other term to survive, it would be necessary that $W^{1_11_2 \cdots n_1n_2}$ contains a term of a form similar to $O^{\alpha^{1}_1} \otimes \sigma^{\alpha^{1}_2} \otimes \cdots \otimes O^{\alpha^{m}_1} \otimes \sigma^{\alpha^{m}_2} \otimes    \id^{\alpha^{m+1}_1 \alpha^{m+1}_2} \otimes \cdots \otimes \id^{\alpha^k_1 \alpha^k_2}  \otimes  Q^{(k+1)_1 (k+1)_2\cdots {n}_1 {n}_2}$ but with a smaller value of $m$ than the one we have chosen, which contradicts the assumption that we have chosen the smallest value. Plugging $\overline{M}^{1_11_2}\otimes \cdots \otimes \overline{M}^{k_1k_2}$ in the expression on the left-hand side of Eq.~\eqref{NSW1} therefore yields $W^{(k+1)_1(k+1)_2 \cdots n_1n_2} + \epsilon Q^{(k+1)_1 (k+1)_2\cdots {n}_1{n}_2}$ for some $\epsilon \geq 0$, which is different from the right-hand side of Eq.~\eqref{NSW1}. This completes the proof of Proposition \ref{prop:3.2}. \\
\end{pr}

\begin{pr}\label{pr:prop:3.3}
\textbf{Proposition \ref{prop:3.3}}. The fact that this form is sufficient for the process matrix to be ECS is obvious because if this is true for each of the individual terms, any extension ${W}_{ecs}^{A_1A_2B_1B_2C_1C_2} \otimes \rho^{A_1'B_1'C_1'} = q_1 W_{ecs; (A,B)\ncp C}^{A_1A_2B_1B_2C_1C_2} \otimes \rho^{A_1'B_1'C_1'}   +q_2 W_{ecs; (A,C)\ncp B}^{A_1A_2B_1B_2C_1C_2}  \otimes \rho^{A_1'B_1'C_1'}  +\ q_3 W_{ecs; (B,C)\ncp A}^{A_1A_2B_1B_2C_1C_2} \otimes \rho^{A_1'B_1'C_1'} $ is also causally separable. The fact that it is necessary can be seen as follows. Let us choose $\rho^{A_1'B_1'C_1'} $ which is a tensor product of three bipartite maximally entangled states of the type used in the `teleportation' argument, one shared between Alice and Bob, the other one between Alice and Charlie, and the third one between Bob and Charlie. For this particular ancilla, it must be possible to write the extended process in the form 
\begin{gather}
{W}_{ecs}^{A_1A_2B_1B_2C_1C_2} \otimes \rho^{A_1'B_1'C_1'} = q_1 W_{1}^{A_1A_2B_1B_2C_1C_2} \otimes \rho^{A_1'B_1'C_1'}   +q_2 W_{2}^{A_1A_2B_1B_2C_1C_2}  \otimes \rho^{A_1'B_1'C_1'}  +\ q_3 W_{3}^{A_1A_2B_1B_2C_1C_2} \otimes \rho^{A_1'B_1'C_1'},
\end{gather} 
where $W_{1}^{A_1A_2B_1B_2C_1C_2} \otimes \rho^{A_1'B_1'C_1'}$ is causally separable and compatible with $ (A,B)\ncp C$, $W_{2}^{A_1A_2B_1B_2C_1C_2} \otimes \rho^{A_1'B_1'C_1'}$ is causally separable and compatible with $ (A,C)\ncp B$, and $W_{3}^{A_1A_2B_1B_2C_1C_2} \otimes \rho^{A_1'B_1'C_1'}$ is causally separable and compatible with $ (B,C)\ncp A$. (This is because the state $\rho^{A_1'B_1'C_1'} $ is pure.) But for each of these terms, we can perform the `teleportation' argument exploiting the respective maximally entangled bipartite state contained in $\rho^{A_1'B_1'C_1'}$, proving that $W_{1}^{A_1A_2B_1B_2C_1C_2} $ has the form we obtained for $W_{ecs; (A,B)\ncp C}^{A_1A_2B_1B_2C_1C_2} $, $W_{2}^{A_1A_2B_1B_2C_1C_2}$ has the form we obtained for $W_{ecs; (A,C)\ncp B}^{A_1A_2B_1B_2C_1C_2}  $, and $W_{3}^{A_1A_2B_1B_2C_1C_2} $ has the form we obtained for $W_{ecs; (B,C)\ncp A}^{A_1A_2B_1B_2C_1C_2} $. This completes the proof.\\
\end{pr}

\end{document}